\documentclass[usenatbib]{aa}
\usepackage{graphicx}
\usepackage[varg]{txfonts}
\usepackage[english,english]{babel}
\usepackage{amsmath}
\usepackage{amssymb,amsfonts,textcomp}
\usepackage{array}
\usepackage{supertabular}
\usepackage{hhline}
\usepackage[usenames]{color}
\usepackage{hyperref}
\hypersetup{colorlinks=true, linkcolor=black, citecolor=blue, filecolor=blue, urlcolor=blue}
\usepackage{xspace}
\bibpunct{(}{)}{;}{a}{}{,}

\def\gtrsim{\lower.5ex\hbox{$\; \buildrel > \frac \sim \;$}}

\newcommand{\msun}{\mbox{M$_\odot$}}

\newcommand{\hagn}{\mbox{{\sc \small Horizon-AGN}}}

\newcommand{\nh}{\mbox{{\sc \small NewHorizon}}}
\newcommand{\nhs}{\mbox{{\sc \small NewHorizon\,}}}
\newcommand{\galactica}{\mbox{{\sc \small Galactica}}}
\newcommand{\galacticas}{\mbox{{\sc \small Galactica\,}}}

\newcommand{\ramses}{\mbox{{\sc \small RAMSES}}}

\newcommand{\PSI}{$\Psi_{\mathrm{star,BH}}$}
\newcommand{\PSIS}{$\Psi_{\mathrm{star,BH}}$\,}
\newcommand{\COSPSI}{cos($\Psi_{\mathrm{star,BH}}$)}
\newcommand{\COSPSIS}{cos($\Psi_{\mathrm{star,BH}}$)\,}
\newcommand{\PSIGAS}{$\Psi_{\mathrm{gas,BH}}$}

\newcommand{\PSISTARGAS}{$\Psi_{\mathrm{star,gas}}$}

\newcommand{\JBH}{$\boldsymbol{J}$$_{\mathrm{\bf{BH}}}$}
\newcommand{\JBHS}{$\boldsymbol{J}$$_{\mathrm{\bf{BH}}}$\,}
\newcommand{\JSTAR}{$\boldsymbol{J}$$_{\mathrm{\bf{star}}}$}
\newcommand{\JSTARS}{$\boldsymbol{J}$$_{\mathrm{\bf{star}}}$\,}
\newcommand{\JGAS}{$\boldsymbol{J}$$_{\mathrm{\bf{gas}}}$}
\newcommand{\JGASS}{$\boldsymbol{J}$$_{\mathrm{\bf{gas}}}$\,}

\newcommand{\MASSLIM}{$2\times$10$^4$ M$_\odot$}
\newcommand{\tuniv}{$T_{\mathrm{Universe}}$}
\newcommand{\fdm}{$f_{\mathrm{BH,merger}}$}
\newcommand{\dsbh}{$d_{\mathrm{star,BH}}$}
\newcommand{\PLP}{P($\lambda$\,|$\Psi$,\,$i_o$)}
\newcommand{\PLPpi}{P($\lambda$\,|$\Psi$,\,$i_o$=$\pi/2$)}

\definecolor{grey}{rgb}{0.75,0.75,0.75}
\definecolor{Orange}{rgb}{1.0,0.5,0.15}
\definecolor{brown}{rgb}{0.7,0.25,0.0}
\definecolor{pink}{rgb}{1.0,0.5,0.5}
\definecolor{darkerred}{rgb}{0,0.5,0.5}
\definecolor{darkerblue}{rgb}{0,0,0.8}
\definecolor{lightblue}{rgb}{0.12, 0.56, 1.0}
\definecolor{Blue}{rgb}{0,0.08,0.65}
\definecolor{Red}{rgb}{0.65,0.08,0.05}
\definecolor{Green}{rgb}{0.15,0.45,0.25}

\begin{document} 

\title{Cosmic evolution of black hole spin and galaxy orientations:\\
  Clues from the NewHorizon and Galactica simulations}

\titlerunning{BH-galaxy spin orientation}
\authorrunning{S. Peirani et al.}

\author{
S\'ebastien Peirani\inst{1,2,3} \and
Yasushi Suto\inst{3,4,5} \and 
Ricarda S. Beckmann\inst{6} \and
Marta Volonteri\inst{2} \and 
Yen-Ting Lin\inst{7}\and
Yohan Dubois\inst{2} \and 
Sukyoung K.\,Yi\inst{8} \and
Christophe Pichon\inst{2,9,10} \and
Katarina Kraljic\inst{11} \and
Minjung Park\inst{12} \and
Julien Devriendt\inst{13} \and 
San Han\inst{8} \and
Wei-Huai Chen\inst{7,14}
}

\institute{
Universit\'e C\^ote d'Azur, Observatoire de la C\^ote d'Azur, CNRS, Laboratoire Lagrange, Bd de l'Observatoire,\\
 CS 34229, 06304 Nice Cedex 4, France \\\email{speirani@oca.eu} 
\and Institut d'Astrophysique de Paris, CNRS and Sorbonne Universit\'e, UMR 7095, 98 bis Boulevard Arago, F-75014 Paris, France
\and Department of Physics, School of Science, The University of Tokyo, 7-3-1 Hongo, Bunkyo-ku, Tokyo 113-0033, Japan
\and Research Center for the Early Universe, School of Science, The University of Tokyo, 7-3-1 Hongo, Bunkyo-ku, Tokyo 113-0033, Japan
\and
Research Institute, Kochi University of Technology, Tosa Yamada, Kochi
782-8502, Japan
\and Institute of Astronomy and Kavli Institute for Cosmology, University of Cambridge, Madingley Road, Cambridge, CB3 0HA, UK
\and Institute of Astronomy and Astrophysics, Academia Sinica, No. 1, Section 4, Roosevelt Road, Taipei 10617, Taiwan
\and Department of Astronomy and Yonsei University Observatory, Yonsei University, Seoul 03722, Republic of Korea
\and IPhT, DRF-INP, UMR 3680, CEA, L'Orme des Merisiers, B\^at 774, 91191 Gif-sur-Yvette, France
\and Korea Institute for Advanced Study, 85 Hoegi-ro, Dongdaemun-gu, Seoul 02455, Republic of Korea
\and Observatoire Astronomique de Strasbourg, Universit\'e de Strasbourg, CNRS, UMR 7550, F-67000 Strasbourg, France
\and Center for Astrophysics | Harvard \& Smithsonian, Cambridge, MA, USA
\and
Astrophysics, University of Oxford, Denys Wilkinson Building, Keble Road, Oxford, OX1 3RH, UK
\and
Bienen School of Music, Northwestern University, 70 Arts Cir Dr, Evanston, IL 60208, USA 
}

   \date{Received ...; accepted ...}

 
  \abstract
  { Black holes (BHs) are ubiquitous components of the center of most
    galaxies.  In addition to their mass, the BH spin, through its
    amplitude and orientation, is a key factor in the galaxy formation
    process, as it controls the radiative efficiency of the accretion
    disk and relativistic jets. 
    Using the recent cosmological high-resolution zoom-in simulations, 
    \nhs and \galactica,
    in which the evolution of the BH spin is followed on the fly, we have tracked
    the cosmic history of a hundred BHs with a mass
    greater than $2\times$10$^4$M$_\odot$. For each of them, we have studied the variations
    of the three-dimensional angle ($\Psi$) subtended between the BH
    spins and the angular momentum vectors of their host galaxies (estimated from the stellar component).
    The analysis of the individual evolution of the most massive BHs
    suggests that they are generally passing by three different
    regimes. First, for a short period after their birth, low-mass BHs
    (M$_{\mathrm{BH}}<$3$\times$10$^4$M$_\odot$) are rapidly spun up by gas
    accretion and their spin tends to be aligned with their host
    galaxy spin. Then follows a second phase in which the accretion of
    gas onto low-mass BHs
    (M$_{\mathrm{BH}}\lesssim$10$^5$M$_\odot$) is quite chaotic and inefficient, reflecting the
    complex and disturbed morphologies of forming proto-galaxies at
    high redshifts.  
    The variations of $\Psi$ are rather erratic during this phase
    and are mainly driven by the rapid changes of the direction of the 
    galaxy angular momentum.
    Then, in
    a third and long phase, BHs are generally well settled in the
    center of galaxies around which the gas accretion becomes much
    more coherent (M$_{\mathrm{BH}}>$10$^5$ M$_\odot$).  In this case, the BH spins
    tend to be well aligned with the angular momentum of their host
    galaxy and this configuration is generally stable even
    though BH merger episodes can temporally induce misalignment.
    We even find a few cases of BH-galaxy spin anti-alignment
      that lasts for a long time in which the gas component is
      counter-rotating with respect to the stellar component.  We
    have also derived the distributions of cos($\Psi$) at different
    redshifts and found that BHs and galaxy spins are generally
    aligned.  Our analysis suggests that the fraction of BH-galaxy
    pairs with low $\Psi$ values reaches maximum at z$\sim$4-3, and
    then decreases until z$\sim$1.5 due to the high BH-merger rate.
    Afterward, it remains almost constant probably due to the fact that BH
    mergers becomes rare, except for a slight increase at late times.
    Finally, based on a Monte Carlo method, we also predict
    statistics for the 2-d projected spin-orbit angles $\lambda$.  In
    particular, the distribution of $\lambda$ traces the
    alignment tendency well in the three-dimensional analysis. Such
    predictions provide an interesting background for future
    observational analyses. }


   \maketitle
%
\section{Introduction}

It is now well established that the evolution of supermassive black holes (BHs) and
their host galaxy is intimately related
\citep[e.g.,][]{silk+98,granato+04}.  The existence of strong
scaling relations between BH masses and galaxy properties, such as the
velocity dispersion of their host bulge
\citep{magorrian+98,merritt&ferrarese01a,merritt&ferrarese01b,ferrarese&merrit00,hu08,mcconnel+13,vandenbosch16,batiste+17,baldassare+20},
the stellar mass of their host bulge
\citep{marconi+03,haring+04,saglia+16,sahu+19,zhao+21}, and the total
stellar mass
\citep{cisternas+11,simmons+11,reines+15,davis+19,sahu+19,ding+20,bennert+21}
does indeed suggest that BHs coevolve with their host galaxy.

Also, due to matter accretion, central BHs in galaxies often exhibit
powerful jets and winds, commonly referred to as active galactic
nuclei (AGNs). The resulting AGN feedback can heat and/or expel the
surrounding gas from the center of galaxies. According to theoretical
works, this mechanism is supposed to play an important role in reshaping
the gas, stellar, and dark matter distributions
in the central and outer parts of their host  
\citep[e.g.,][and references therein]{peirani+08,duffy+10,martizzi+13,  dubois-seb16,peirani+17,peirani19,felipe+21} and in quenching the star formation in massive galaxies,
producing more realistic galaxy population that are broadly consistent
with observations
\citep[e.g.,][]{dimatteo+05,springel+05,schawinski+06,croton+06,sijacki+07,booth-schaye09,dubois+12,choi+15,kurinchi+23}.

Black holes are generally characterized by two classical
properties\footnote{A third one, the electric charge, is assumed to be
negligible in most astrophysical settings \citep{gurlebeck15}.}: their mass $M_{\rm BH}$  and their angular momentum vector 
\JBH. In addition, it is customary to introduce the
dimensionless spin parameter $a$ assuming a Kerr metric for rotating
bodies:
\begin{equation}
    a \equiv \frac{cJ_{\mathrm{BH}}}{GM^2_{\mathrm{BH}}},
\end{equation}
where $c$ is the speed of light and $G$ is the gravitational constant.
Although one naturally expects that $a$ ranges from 0 to 1, it is not
rare to find in the literature $a$ ranges from -1 to 1 which gives
additional information on how the surrounding gas is accreted onto the
BH. If the gas accretion disk does indeed have a prograde rotation, a>0, while a<0 for retrograde rotation.

From their initial values, both the BH the mass and spin 
evolve according to successive gas accretion phases and BH merger episodes.
While the BH mass increases through cosmic time
(though a small amount of mass is lost in the emission of gravitational waves during BH-BH mergers), the spin magnitude may
decrease via different mechanisms
such as gas accretion \citep{bardeen70},
BH-mergers \citep{rezzolla+08,barausse+2009}
and energy extraction by feedback \citep{blandford+77}.
However, the long-term evolution of the BH mass and spin
is not independent but generally connected.

The amplitude and orientation of the BH spin control the efficiency of
the conversion of the accreted gas into energy as
\begin{equation}
\label{equ:equ1}
    L_{\mathrm{BH}} = \epsilon_r(a)\dot{M}_{\mathrm{BH}}c^2 ,
\end{equation}
where $L_{\mathrm{BH}}$ is the BH accretion luminosity, $\dot{M}_{\mathrm{BH}}$ the BH mass
accretion rate, $c$ the speed of light, and $\epsilon_r(a)$ the
radiative efficiency of the accretion disk which depends on the BH
spin $a$.  Then the accreted mass that has not been converted into
radiation, $\Bigl(1-\epsilon_r(a)\Bigr)\dot{M}_{\mathrm{BH}}$,
contributes to the BH mass growth directly.
Thus, the BH spin is also a fundamental parameter for the jet efficiency
and AGN feedback, which can, in turn,  alter and self-regulate the BH accretion.

The BH spin is therefore a key component, not only for BH physics but
also and more generally in the galaxy formation process.  In addition,
the spin of merging BHs (especially their relative orientation) has a
strong influence on the gravitational wave emission and therefore on
the expected recoil velocity of the merger remnant
\citep{campanelli+07,gonzalez+07,lousto+11,lousto+13}.  
Thus, modeling and characterizing the long-term evolution of the BH spin and its impact on the host galaxies are
a valuable effort for many domains in modern astrophysics.

For this purpose, semi-analytical models and, above all, cosmological
simulations are vital tools in characterizing BH mass growth and spin
evolution.  The task has however proven to be difficult.  The different
physical processes that govern the BH evolution, in particular the gas
accretion and the associated feedback processes, are still poorly
understood and occur at physical scales (<pc) much lower than the
resolution limit of the simulations. Therefore, one needs to employ
empirical prescriptions and sub-grid physics.

While BHs are now commonly incorporated in hydrodynamical simulations
generally as sink particles of a given mass,
their spin is often neglected \citep[e.g.,][]{dimatteo+08,
  booth-schaye09, hagn, illustris, eagle, thesan}.  Nevertheless, 
increasing effort in the last decade has tried to fill the gap by
proposing improved modeling including spin evolution in traditional
cosmological simulations \citep[e.g.,][]{dotti+13, dubois+14a,fiacconi+18, bustamante+19, talbot+21,nhz, hopkins+22,
dong-paez+23,husko+23, rennehan+23} as well as in general
relativistic magneto-hydrodynamical simulations \citep[e.g.,][and
  references therein]{fedrigo+23, cui+23, koudmani+23}.

The bulk of the past effort on the cosmic evolution of black hole spin has
focused almost exclusively on its amplitude and on its relative 
alignment with the gas disk  
using semi-analytical
models \citep[e.g.,][]{volonteri+05,shapiro,king+08,berti+08,
  fanidakis,dotti+13,volonteri+13,sesana+14,griffin+19,izquierdo+20}
or cosmological simulations \citep[e.g.,][]{dubois+14_II,dubois+14a,
  bustamante+19, beckmann+23, sala+23}. To date, there are only a few attempts
that compute the relative orientation of the BH spin orientation with
respect to that of their host galaxy.

A first detailed analysis of the evolution of the BH spin and orientation
from numerical study in a cosmological context was presented in
\cite{dotti+13}.  They found the occurrence of two regimes.  An early
phase (M$_{\mathrm{BH}}\leq$10$^7$ M$_\odot$) in which rapid alignment of the
BH spin direction to the gas accretion disk angular momentum in each
single episode leads to erratic changes in the BH spin orientation.
For more massive BHs (>10$^7$ M$_\odot$), a single accretion episode
does not modify significantly the BH spin direction, and the BH spin
tends to align with the direction of the angular momentum of the
accreting material.

Also, a pioneering work based on cosmological simulation was done by
\cite{dubois+14a}.  They found that BHs with mass 10$^7$ $\leq$
M$_{\mathrm{BH}}$ $\leq$ 10$^8$M$_\odot$ show a high level of alignment with
their host galaxy, as gas accretion is mostly responsible for both the BH
and stellar mass growth in this mass range. As BHs become more
massive, spins are more randomly oriented with respect to their host
galaxy angular momentum (see their Fig.~10). Moreover, mergers rapidly
change the orientation of galaxies, in addition to coalescence for
BHs, but the latter does not necessarily follow the same evolution
while they merge. They also highlighted an early phase of misalignment when gas was turbulent and the galaxy dis not have coherent angular momentum structures.
More recently, \cite{beckmann+23} used the \hagn\,
simulation from which black hole spin evolution was post-processed
following \cite{dubois+14a}. One of the main results is that
merger-free galaxies tend to have higher BH spins which are
preferentially aligned with their host galaxy's spin.

The aim of the present paper is to contribute to such theoretical
efforts by presenting the most up-to-date detailed statistical study
of the evolution of the BH spins relative to the galaxy angular momentum
vectors.  This is a complementary paper to Beckmann et al. (in prep)
which
focuses on the evolution of the amplitude of the BH spin.  Both analyses
rely on the \nhs simulation \citep{nhz} that presents several
advantages over \hagn. First, it includes a sophisticated
prescription of the BH model, wherein the BH spin is evolved for all BHs
using accretion disk models, and the jets are launched along the
direction of the BH spin vectors, with spin-dependent efficiencies.
Second, it follows the evolution of a statistical number of galaxies
and black holes. Third, it has a high spatial resolution ($\sim$34
pc) that accurately resolves a typical scale height of
galactic disks \citep[see, for instance,][]{park+21}, and therefore
allows for a reliable estimation of their stellar angular momentum.
\nhs has also a sufficient resolution to capture the injection scale of gas turbulence, and has a multiphase interstellar medium.
Furthermore, existing observations of outflows accelerated by active galactic nuclei suggest that some radio jets are inclined with respect to the galaxy disk \citep[e.g.,][]{morganti+15,venturi+21},
which strongly suggests misalignment between the central BH  and
the galaxy angular momentum. However, a recent observational
analysis of a sample of 3682 radio AGNs (with reliable radio and optical 
position angle measurements) suggests a tendency of BH/galaxy spin alignment, especially for lower radio luminosity \citep{zheng+24}.
These observational trends give therefore additional motivation for this work.

It is interesting to note the similarity of the spin-orbit
  orientation between the BH-galaxy and star-planet systems.  Planets
  are occasionally found to have misaligned or even
  retrograde orbits with respect to the direction of stellar rotation.
  Specifically, \citet{kamiaka+19} found that a non-negligible
  fraction ($\sim$20\%) of hot Jupiters exhibits the projected
  spin-orbit angle $\lambda>30^\circ$ \citep[see also,
    e.g.,][]{Ohta2005,winn+15,Albrecht2022}. The origin of those
  misalignment between the stellar spin and planetary orbits is not
  yet well-understood. \citet{takaishi+20} performed a series of
  hydrodynamical simulations of the collapse of turbulent molecular
  cloud cores, and suggested the presence of two regimes: at the
  initial epoch when the protostar formed, $\Psi$ (i.e., the angle
  between the protostellar spin and the protoplanetary disk rotation
  axes) is very broadly distributed within $\sim 130^\circ$. With the
  subsequent mass accretion from the proto-planetary disk to the
  protostar, however, the part of the angular momentum of the disk is
  transfered to the stellar spin, and $\Psi$ gradually decreases and
  tends to be aligned ($\lesssim 20^\circ$).

  Thus, they concluded that the isolated planetary disks formed in the
  turbulent cloud cores are unlikely to explain the observed
  misalignment, implying the importance of the interacton with the
  nearby stellar systems, or the gravitational planet-planet
  scattering after the gas disk dispersal.

  Those results inspired us to ask the same question for the
  orientation between the central BH spin and the host galaxy angular
  momenta. Disk galaxies and planetary systems exhibit several
  similarities: they are both dominated by a central object (BH/star)
  and surrounded by either a galactic disk of stars and gas or a
  planetary disk of accreting gas.  One may naturally ask whether the
  same physical mechanisms (inherent or external) operate in the
  formation process of the two astrophysical objects.  It is 
  definitely interesting to explore this possibility by exploring
  the connection of BH-galactic disk and central star-planetary disk
  as well as their mutual coevolution, to constrain the formation of
  planetary systems and galaxies, two major topics of contemporary
  astrophysics.

The paper is organized as follows. Section~\ref{sec:simu} briefly
introduces the \nh\, simulation and the numerical modeling used in
this work (simulations and post-processing).
Section~\ref{sec:3Danalysis} presents our main results on the
3-d statistics of the BH-galaxy spin angles while in
section~\ref{sec:progectedanalysis} we derive statistics on the 2-d
projected angles to make predictions for future observational studies.
We summarize our results and conclusions in
section~\ref{sec:conclusions}.

\section{Simulation data}
\label{sec:simu}


Throughout this paper, we analyze the results of the 
\nh\footnote{https://new.horizon-simulation.org/} simulation.
The details of the simulation has been described in detail in \cite{nhz}, so 
we only summarize here its main features.
Those who are interested in the BH spin results may skip this section 
and move to section~\ref{sec:3Danalysis}.

\subsection{General}

\nh\, is a high-resolution zoom-in simulation from the \hagn\, simulation \citep{hagn}, which extracts a spherical sub-volume with a radius of 10 comoving Mpc.
A standard $\Lambda$CDM cosmology was adopted with the total matter density
$\Omega_m$ = 0.272, the dark energy density $\Omega_\Lambda$ = 0.728,
the baryon density $\Omega_b$ = 0.045, the Hubble
constant $H_0$=70.4 km s$^{-1}$ Mpc$^{-1}$, 
the amplitude of the matter power spectrum $\sigma_8$ = 0.81
and the power-law index of the primordial power spectrum $n_s$ = 0.967,
according to the WMAP-7 data \citep{komatsu+11}. 
The initial conditions have been generated with 
{\mbox{{\sc \small MPgrafic}}} \citep{mpgrafic} at the resolution of 4096$^3$ for \nh\, in contrast to 1024$^3$ for \hagn.
The dark matter mass resolution reaches 1.2$\times$10$^6$ M$_\odot$ compared to
8$\times$10$^7$ M$_\odot$ in \hagn.

Both simulations were run with the \ramses\, code \citep{ramses} in
which the gas component is evolved using a second-order Godunov scheme
and the approximate Harten-Lax-Van Leer-Contact~\citep[HLLC,][]{toro}
Riemann solver with linear interpolation of the cell-centered
quantities at cell interfaces using a minmod total variation
diminishing scheme.  In \nh, refinement is performed according to a
quasi-Lagrangian scheme with the highest resolution of $\Delta x=$34
pc at $z=0$. The refinement is triggered in a quasi-Lagrangian manner,
if the number of DM particles becomes greater than 8, or the total
baryonic mass reaches 8 times the initial DM mass resolution in a
cell. Extra levels of refinement are successively added at $z=$ 9, 4,
1.5 and 0.25 (i.e., $a=$ 0.1, 0.2, 0.4 and 0.8 respectively). 
The \nhs simulation is currently completed at redshift $z=0.18$.
Additionally, we have analyzed two other zoom simulations (nicknamed \galactica) focusing on isolated galaxies. For
them, we have used exactly the same physics and mass resolution as \nh\, but
they are located in different regions of \hagn\, (see, for instance,
\citealt{park+21}).

\subsection{Gas and stellar physics}

The gas follows an equation of state for an ideal monoatomic gas with
an adiabatic index of $\gamma_{ad}$=5/3. Gas cooling is modeled
assuming equilibrium chemistry with rates tabulated by
\cite{sutherland&dopita93} above 10$^4$K and by \cite{rosen&bregman1995}
below 10$^4$K.  Gas is also heated via a uniform ultraviolet radiation
after the reionization epoch at $z=10$ following
\cite{haardt&madau96}.  Star formation (SF) is also included: stars
can form out of gas cells with a hydrogen number density greater than
n$_0$$=$10\;cm$^{-3}$ and a temperature lower than 2$\times$10$^4$
K, following a Schmidt relation:
\begin{equation}
  d\rho_*/dt =\epsilon_* \rho_{\mathrm{gas}}/t_{\mathrm{ff}} ,
\end{equation}
where $\rho_{\mathrm{gas}}$ is the gas density, $t_{\mathrm{ff}}$ is the
free-fall time of the gas, and $\epsilon_*$ is the efficiency of star
formation per free-fall time.

Contrary to \hagn, \nh\ adopts the efficiency of star formation that
depends on the local turbulent Mach number and Jeans
length~\citep{kimmetal17,trebitschetal17,trebitschetal20}.  The
initial mass function follows a Chabrier functional form
\citep{chabrier05} with cutoffs at 0.1 and 150 M$_{\odot}$.  Finally,
a model of type II supernovae is based on the amount of linear
momentum injected at the adiabatic or snow-plow phase
\citep{kimm&cen14,kimmetal15} .  The typical mass resolution for star
particles in \nhs is $\sim$10$^4$ M$_\odot$.

\subsection{Black hole physics}

Here we summarize the BH sub-grid physics which is implemented in \nh,
and particularly relevant to our current investigation of BH-galaxy
spin relative orientation.  It includes a model for BH mass growth and
AGN feedback in alternating radio/quasar (jet/heating) mode
\citep{dubois+12} coupled to a model of BH spin evolution
\citep{dubois+14a}.

\subsubsection{Formation, mass growth and dynamics}
\label{sec:BH_formation}

First, supermassive BH seeds are allowed to form within any cells
satisfying the following criteria:
\begin{description}
\item[i)] both the stellar and gas densities exceed the threshold for star formation
\item[ii)] the local stellar velocity dispersion is larger than
  $20\,\rm km\,s^{-1}$
\item[iii)] no preexisting BH can be found within a distance of 50
  comoving kpc from the cell.
\end{description}

Then BHs form with an initial mass of 10$^4$M$_\odot$ and a spin
parameter of $a=0$. Then, their mass grows at a rate
$\dot{M}_{\mathrm{BH}}$ over time by accreting gas following an
un-boosted Bondi-Hoyle-Lyttleton accretion rate
$\dot{M}_{\mathrm{Bondi}}$ and a spin-dependent radiative efficiency
$\epsilon_r$:
\begin{eqnarray}
   \dot{M}_\mathrm{BH} &=& (1-\epsilon_r)\dot{M}_\mathrm{Bondi}, \\
   \epsilon_r &=& f_{\rm att}(1-e_{\rm isco})=f_{\rm att}\left(1-\sqrt{1-2/(3r_{\rm isco})}\right), \\
   \dot{M}_\mathrm{Bondi}  &=&
   4\pi\bar{\rho}\frac{(GM_\mathrm{BH})^2}{(\bar{u}^2 + \bar{c_s}^2)^{3/2}},
\end{eqnarray}
where $e_{\rm isco}$ is the energy per unit rest mass energy of the innermost stable circular orbit (ISCO) of radius $r_{\rm isco}$,
$\bar{u}$ is the average BH-to-gas relative velocity,
$\bar{c_s}$ the average gas sound speed and $\bar{\rho}$ the average
gas density.  Those quantities are calculated by averaging over a
sphere of radius $4 \Delta x$ ($\sim$150 pc) of the considered
BH and using mass and kernel weighting \citep{dubois+12}.
$r_{\rm isco}$ is in units of the half of the Schwarzschild radius and depends on the BH spin magnitude and its orientation with respect to the gas accretion disk angular momentum.

Finally, the Bondi-Hoyle-Lyttleton accretion rate is capped at
the Eddington luminosity rate for the appropriate $\epsilon_r$:
\begin{equation}
  \dot{M}_\mathrm{Edd}
  = \frac{4\pi G M_\mathrm{BH}m_\mathrm{p}}{\epsilon_r\sigma_Tc} ,
\end{equation}
where $\sigma_T$ is the Thomson cross-section and $m_p$ the proton
mass.  According to eqn.~\ref{equ:equ1}, a fraction of the mass
$\epsilon_r$ accreted is radiated away, while the rest of the mass
(1-$\epsilon_r$) is accreted onto the BH and increases the BH mass.
We introducce the Eddington ratio $\chi$:
\begin{equation}
\chi = \left\{
    \begin{array}{ll}
      {\dot{M}_\mathrm{Bondi}}/{\dot{M}_\mathrm{Edd}}
      & \mbox{if\,\,\,\,} \dot{M}_\mathrm{Bondi} < \dot{M}_\mathrm{Edd} \\ 
      1 & \mbox{if\,\,\,\,} \dot{M}_\mathrm{Bondi} > \dot{M}_\mathrm{Edd}
    \end{array}
\right.
\label{equ:eddington}
\end{equation}
and distinguish the different AGN feedback
at the threshold value of $\chi_{\mathrm{trans}}=0.01$
as discussed in section~\ref{sec:agn_feedback}.
We note that in the case of the radio mode (see section~\ref{sec:agn_feedback}), $\epsilon_r$ used to estimate the effective growth of the BH is attenuated by a factor
$f_{\rm att}$ =min($\chi/\chi_{\rm trans}$,1)
following \cite{benson&babul09}. We also impose 
a maximum value of the BH spin at $a_{\rm max}=0.998$ due to the emitted photons by the accretion disk captured by the BH 
\citep{thorne74}.

Furthermore, BHs merge when they get closer than $4 \Delta x$ and the
relative velocity of the pair is smaller than the escape velocity of
the binary.  The less massive BH of the binary is absorbed into the
more massive one. It is then possible to know for each (existing)
black hole, the percentage of its mass that has been gained through BH
mergers at a specific redshift, \fdm. \fdm=0\% means no mass
contribution from BH mergers while 
high values of \fdm\, indicate that a large percentage of mass has been gained through mergers over the BHs history.
It is also worth mentioning that no recoil velocities have been applied to BHs during the merger process.

Finally, due to the finite force resolution effect, an explicit drag
force is introduced for the gas onto the BH following
\cite{ostriker99}, in order to avoid any spurious motion that can
arise especially around high density gas regions.  A detailed analysis
of BH mergers in \nh\, is presented in \cite{volonteri+20}, and an
analysis of the population statistics of intermediate mass BHs
in dwarf galaxies is found in \cite{beckmann+23a_pop}.

\subsubsection{Spin evolution model}

The BH spin is modeled on-the-fly in \nh\, and updated according to
the gas accretion and BH-BH mergers.  This is a major improvement
relative to \hagn\, in which the evolution of the BH spin was not included
and required some post-processing afterwards
\citep{dubois+14a,beckmann+23}.  The spin evolution follows
specifically the model detailed in \cite{dubois+14_II, dubois+14a} to
which we refer the readers for a full description and technical
details.  The only difference is that a new model of spin evolution is
adopted for low accretion rates $\chi<\chi_{\mathrm{trans}}$.  We summarize
the main points here.

We stress that the accretion disk is not resolved in the simulation.
Thus, at high accretion rates, a thin accretion disk solution 
by \cite{shakura+73} is adopted. Then the
evolution of the amplitude of the BH spin $a$
through gas accretion is estimated following  \cite{bardeen70}:
\begin{equation}
a_{\rm n+1} = \frac{1}{3}r^{1/2}_{\rm isco}\frac{M_{\rm BH,n}}{M_{\rm BH,n+1}}\left[4-\left(3r_{\rm isco}\left(\frac{M_{\rm BH,n}}{M_{\rm BH,n+1}} \right)^2-2  \right)^{1/2}\right],
\label{equ:bardeen}
\end{equation}
where $n$ refers to the value of the different variables at the $n$-th
timestep.

We note that eqn.~\ref{equ:bardeen} assumes that BH spin and accretion disk
angular momentum are perfectly aligned or anti-aligned (i.e., only the
BH spin amplitude changes but not its direction), which is not the
case in general.  A misaligned accretion disk experiences a torque due
to the Lense-Thirring effect, and precesses the spin axis of the BH
and warps the innermost parts of the disk (for large enough
viscosity). The result of the Lense-Thirring effect is that the BH
spin and the accretion disk angular momentum tend to align (or
anti-align) with the total angular momentum.  If one defines the total
angular momentum of the system {BH+disk} at the n-th timestep by
$\boldsymbol{J}_\mathbf{tot,n} = \boldsymbol{J}_\mathbf{BH,n} + \boldsymbol{J}_\mathbf{d,n}$, the direction
of the new BH spin is obtained by
$\boldsymbol{J}_\mathbf{BH,n+1}=\boldsymbol{J}_\mathbf{tot,n}=\boldsymbol{J}_\mathbf{BH,n}+\boldsymbol{J}_\mathbf{d,n}$,
due to the conservation of total angular momentum.  Its co- or
counter-rotation with respect to the accretion disk is decided
following criterion from \cite{king+05}.
In \nhs, the orientation of the BH spin is updated first, and then its
amplitude (using eqn.~\ref{equ:bardeen}). 

Finally, at lower accretion rates, $\chi$<0.01,
jets are assumed to be powered by energy extraction from black
hole rotation \citep{blandford+77}. Consequently, the black hole spin
only decreases and the variations of $da/dt$ are obtained using the
polynomial fits in \cite{mckinney+12}.  The orientation of the BH spin
is updated following the same procedure as for high accretion rates.

The BH-BH mergers are also taken into account in the simulation.  The
spin of the remnant is calculated according to the spin of each BH
prior the merger as well as the orbital angular momentum of the binary
system. Specifically, we adopt the
analytical fit of \cite{rezzolla+08},
assuming a random orientation between spins and the orbital angular momentum.

\subsubsection{AGN feedback}
\label{sec:agn_feedback}

The AGN feedback follows two different prescriptions, depending on the
Eddington ratio $\chi$ defined in eqn.~\ref{equ:eddington}.  Energy is
released with an efficiency $\eta$ after each accretion episode in the
form of
\begin{equation}
L_{AGN,R,Q}=\eta_{R,Q}\dot{M}_{\mathrm{BH}}c^2,
\end{equation}
where $R$ and $Q$ stand for the radio and quasar heating mode, 
respectively.

\begin{description}
\item[$\chi<\chi_{\mathrm{trans}}$ (``radio mode'')] BHs 
power jets that continuously release mass, momentum and
energy. Bipolar jets are assumed as a cylinder of size $\Delta x$
in radius and semi-height, centered on the BH \citep{dubois+10}.
The AGN axis is aligned to the spin of the BH direction (without any opening
angle). The jets are launched with a speed of 10$^4$ km/s. 
We note that $\eta_R$ is not a free parameter but computed
from the BH spin following the results of \cite{mckinney+12}
\citep[see also][for the interpolating function]{nhz}.
\item[$\chi>\chi_{\mathrm{trans}}$ (``quasar mode'')] BHs release only thermal
  energy into the gas \citep{teyssier+11} within a sphere of radius
  $\Delta x$ (isotropic and uniformly distributed).  The efficiency of
  the feedback in quasar mode is given by $\eta_Q=\epsilon_r\eta_c$,
  where $\eta_c=0.15$ is calibrated 
on the local $M_{\mathrm{BH}}$-galaxy mass in lower resolution ($\sim$kpc)
simulations  \citep{dubois+12}.
\end{description}

\subsection{Galaxy and black hole catalogs}

Dark matter halos and galaxies are identified using the
{\mbox{{\sc \small Adaptahop}}} structure finder
\citep{aubertetal04, tweed+09} at different redshifts using a local density threshold
  of 178 and 80 times the average DM and stellar densities,
  respectively.
{\mbox{{\sc \small Adaptahop}}} allows to separate substructures from
  their host halos/galaxies.  Since \nh\, is a zoom simulation, 
  low-mass resolution dark matter particles might ``pollute'' some halos,
  especially when they are located close to the boundary of the high
  resolution area.  We remove those DM halos and their embedded
  galaxies in the following statistical analysis, except in
  Appendix~\ref{appendix1} where we present individual evolution histories of
  several BHs in ``contaminated'' DM halos in which low resolution DM particles
  represent less than 0.1\% of the total mass of the halos.
  Also, to give an order of magnitude, $\sim$35\% of galaxies with
  a stellar mass greater than 10$^6$M$_\odot$ at $z=0.18$ lies in 
  non contaminated halos.
  In the following, we define the galaxy mass by the value returned
  by {\mbox{{\sc \small Adaptahop}}}.

To link BHs to galaxies, we adopt the same methodology developed in
previous studies using either \hagn\, or \nh\,
\citep[e.g.,][]{volonteri+16,smethurst+23,beckmann+23}.  In the first
step, we loop over all galaxies, from the most to the least massive
ones, and identify and associate for each of them the most massive BH
to be contained within 2 effective radii of the galaxy's center. Such
objects are then flagged as a primary BH, and removed from the list of
non-allocated BHs.  Then we repeat the second loop, and label all BHs
within two effective radii as secondary BHs.  All non-allocated BHs
are finally removed from the sample since they are considered too far
(``wandering'').  In other words, a galaxy can contain multiple
BHs, but any BH is associated uniquely to a single galaxy.  In this
scheme, we use a shrinking sphere approach \citep{power+03} to
determine precisely the galaxy center. We also estimate the effective
radius $R_e$ of each galaxy by taking the geometric mean of the
half-mass radius of the projected stellar densities along each of the
simulation's Cartesian axes.

Initially, our sample extracted from \nhs contains 572 BHs (379 primary and 193 secondary BHs) at $z=0.18$. This sample is 
completed with two additional primary BHs provided by the two \galacticas zoom-in simulations.
However, BHs with mass below 2$\times$10$^4$M$_\odot$
are discarded from our statistical analysis because they are too close
to the initial seed BH mass and likely to suffer from mass
resolution. Furthermore, we mainly focus on primary black holes.
All of these constraints finally lead to a sample of
102 primary BHs at $z=0.18$.

\section{Three-dimensional statistics}
\label{sec:3Danalysis}

\begin{figure}
\begin{center}
\rotatebox{0}{\includegraphics[width=\columnwidth]{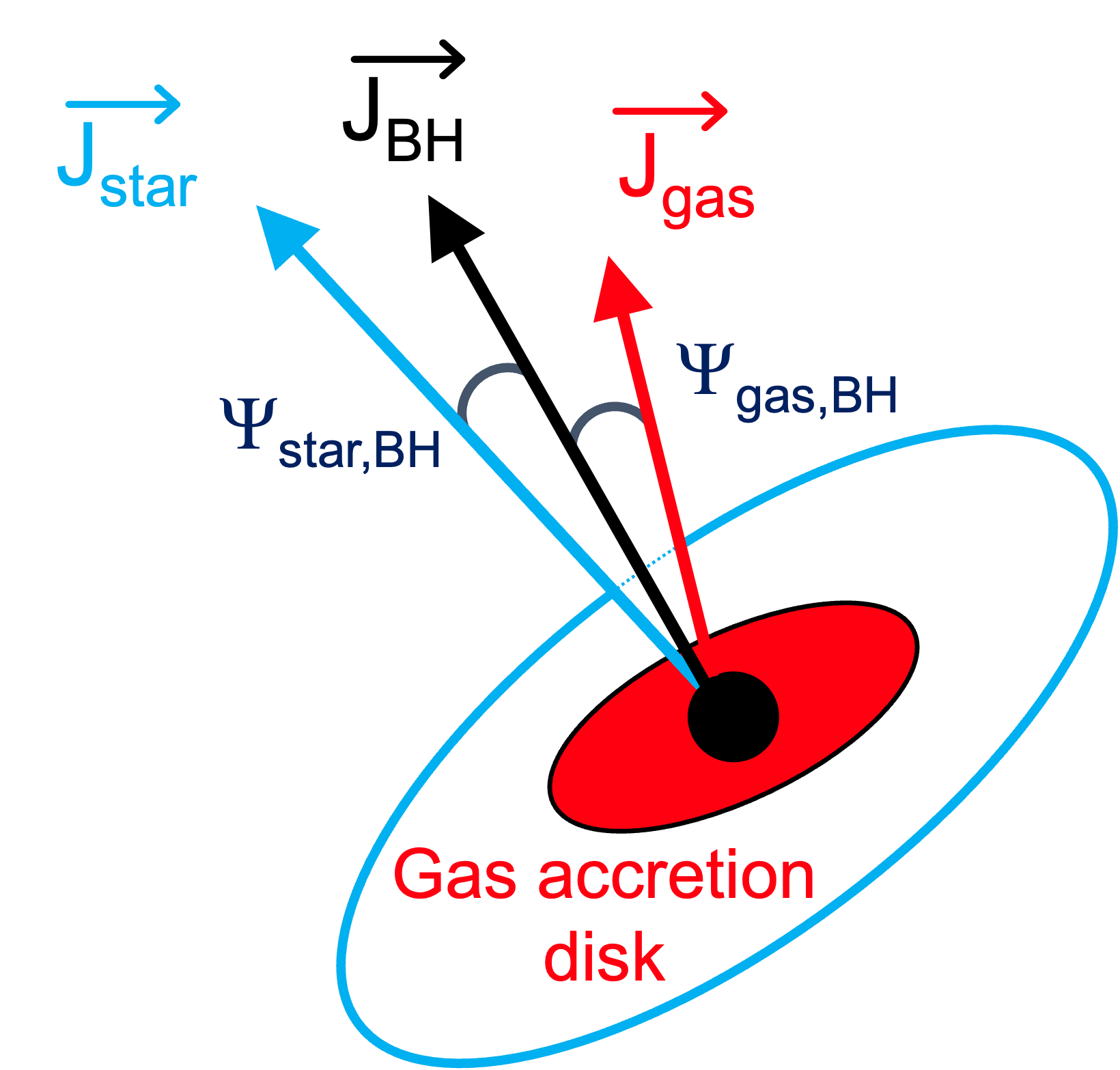}}
\caption{Schematic view of the different galaxy components and the
  relevant angles studied throughout the paper. We note that \PSIS is the 3-d angle
  between the BH spin and the angular momentum of the host
  galaxy. The latter was estimated by considering all star particles
  within a sphere of radius $R_e$. Furthermore, \PSIGAS\, is the 3-d angle between
  the BH spin and the gas accretion disk gas (\JGAS)  estimated within the four
  closest cells (in radius) from the BH position and using mass and
  kernel weighting.  }
\label{fig_scheme1}
\end{center}
 \end{figure}

\subsection{General}

In this section, we mainly focus on the cosmic evolution of the 3-d
angle (hereafter $\Psi$) between the BH spin vector, \JBH, and the
angular momentum vector of the host galaxy stellar component, \JSTAR, as well as its
correlations to specific galaxy and BH properties.  If not specified,
the galaxy spin is estimated from the total angular momentum vector of
all star particles associated with the galaxy within one effective
radius $R_e$.
The angle \PSIS between the two vectors is then simply computed by
\begin{equation}
\mathrm{cos}\, (\Psi_{\mathrm{star,BH}}) = \frac{J_{\mathrm{star}}\cdot J_{\mathrm{BH}}}{||J_{\mathrm{star}}||\,||J_{\mathrm{BH}}||} .
\end{equation}
As such, \PSIS can only lie between 0 and 180 degrees.

We similarly define the angles, \PSIGAS\ and \PSISTARGAS\, between the
angular momentum of the gas accretion disk (\JGAS) and the BH spin and the stellar
spin, respectively.  We recall that the gas accretion disk properties
are estimated within the four closest cells in radius from the BH
position and using mass and kernel weighting.
Fig.~\ref{fig_scheme1} presents a schematic view of the different
galaxy components as well as the relevant angles
we are particularly interested in.   

\begin{figure}
\begin{center}
\rotatebox{0}{\includegraphics[width=\columnwidth]{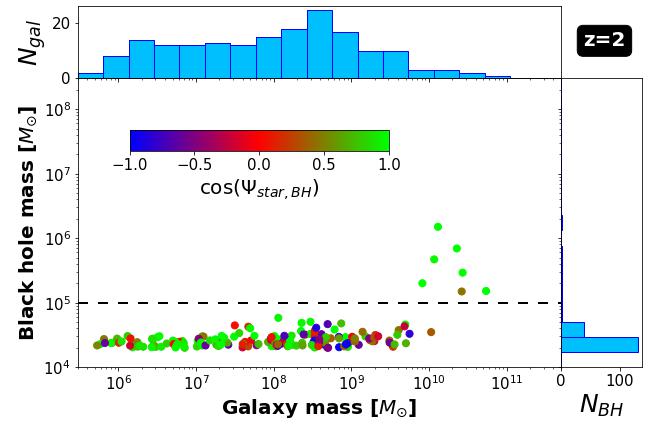}}
\rotatebox{0}{\includegraphics[width=\columnwidth]{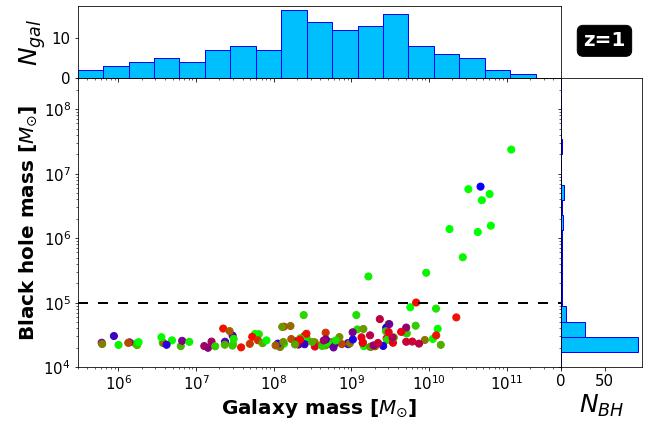}}
\rotatebox{0}{\includegraphics[width=\columnwidth]{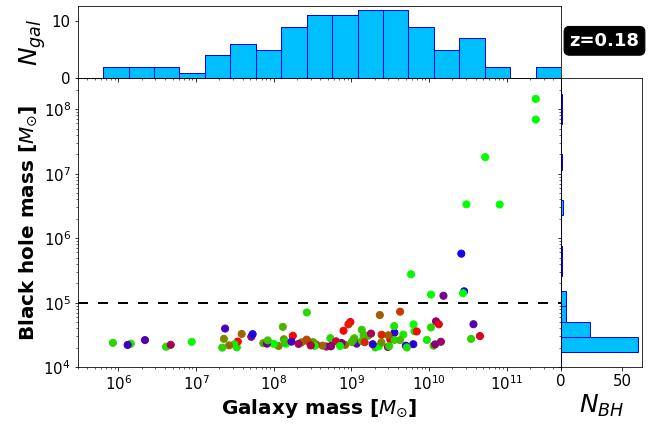}}
\caption{Variations of the primary BH mass with respect to
their host (stellar) galaxy mass at three different redshifts. 
In each panel, the color code indicates the value of the corresponding
angle between the BH spin and the angular momentum of
the stellar component computed within one effective
radius. The most
massive BHs (>10$^{5}$ M$_\odot$) tend to have a
spin well aligned to \JSTARS at all redshifts. }
\label{fig2}
\end{center}
 \end{figure}

Let us begin the analysis by taking a look at the BH population
properties as well as host galaxies at different redshift.  In
Fig.~\ref{fig2}, we plot the primary BHs mass against their host
galaxy mass at three different redshift ($z=2$, $1$ and $0.18$). We
also use a color code to indicate the associated value of cos(\PSI).
Since the volume covered by \nh\, is relatively small, and due to our
tight constraints on BH and galaxy selection, the total number of BHs
in our analysis with M$_{\mathrm{BH}}>$10$^5$ M$_\odot$ is quite low: at $z=2$,
only $7$ BHs has been identified, and this number goes up to $11$
at lower redshifts.  As already pointed out in \cite{nhz}, on
average, central massive BHs in \nh\, grow significantly only above a
stellar mass threshold of a few $10^{10}$ M$_\odot$.  For galaxies with
stellar masses lower than 5$\times$10$^9$ M$_\odot$, however, BHs
growth is in general regulated and limited by supernovae feedback
 \cite[see also][]{dubois+15, habouzit+17,trebitschetal17, lapiner+21}.
Moreover, due to their low seed mass, BHs may not 
remain strongly attached to the centers of their host galaxies,
especially in the dwarf galaxy regime 
as studied in \cite{beckmann+23a_pop}.
Consequently, most of the BHs
population grows little over the course of the simulation.

As far as the BH-galaxy spin orientation is concerned, two trends
seem to emerge.  First, at high redshift ($z=2$), most of BHs tend to
have their spin aligned with the host galaxy spin.  Second, the most
massive BHs (>10$^{5}$ M$_\odot$) at all redshifts also tend to have a
spin aligned to \JSTARS.  In the next sections, we investigate
the origin of these different trends.

Finally, since we estimate the angular momentum of each galaxy within
one effective radius $R_e$, it is instructive to derive the variations
of $R_e$ as a function of the total stellar mass.
One major advantage of \nhs is its high resolution ($\sim$34 pc), enabling to
describe even the size of low-mass galaxy (10$^6$ - 10$^7$M$_\odot$)
with several resolution elements.  As reported previously in
\cite{nhz}, \nhs reproduces fairly well the size-mass
relation similar to observations at all redshifts
(see their Fig.~12).

\subsection{Cosmic evolution of \PSI}

\subsubsection{Individual histories}
\label{sec:individual}

\begin{table}
\centering
\begin{tabular}{ | c | c | c | c | c |}
\hline
   BH id & BH mass &z & Fig. & Comments\\
\hline   
   166  & 1.4$\times$10$^8$M$_\odot$ && \ref{fig_166_map}, \ref{fig_bh166} & BH-mergers\\
   455  & 6.9$\times$10$^7$M$_\odot$ && \ref{fig_appendix1} & BH-mergers\\
   1049 & 1.8$\times$10$^7$M$_\odot$ && \ref{fig_1049_map}, \ref{fig_bh1049}, \ref{fig_color_map} & anti-aligned spin\\
   936  & 3.3$\times$10$^6$M$_\odot$ && \ref{fig_appendix1} & smooth accretion\\
   549  & 5.8$\times$10$^5$M$_\odot$ &0.18& \ref{fig_549_map}, \ref{fig_appendix1} & anti-aligned spin \\
   132  & 1.5$\times$10$^5$M$_\odot$ && \ref{fig_appendix1} & Off centered BH \\
   146  & 7.8$\times$10$^7$M$_\odot$ && \ref{fig_appendix2} & <1\% LR DM cont. \\
   348  & 7.1$\times$10$^7$M$_\odot$ && \ref{fig_appendix2} & <1\% LR DM cont. \\
   541  & 3.5$\times$10$^7$M$_\odot$ && \ref{fig_appendix2}, \ref{fig_color_map} & <1\% LR DM cont. \\
\hline
   G9685 & 4.8$\times$10$^6$M$_\odot$ &0& \ref{fig_G9685_map}, \ref{fig_g9685}, \ref{fig_color_map} & smooth accretion\\
   G648  & 1.6$\times$10$^7$M$_\odot$&0.26&  \ref{fig_appendix2}  & smooth accretion\\   
\hline
\end{tabular}
\vspace{0.08cm}
\caption{
\small{
Simulated BHs that are examined in detail throughout this
  paper. The upper part of the table indicates primary BHs extracted
  from \nhs at $z=0.18$.  The first five lines correspond to the most
  massive BHs of our fiducial sample while BH-549 and BH-132 present
  specific features namely an anti-aligned BH-galaxy spin
  configuration and a off-centered BH respectively.  The last three
  lines gathers three other massive primary BHs but with a ``contamination'' of
  <0.1\% in mass from lower resolution (LR) DM particles.  For
  these latter, we just show their evolution in the
  appendix~\ref{appendix1} while they are discarded from any
  statistical analysis.  The lower part of the table is related to the
  additional two BHs from the \galacticas zooms, G9685 et G648.
  }
  }
\label{tab1}
\end{table}

\begin{figure}
\begin{center}
\rotatebox{0}{\includegraphics[width=\columnwidth]{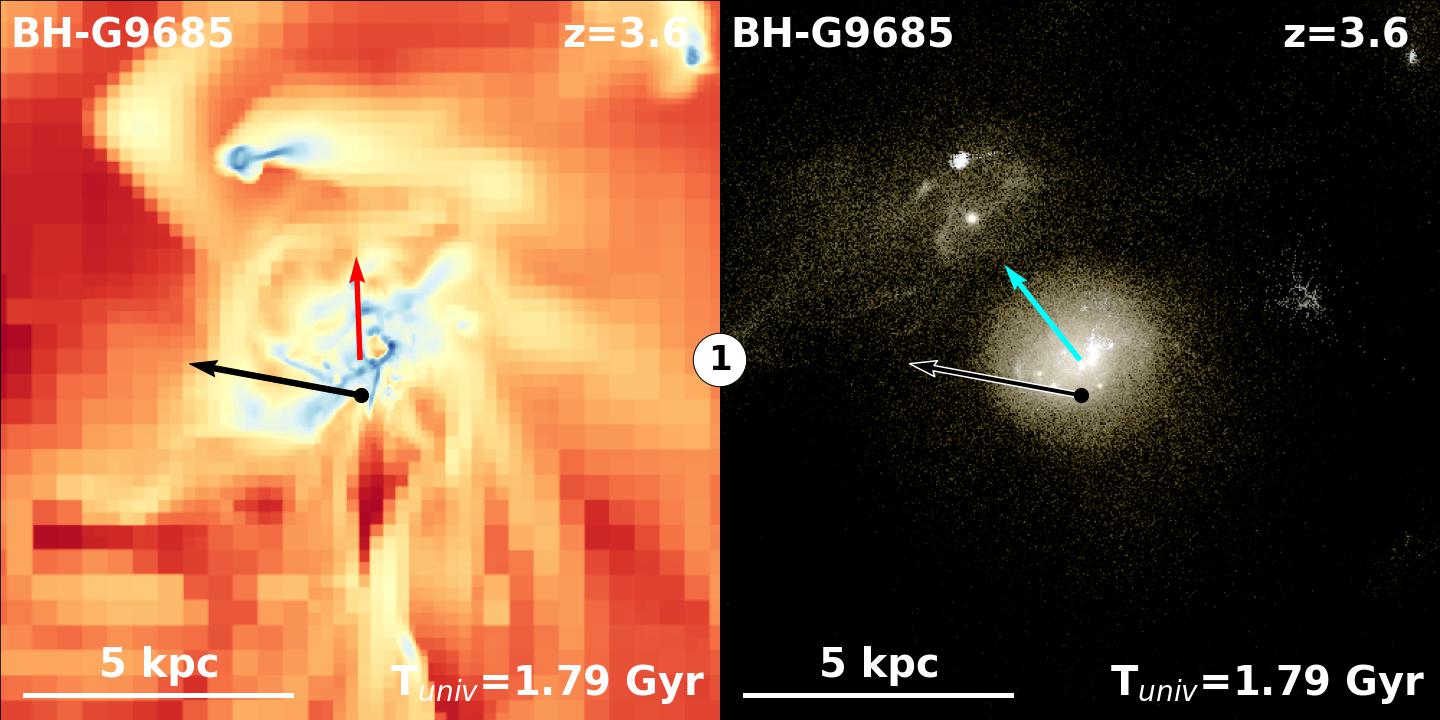}}
\rotatebox{0}{\includegraphics[width=\columnwidth]{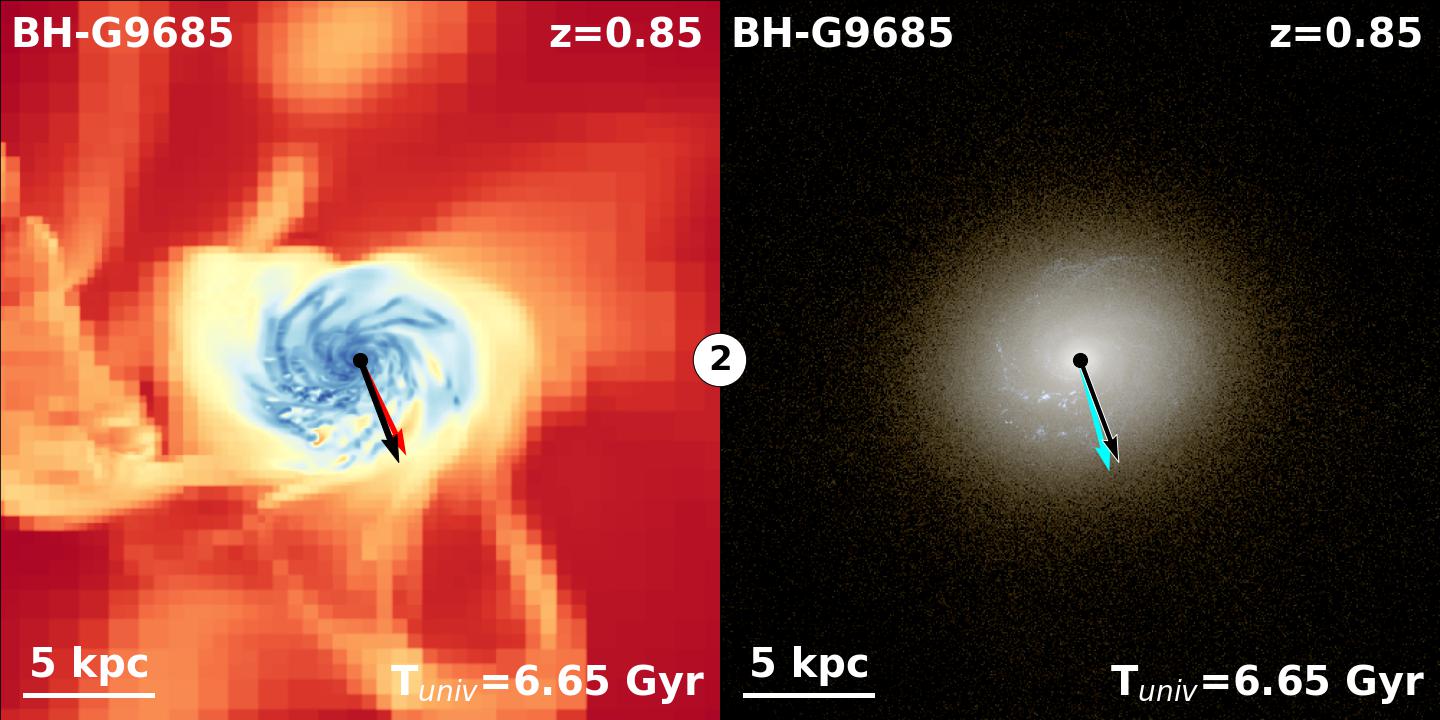}}
\rotatebox{0}{\includegraphics[width=\columnwidth]{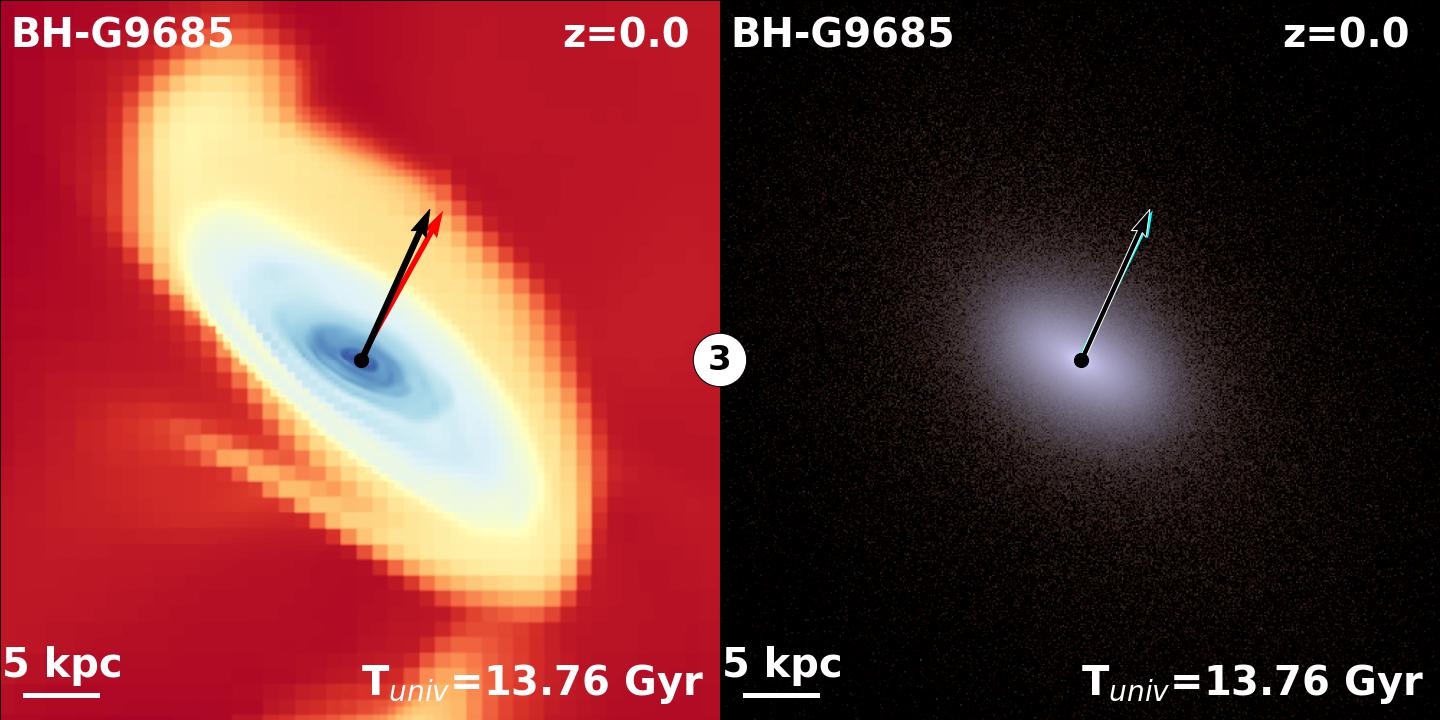}}
\caption{Projected distribution of the gas density (left column) and
  stars (u-g-r band images using {\mbox{{\sc \small sunset}}}, right column) for BH-G$9685$ and its host
  galaxy at three different epochs. Vectors correspond to the spin of
  the stellar (cyan), gas accretion disk (red) and BH (black). 
  \textcircled{\tiny{1}} At
  high redshift, a proto-galaxy is forming and is displaying a
  disturbed morphology.  \textcircled{\tiny{2}} Once the galactic disk
  is formed, the three spins are well aligned.  \textcircled{\tiny{3}}
  This Galaxy has a very quiet evolution and turns to a S0 galaxy due
  to the quenching of star formation. The three spins remains aligned.
}
\label{fig_G9685_map}
\end{center}
 \end{figure}

In order to have a better view on the cosmic evolution of 
relevant BH properties and \PSIS, we explore further in
Fig.~\ref{fig_G9685_map} to Fig.~\ref{fig_bh1049} the evolution of
four BHs that represent different histories.  We also present in
Appendix~\ref{appendix1} the evolution of other BHs (see
Table~\ref{tab1}).

\begin{figure*}
\begin{center}
\rotatebox{0}{\includegraphics[width=18cm]{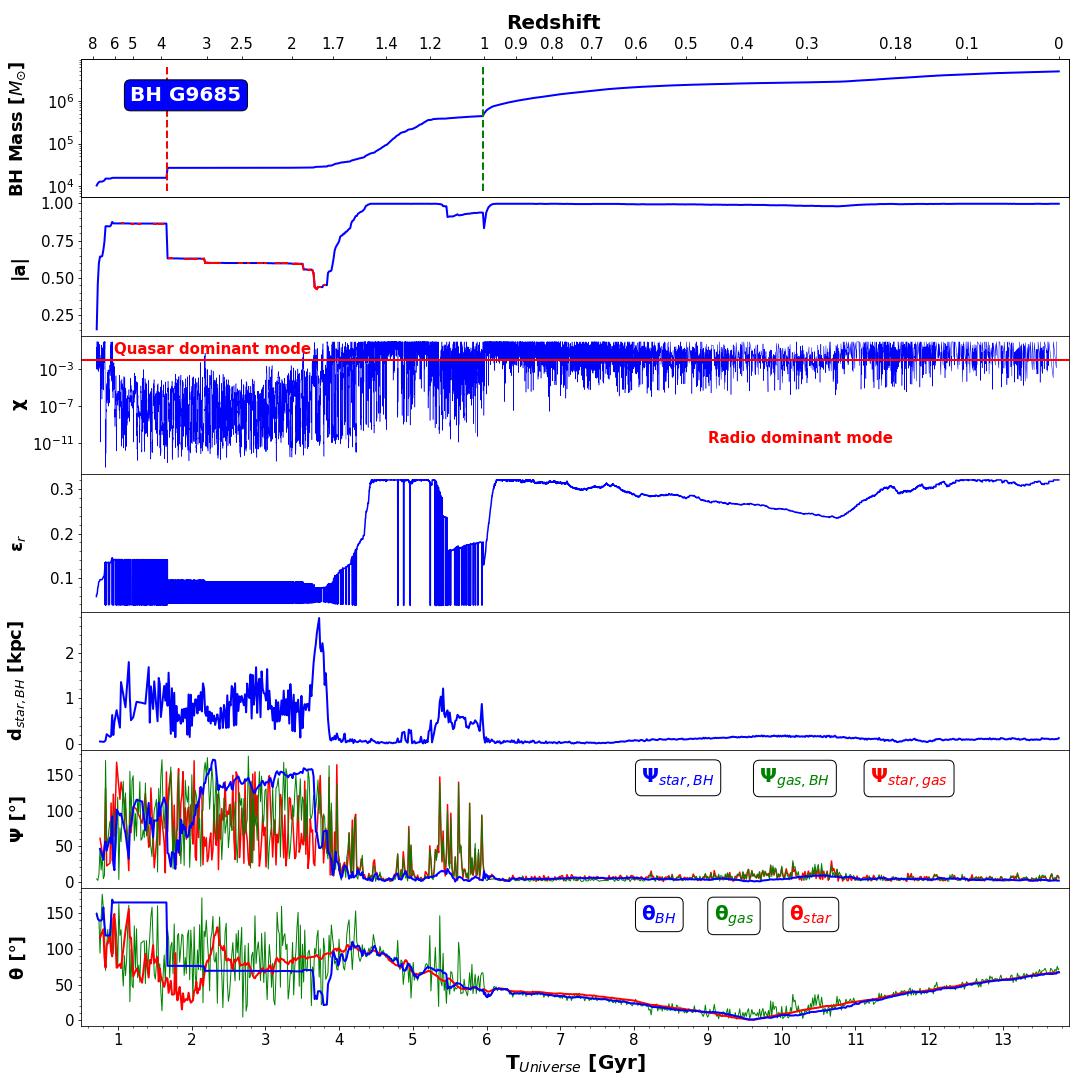}}
\caption{ Evolution of relevant properties of the central black
  hole in the \galacticas run G9685.  \textit{From top to bottom:}
  {\bf 1)} BH mass evolution. The red and green dashed lines indicate
  epochs of major and minor mergers episodes respectively; {\bf 2)} BH
  spin evolution |a|. Red colors indicates when the gas accretion disk is in
  counter-rotation (i.e., a<0); {\bf 3} The Eddington ratio evolution
  ($\chi$). The red line delimits the quasar dominant mode
  ($\chi>$0.01) and the radio dominant mode ($\chi<$0.01); {\bf 4)}
  The evolution of the radiative efficiency of the gas accretion disk
  ($\epsilon_r$); {\bf 5)} The variation of the distance separation
  between the location of the BH and the center of its host galaxy
  (\dsbh); {\bf 6)} the evolution of the angle \PSIS between the BH
  spin and the angular momentum of the stellar component (blue)
  estimated within one effective radius.  We also plot the angle
  between the BH spin and the angular momentum of the gas accretion disk
  (green) as well as the angle between the stellar and the accreted
  gas component (red); {\bf 7)} The evolution of the polar angle
  $\theta$ describing the orientation of the BH spin (blue), the
  stellar (red) and gas accretion disk (green) angular momenta relative to a
  fixed reference frame of the simulation.  The fraction of mass
  gained through BH mergers is \fdm=1.53\%. The cosmic evolution of
  \PSIS follows three different regimes.}
\label{fig_g9685}
\end{center}
 \end{figure*}

The first BH, BH-G$9685$, 
is extracted from a \galacticas zoom.  Its
host galaxy is selected from the \hagn\, volume, and exhibits a very
quiet evolution, dominated by a smooth accretion and no major
merger. Before $z=1$, the galaxy has a clear disk with spiral arms but
turns into an S0 galaxy at a lower redshift mainly due to the quenching
of star formation. Its final stellar mass is 2.4$\times$10$^{10}$M$_\odot$ at
$z=0$.  The projected distribution of the gas density and stars (using
u-g-r bands) at different redshifts is shown in
Fig.~\ref{fig_G9685_map}.  The associated BH is seeded at redshift
$z=7.5$ (\tuniv=0.7 Gyr) and ``survives'' until $z=0$ (i.e., it did not
merge with a more massive BH). It reaches the final mass of
4.8$\times$10$^{6}$ M$_\odot$.  It exhibits very quiet evolution as
well, as indicated in the upper panel of Fig.~\ref{fig_g9685}.

This BH underwent only two merger episodes with other BHs in its life; a major
merger at $z\sim4$ and a minor one at $z\sim1$ (minor and  major merger are delimited by a 1:4 mass ratio). The percentage of its mass gained through the BH mergers is \fdm=1.53\% at z=0.  For a very short period
right after its birth, the BH is rapidly spun up ($|a|$ reaches
0.865), driven by gas accretion.  During this early phase, its spin
tends to be aligned with its host (\PSI<50$^\circ$). Then, the evolution of
the BH encompassed two long phases.  The first one lasts until
\tuniv$\sim3.9$ Gyr (z$\sim$1.7), and is characterized by accretion of gas that is rather chaotic, suggested by the evolution 
of the \PSIGAS, and inefficient (i.e., low Eddington rates), mainly caused by
the complex and disturbed morphology of the forming proto-galaxy at
high redshifts (see for instance the upper panel of
Fig.~\ref{fig_G9685_map}).  Note
also that the BH tends to be in radio mode and the amplitude of its
spin generally decreases because the extracted spin energy powers the
jets.  Consequently, its mass does not grow significantly.  During
this phase, the BH is in general not settled at the center of its host
\citep{bellovary+19, pfister+19}.
Also, the time evolution of the polar angle $\theta_{\rm BH}$ 
 indicates that
the orientation of the BH spin does not vary significantly
but instead, is almost fixed for a long period.
On the contrary, the orientation of the angular momentum
of the star component, through the evolution of $\theta_{\rm star}$, 
exhibits rapid variations which lead to the erratic variations 
seen in the evolution of \PSI.
This is consistent with  \cite{lodato+06} and \cite{lodato+13} 
who showed 
that for low accretion rates, if the Eddington ratio
is sufficiently low, the alignment time between BH spin
and disk becomes very long.

The second long phase starts once the BH is well settled in the center
of the galaxy with a mass close to 10$^5$ M$_\odot$, where the spatial
separation of its position and the center of its host galaxy tends to
zero.  The BH mass and spin start to increase again mainly
through the accretion of gas that is now more coherent,
characterized by high Eddington ratios.  The BH spin is also well aligned with the spin
of the galaxy and this configuration continues down to $z=0$.  The
lower panel of Fig.~\ref{fig_g9685} shows the evolution of the polar
angle $\theta$ that describes the BH/gas/star spins direction relative to
a fixed coordinate system of the simulation.  This angle contains
information on how much redirection the BH/gas/star spin vectors have
experienced.  We note that the 3-d orientation of
these three vectors are coherently changing over the time during the
third phase.
The correlation between \JSTARS and \JGASS can be indeed explained by the fact that, on one hand, 
 \JSTARS is correlated with the angular momentum of the gas component at galactic scales (a few kpc), which is expected for instance in regular S0 or disk galaxies.
On the other hand, we also found that the gas angular momentum vector at those galactic scales shows a significant degree of correlation with the gas angular momentum at scales 
$4\Delta x\sim$136 pc (i.e., within which the properties of the gas accretion disk are estimated in the simulation due to the spatial resolution limit), in agreement with \cite{dubois+14_II}.

\begin{figure}
\begin{center}
\rotatebox{0}{\includegraphics[width=\columnwidth]{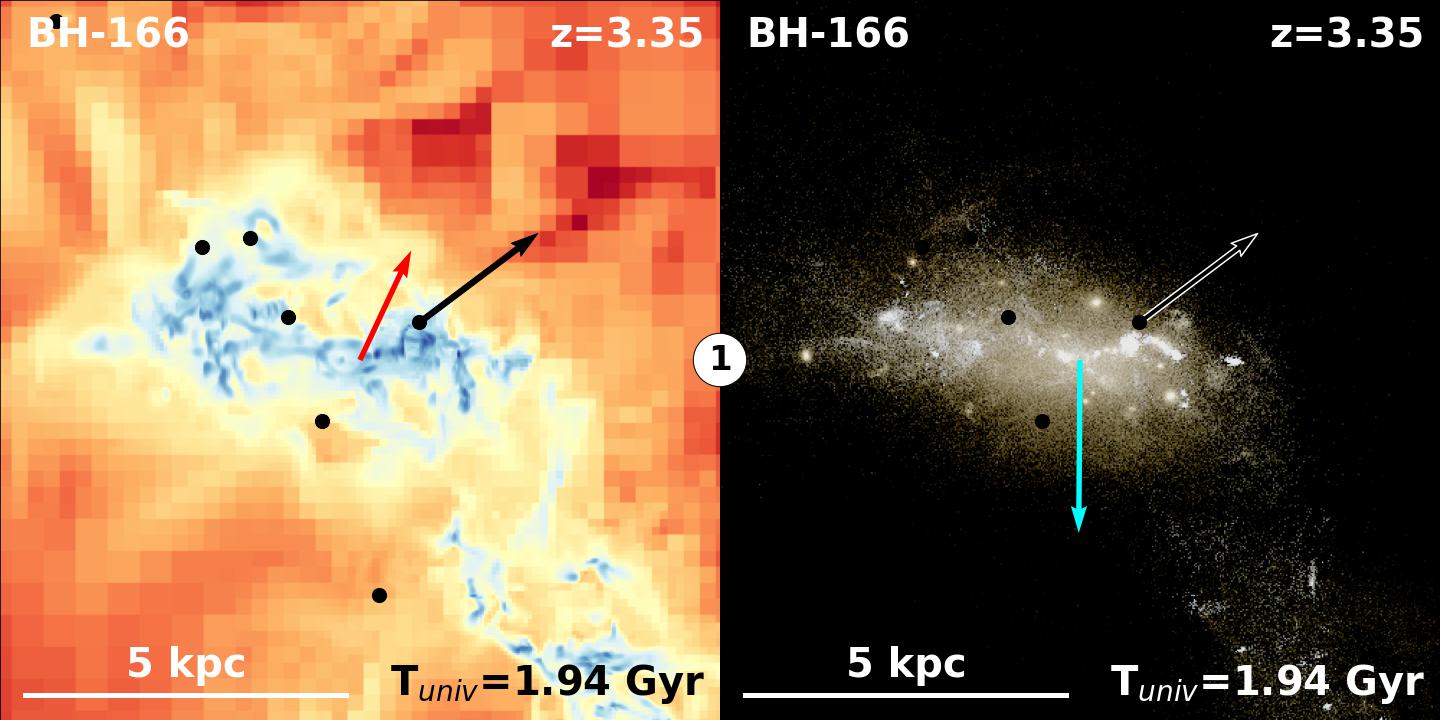}}
\rotatebox{0}{\includegraphics[width=\columnwidth]{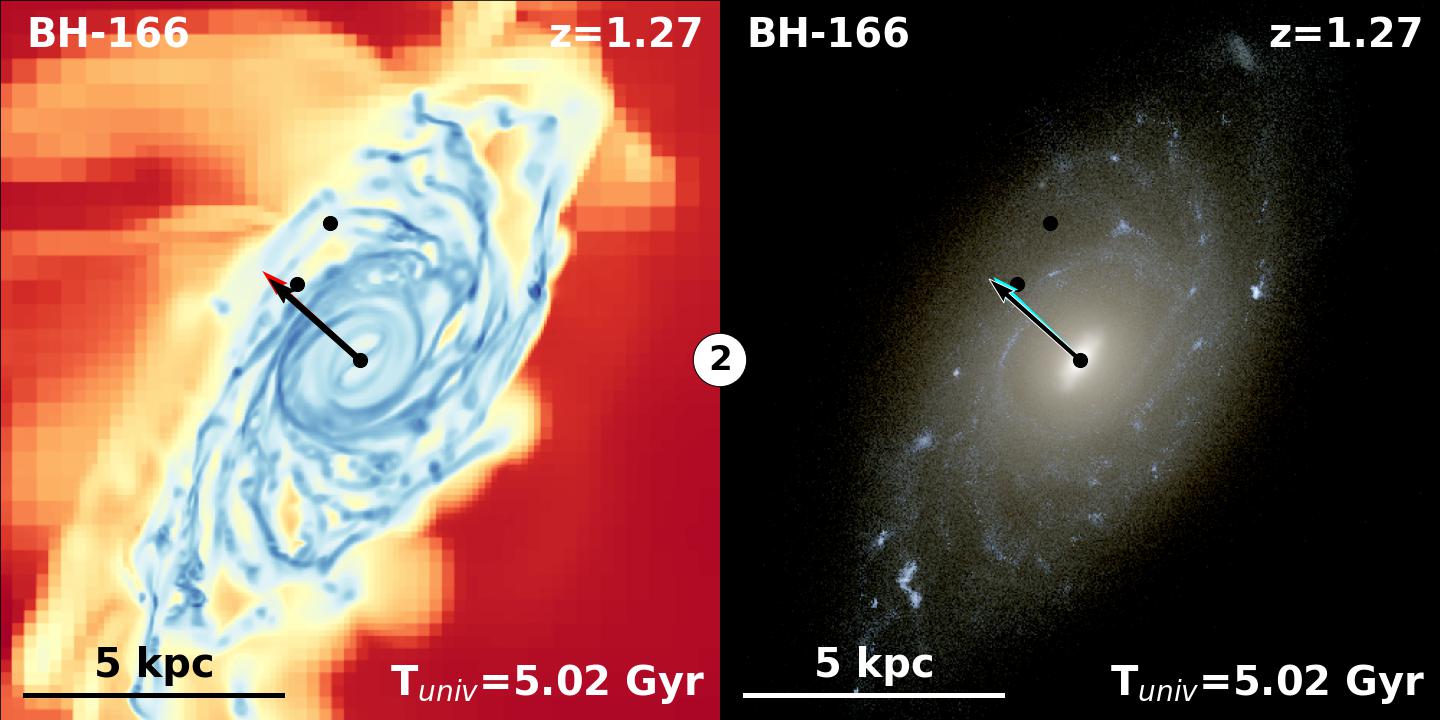}}
\rotatebox{0}{\includegraphics[width=\columnwidth]{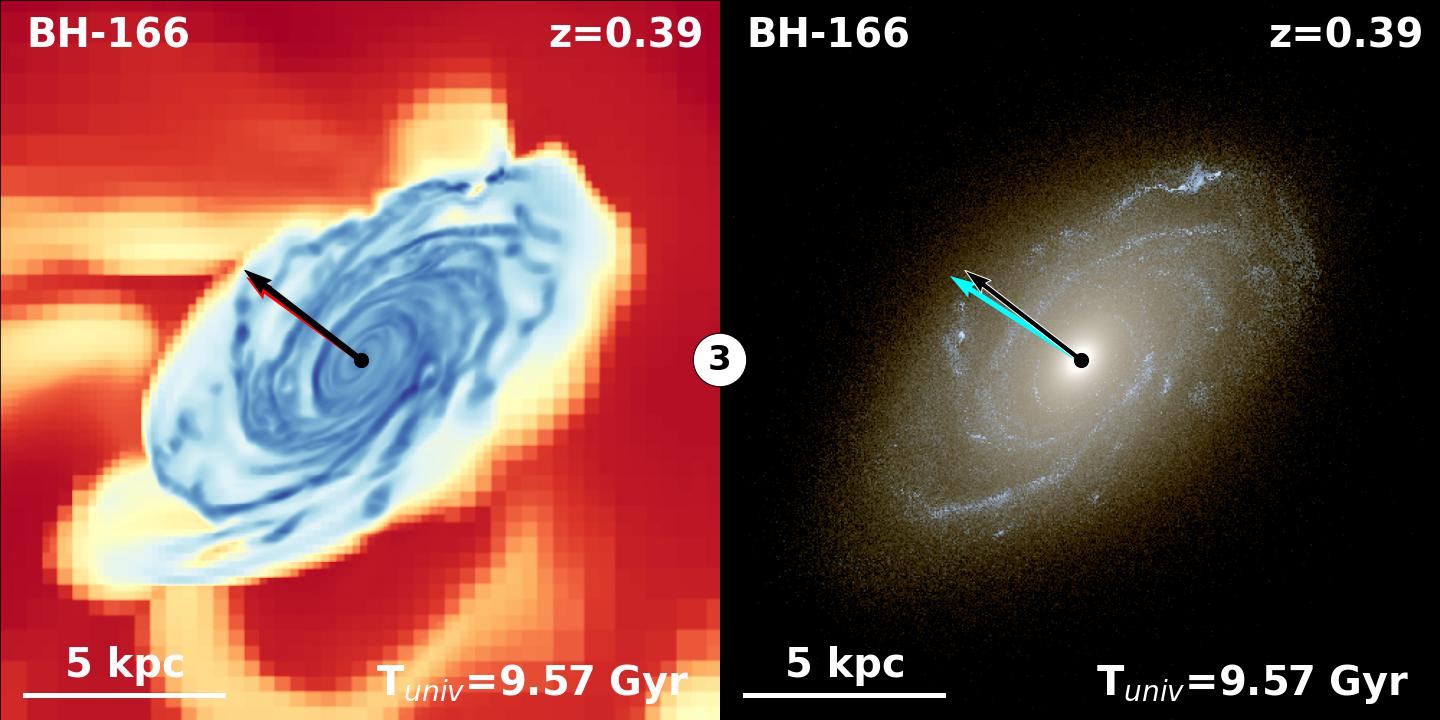}}
\caption{Same as Fig.~\ref{fig_G9685_map} but for BH-$166$ and its host
  galaxy. We can notice
  the presence of other BHs (black dots) that are going to merge soon with the main
  one (associated with the spin vector).
  \textcircled{\tiny{1}} At high redshift, the proto-galaxy
  has again a disturbed morphology. The accretion of gas onto the BH
  is rather chaotic.  \textcircled{\tiny{2}} Here also, once the
  galactic disk is formed, the three spins are well aligned.
  \textcircled{\tiny{3}} Although the main BH experience several mergers
  and its host galaxy one minor merger,  the galactic disk is stable with
  all spin remaining aligned.  }
\label{fig_166_map}
\end{center}
 \end{figure}

\begin{figure*}
\begin{center}
\rotatebox{0}{\includegraphics[width=18cm]{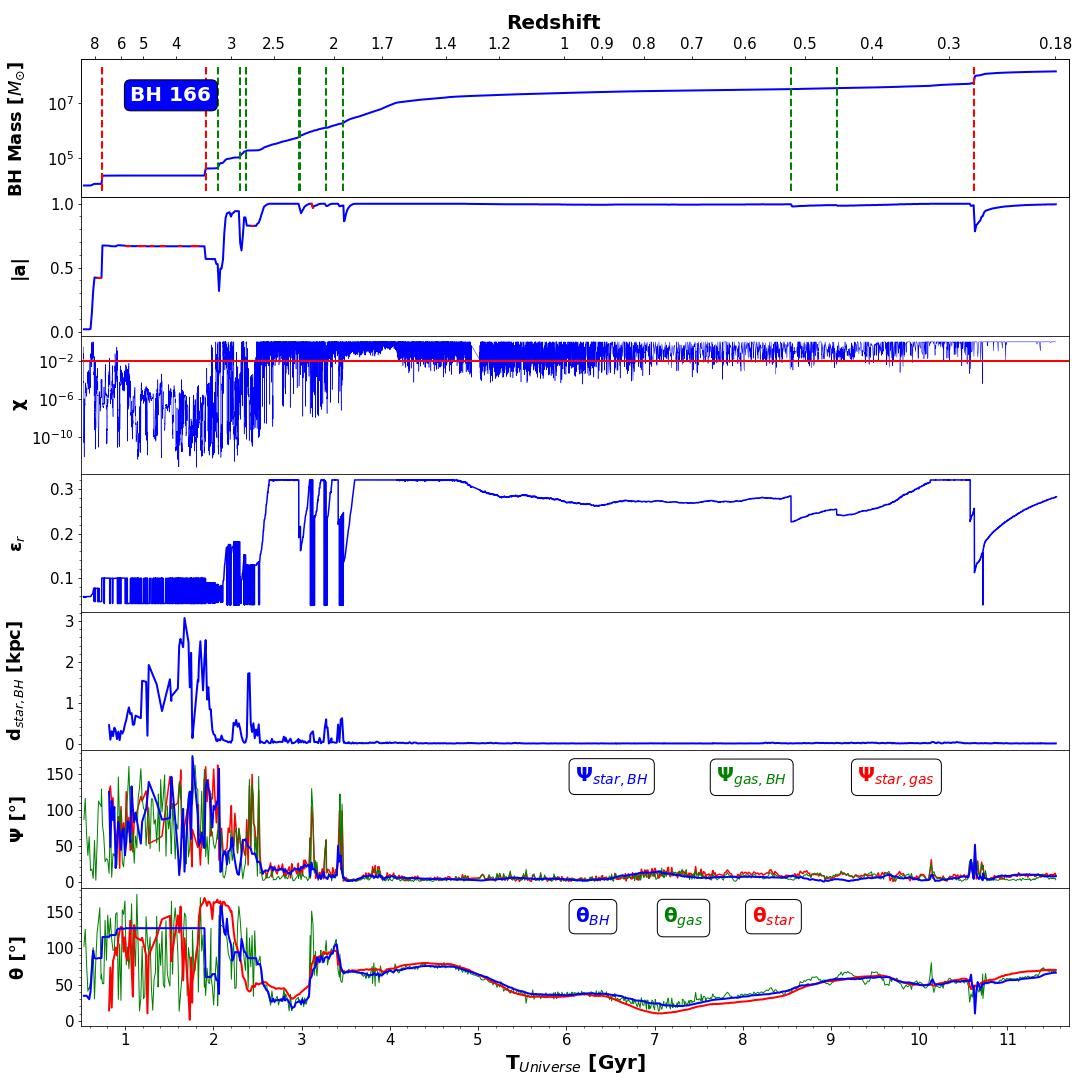}}
\caption{Same as Fig.~\ref{fig_g9685} bur for thee Cosmic evolution of BH-$166$.
  Its evolution differs from BH-G9685 mainly by a higher
  number of encountered major and minor mergers during its life,
  respectively 3 and 9. Hence, the mass gained through BH mergers is
  here much higher, with a value of \fdm=27.3\% at $z=0.18$.  Note
  that from its birth to \tuniv=0.81 Gyr, we were unable to estimate
  the variations of \PSIS as this BH is not affected to a galaxy yet
  following our selection criteria (it is a  ``wandering'' BH).  }
\label{fig_bh166}
\end{center}
\end{figure*}

\begin{figure}
\begin{center}
\rotatebox{0}{\includegraphics[width=\columnwidth]{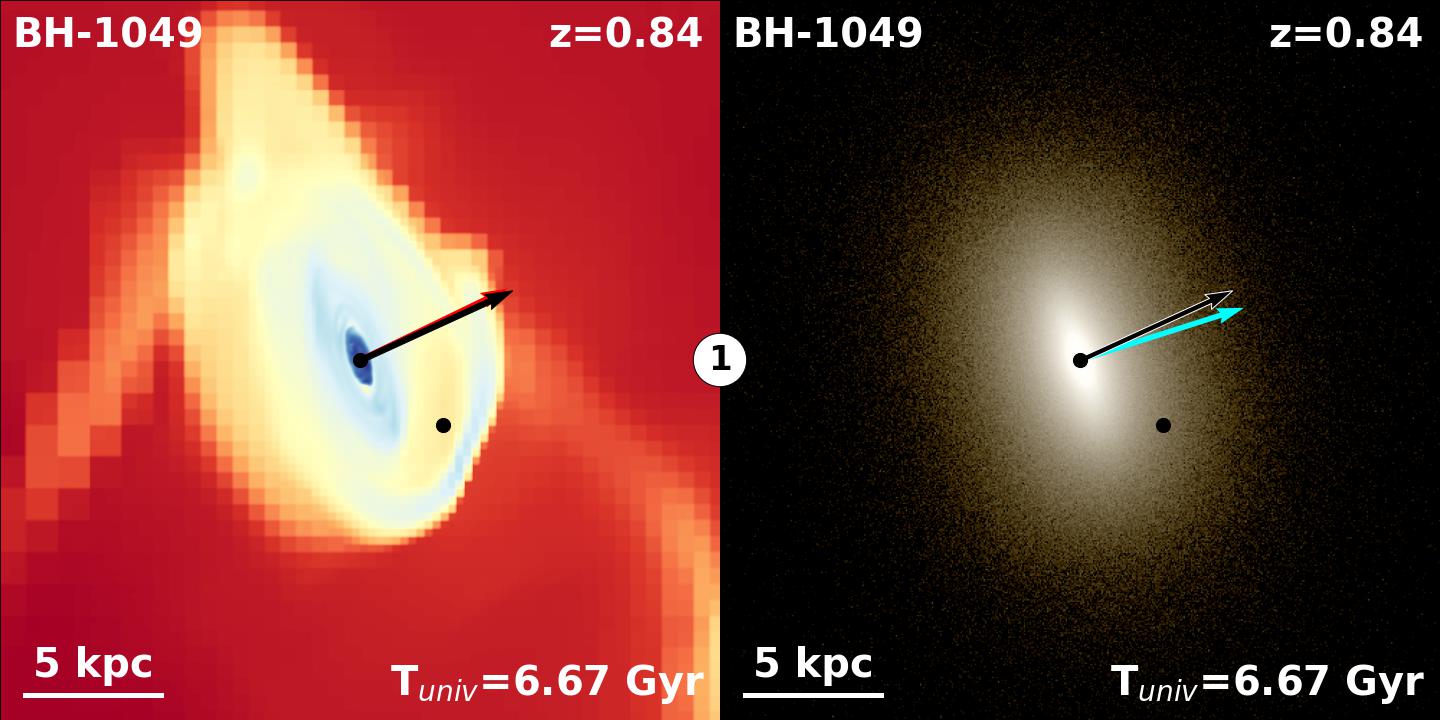}}
\rotatebox{0}{\includegraphics[width=\columnwidth]{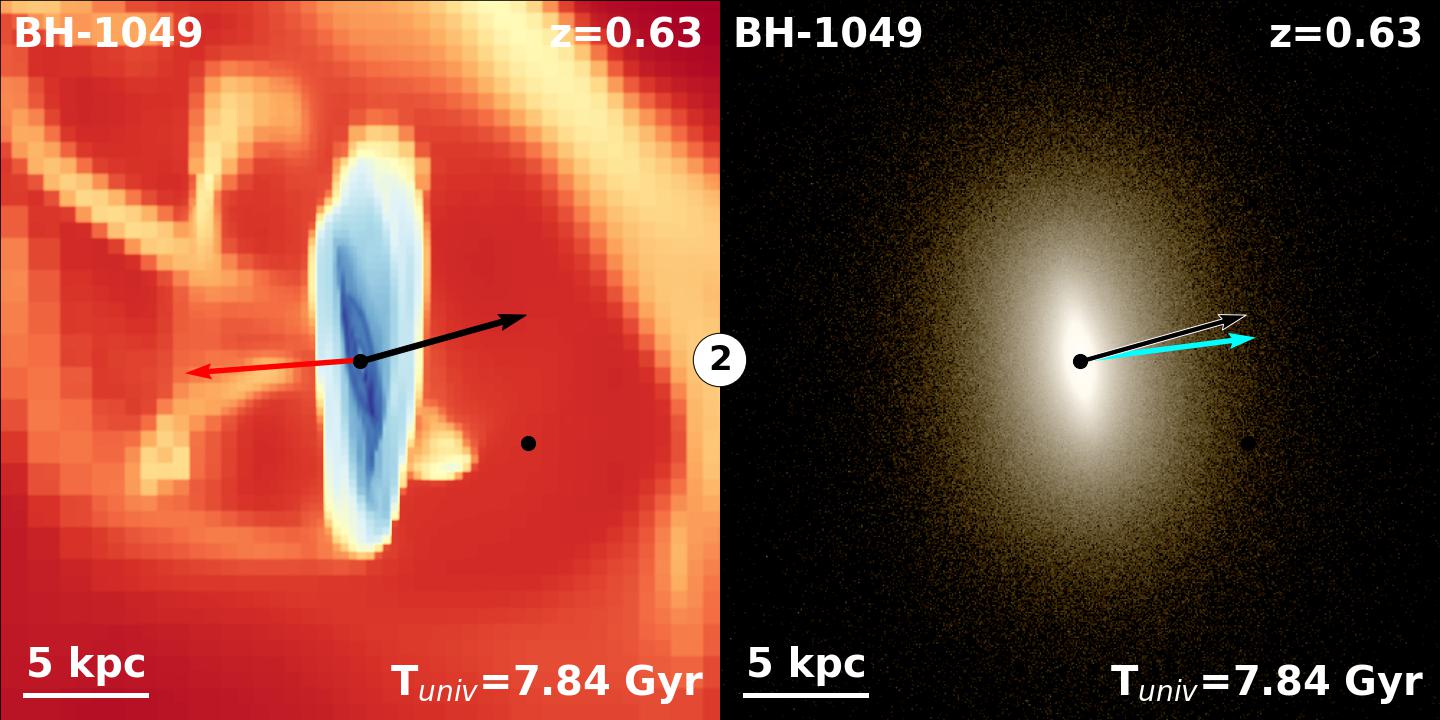}}
\rotatebox{0}{\includegraphics[width=\columnwidth]{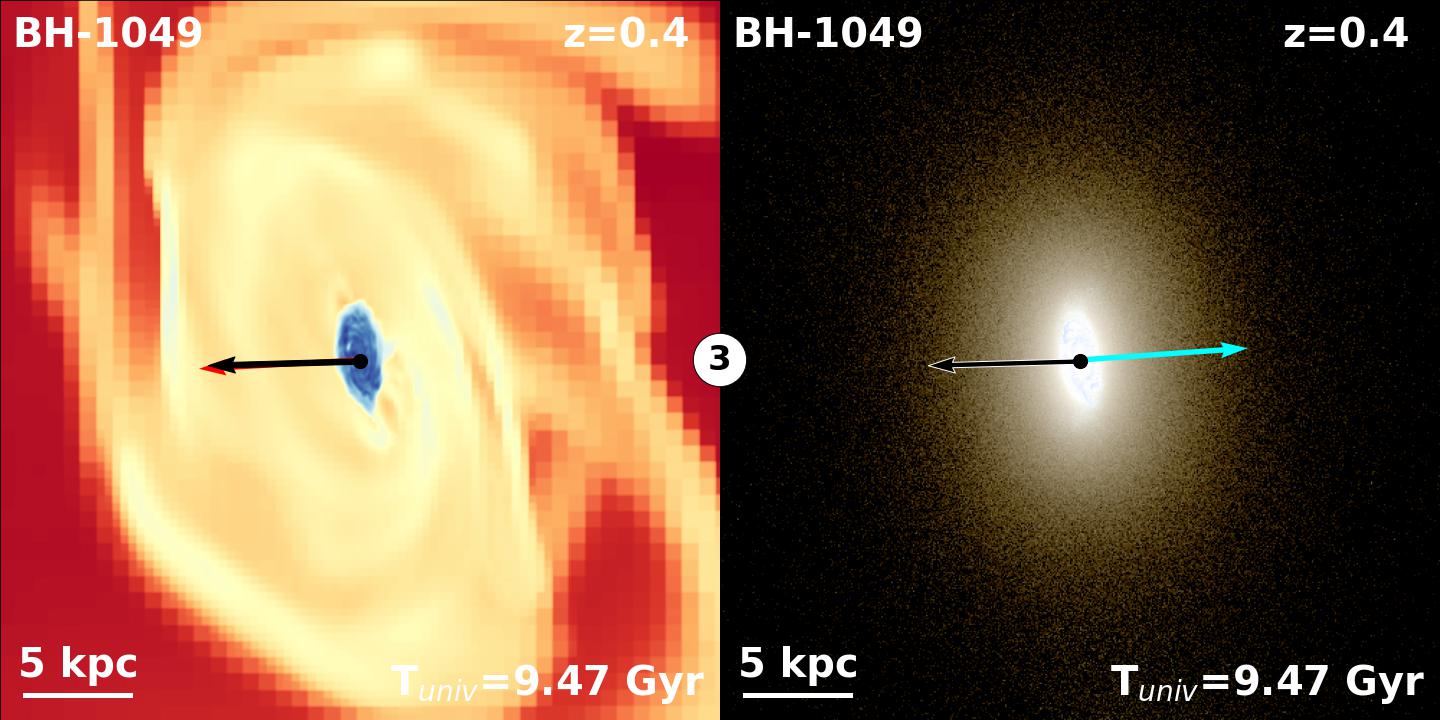}}
\caption{Same as Fig.~\ref{fig_G9685_map} but for BH-$1049$ and its host
  galaxy.  This system has a much more complex evolution than
  BH-G9685 and BH-$166$ as the host galaxy is passing by two
  successive phases of cosmic gas accretion with a retrograde rotation
  with respect to the galactic disk. We illustrate here the first
  phase: \textcircled{\tiny{1}} Due to a cosmic inflow of gas, there
  is the coexistence of two disks of gas rotating in opposite
  direction. The central and original one (dark blue) is progressively
  disappearing and replacing by the new disk (ligh blue) from the outer parts. All
  the spins are still aligned.  \textcircled{\tiny{2}} Once the
  original disk has completely disappeared, The spin of the accreted
  gas (red) flips to follow the rotation of the newly formed
  disk. This latter one is counter-rotating with respect to the
  stellar component. The BH is still aligned with the spin of the
  galaxy.  \textcircled{\tiny{3}} After some time, the BH aligns with
  the spin of the gas accretion disk and becomes anti-aligned with the
  galaxy spin.  We note that a new episode of (retrograde) cosmic gas
  accretion is already in progress and will repeat the same scenario
  }
\label{fig_1049_map}
\end{center}
 \end{figure}

\begin{figure*}
\begin{center}
\rotatebox{0}{\includegraphics[width=18cm]{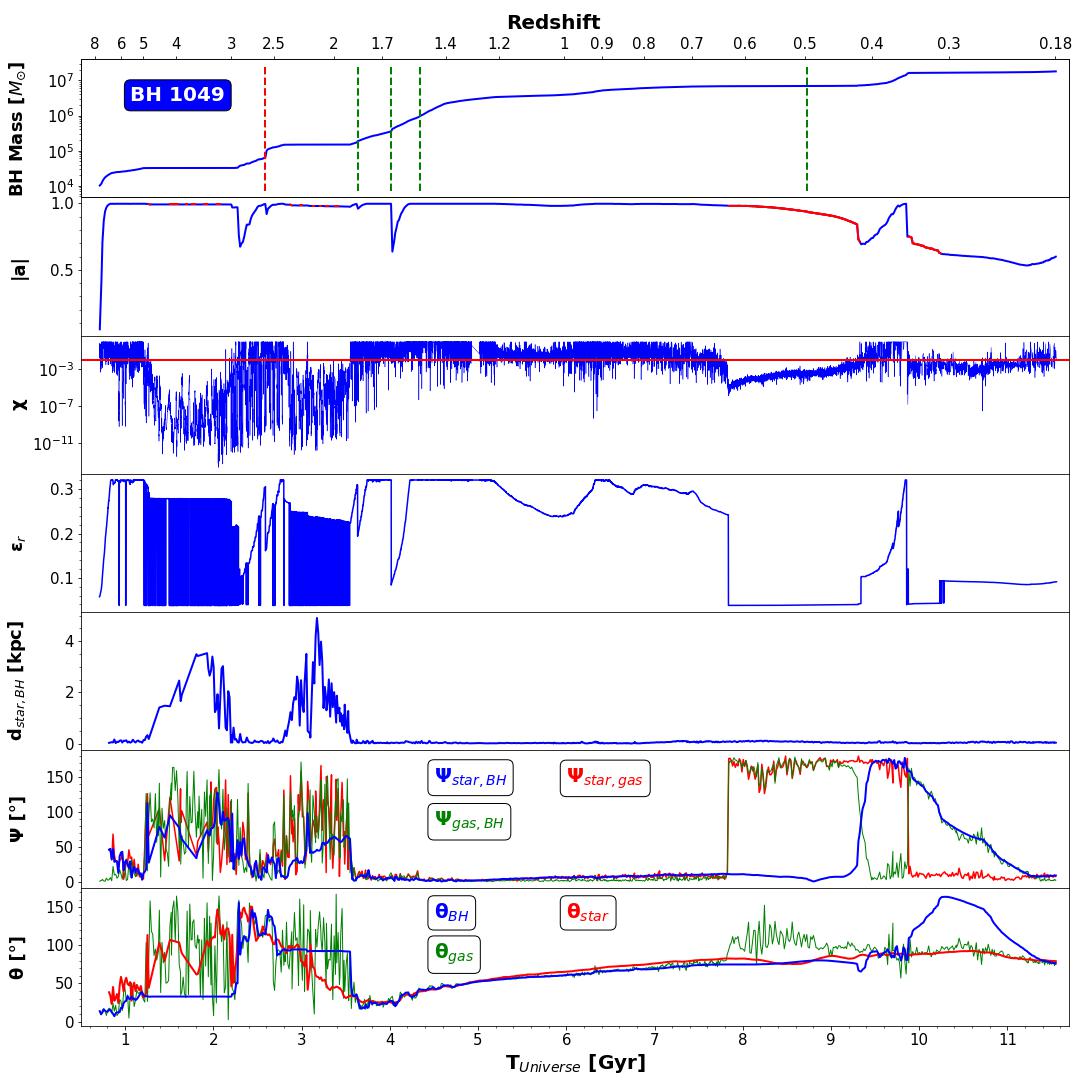}}
\caption{Same as Fig.~\ref{fig_g9685} but for the cosmic evolution of
BH-$1049$.
This example displays a phase in which the accretion of gas onto the BH has
a retrograde accretion (7.8$\lesssim$ \tuniv$\lesssim$ 9.8 Gyrs) 
which drives the orientation of \JBHS toward anti-alignment with respect
to the stellar component.
}
\label{fig_bh1049}
\end{center}
 \end{figure*}

The second example, BH-$166$, is extracted from the \nh\, simulation.
The projected images of the stellar and gas distributions of its host galaxy
at different
redshifts are displayed in Fig.~\ref{fig_166_map}. The evolution of
its main properties is plotted in Fig.~\ref{fig_bh166}.  The BH was
born at $z=9.4$ (\tuniv=0.53 Gyr) and became quite massive,
M$_{\mathrm{BH}}$=1.4$\times$10$^{8}$M$_\odot$ at $z=0.18$ (the last output of
simulation).  It is hosted by a galaxy with a final mass of
2.4$\times$10$^{11}$M$_\odot$.  Its evolution significantly differs
from BH-G$9685$ due to the frequent mergers experienced (3 major and 9
minor mergers). As a result, it acquired a high fraction of the mass
through the BH mergers, \fdm=27.3\% at $z=0.18$.

Nevertheless, the evolution of the main properties of BH-166 is quite
similar to BH-G$9685$.  Just after its birth, the amplitude of its
spin shortly reaches the value $a=0.42$.  We note that from its birth to
\tuniv=0.81 Gyr, we were unable to estimate the variations of \PSIS as
this BH is not associated to a galaxy yet following our selection
criteria (i.e., it is a ``wandering'' BH).  BH-$166$ also experienced a long phase
of an inefficient gas accretion until \tuniv$\sim$2.5 Gyr.
During
this period, its mass grows merely through BH mergers, and the BH spin
magnitude, along with its direction, does not evolve significantly. 
Yet, the variations of \PSIS are erratic simply due to
the rapid changes of the stellar component spin, as suggested
by the evolution of $\theta_{\rm star}$.

After its mass exceeds
10$^5$M$_\odot$ at $z\sim1.9$ (\tuniv$\sim$3.5 Gyr), the BH starts to
be settled in the host galaxy as suggested by the evolution of \dsbh,
and evolves through a smooth gas accretion and minor merger episodes.

At $z=0.27$ (\tuniv$\sim$10.6 Gyr), BH-166 experienced a major merger with
BH-$796$ (merger mass ratio: 1-1.4). This resulted in the decrease of
the spin amplitude and misalignment of the spin orientation with
respect to the galaxy spin (up to 50$^\circ$).  However, this
misalignment lasts only for a short period ($\Delta t\sim$100 Myr), after which the BH spin is rapidly realigned to the spin of the host galaxy.  As in the case of BH-G$9685$, the spin orientations of the BH, accreted
gas, and the galaxy are coherently changing over time during this last
phase.

\begin{figure}
\begin{center}
\rotatebox{0}{\includegraphics[width=\columnwidth]{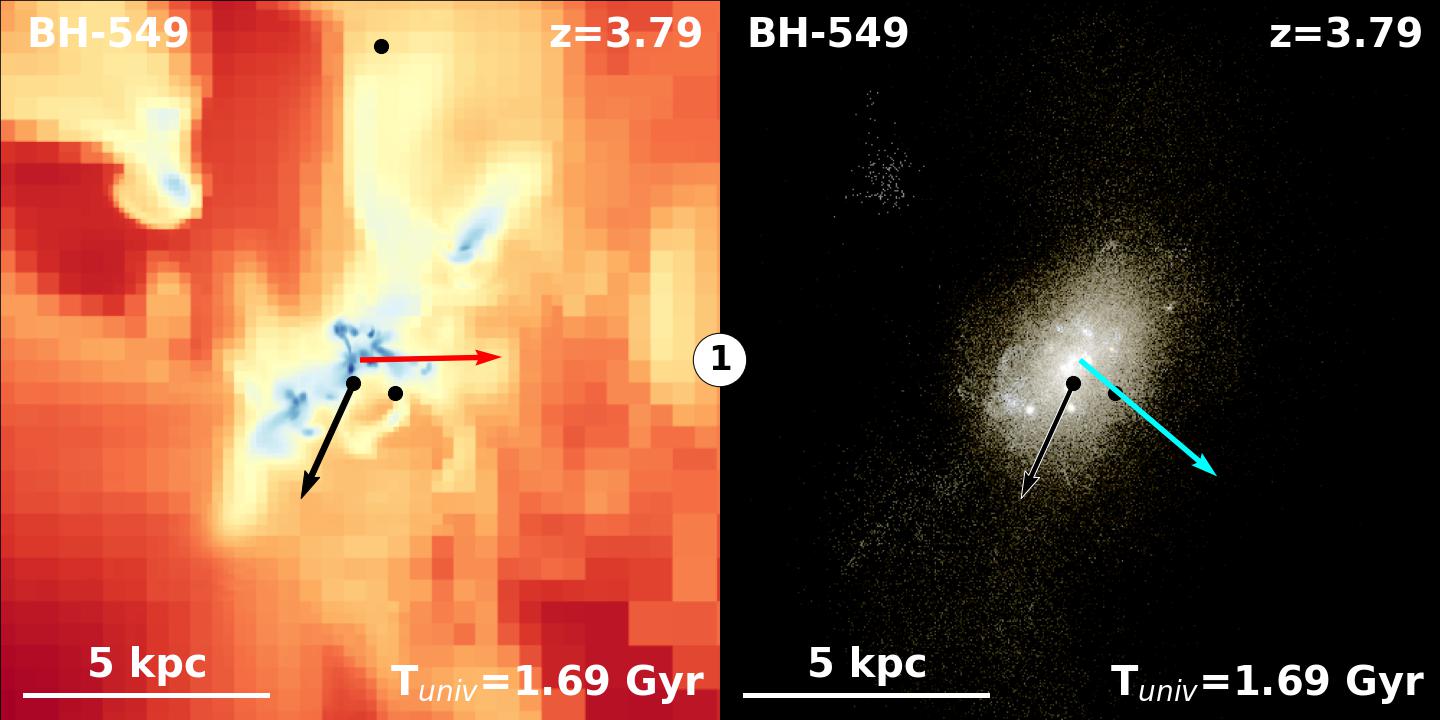}}
\rotatebox{0}{\includegraphics[width=\columnwidth]{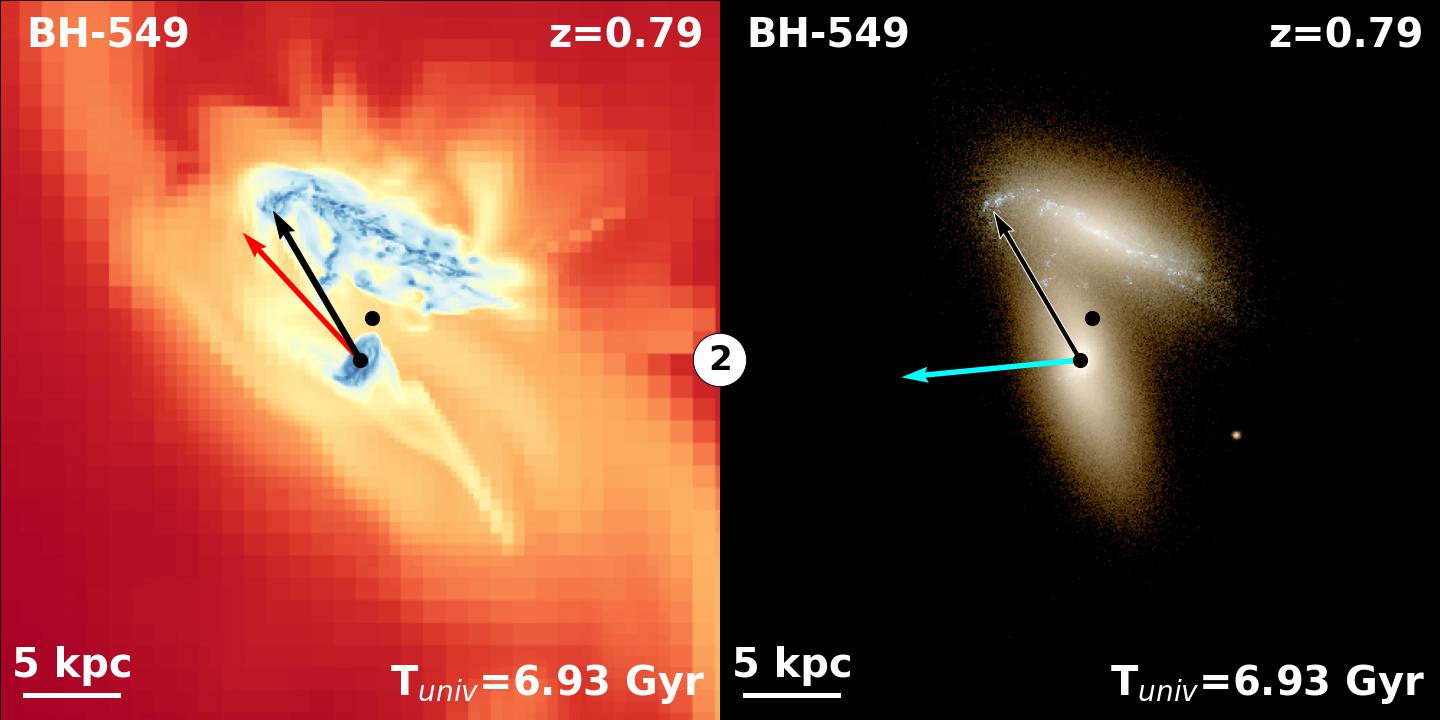}}
\rotatebox{0}{\includegraphics[width=\columnwidth]{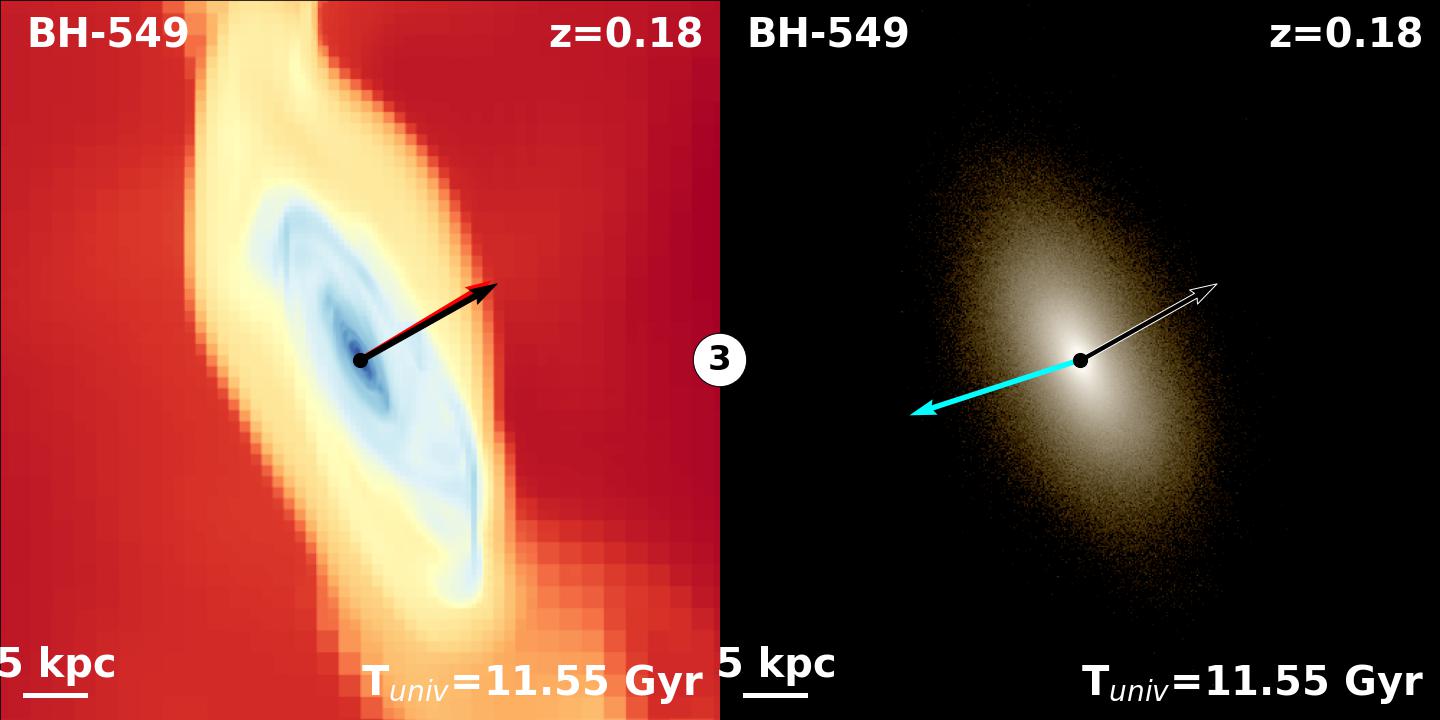}}
\caption{Same as Fig.~\ref{fig_G9685_map} but for BH-$549$ and its host
  galaxy.  This system is displaying an interesting feature since from
  high redshift to z$\sim$0.7, there is no stable disk and coherent
  gas accretion onto the BH.  However, the host galaxy merges with
  another galaxy at this specific epoch as illustrated in the
  panel~\textcircled{\tiny{2}}.  After the merger, a gas of disk if
  forming but rotating in counter-rotation with respect to the stellar
  component.  As a result, the BH spin aligns with the accreted
  material angular momentum vector and becomes anti-aligned with the
  galaxy spin as illustrated in \textcircled{\tiny{3}}. }
\label{fig_549_map}
\end{center}
 \end{figure}

\begin{figure}
\begin{center}
\rotatebox{0}{\includegraphics[width=\columnwidth]{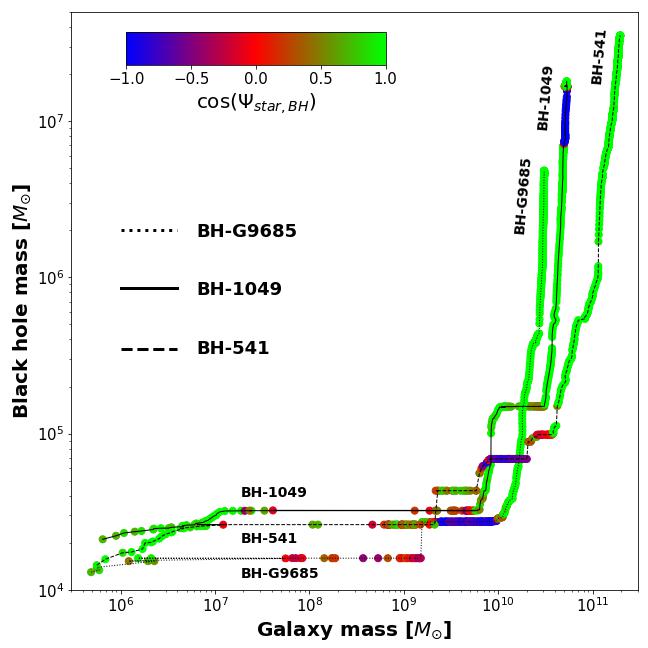}}
\caption{Variations of the BH mass v.s. the galaxy mass with color
  coding according to the value of \COSPSI.  We show three typical
  evolutions of intermediate mass BHs to illustrate the three different
  regimes followed by the variations of \PSI: 1) a tendency to
  BH-galaxy spin alignment right after their birth and for relatively
  short periods (M$_{\mathrm{BH}}$<3$\times$10$^4$ M$_\odot$ and
  M$_{\mathrm{star}}\lesssim$2$\times$10$^7$ M$_\odot$) 2) erratic variations
  (M$_{\mathrm{BH}}$$\lesssim$10$^5$ M$_\odot$)
3) new tendency to alignment at late time (5$\times$10$^9$ M$_\odot$
$\lesssim$ M$_{\mathrm{star}}$).  Note that in the case of BH-1049, the blue
region at late time corresponds to the BH spin anti-alignment
situation.  }
\label{fig_color_map}
\end{center}
 \end{figure}

The third example, BH-$1049$, is also a massive object with
M$_{\mathrm{BH}}$=1.8$\times$10$^7$M$_\odot$ at $z=0.18$, hosted by a galaxy with a
final mass of 5.3$\times$10$^{10}$ M$_\odot$.  Images of the projected
distribution of stars and gas at different redshifts are displayed in
Fig.~\ref{fig_1049_map}, and the evolution of its main properties is
presented in Fig.~\ref{fig_bh1049}.  Its mass evolution is dominated
by a smooth accretion with one major merger at $z\sim$2.6 and four
subsequent minor mergers, leading to a value of \fdm=0.79\% at
$z=0.18$.
Right after being seeded at $z=7.5$ (\tuniv=0.7 Gyr), its spin
magnitude increases until it reaches a maximum value (a=0.998) with
\PSI<50$^\circ$. Similarly to the previous two cases, BH-$1049$ crosses
a regime dominated by chaotic gas accretion until it is completely
settled at the center of its host galaxy at \tuniv$\sim$3.6 Gyr
(M$_{\mathrm{BH}}$>10$^5$M$_\odot$).  Since then, its spin is well aligned
with \JSTARS, and this configuration remains for about 5.6 Gyr until
\tuniv$\sim$9.2 Gyr.

Later, its evolution exhibits a much more complex behavior.  One
can notice from the time evolution of \PSISTARGAS\, and \PSIS, at
\tuniv$\sim$7.8 Gyr, that the gas accretion disk does indeed start a retrograde rotation
relative to the BH spin and stellar component. This state lasts until
\tuniv$\sim$9.8 Gyr.  During this phase, the BH spin also starts to be
misaligned with that of the galaxy, and eventually becomes fully
anti-aligned. Then, it becomes gradually realigned with the galaxy
spin down to $z=0.18$.  In fact, the host galaxy undergoes two
successive phases of the cosmic gas accretion from larger scales.  As
a result, two gas disks with different rotations emerged and coexisted:
the original one in the inner parts (and corotating with the stellar
disk) and a second one in the outer parts which is
counter-rotating. The original one is progressively replaced by the new
counter-rotating disk until it totally disappears at \tuniv$\sim$7.8 Gyr when
the gas accretion disk spin flips.  Meanwhile, such retrograde accretion
decreases the spin magnitude toward $a=0$ and then grows increasingly
negative $a<0$, which drives the BH spin orientation toward
anti-alignment. 

Once the BH spin and the angular momentum of the gas are re-aligned
(9.4$\lesssim$\tuniv$\lesssim$9.8 Gyr), the BH spin, radiative
efficiency and mass significantly increase.  However, due to the
second phase of cosmic accretion of gas, the same phenomenon happens
(coexistence of two gas disks) and the situation reverses: once the
inner disk has been replaced by the new gas disk (\tuniv$\sim$9.8
Gyr), the angular momentum of the gas accretion disk becomes aligned with
that of the stellar component (second spin flip) until the end of the
simulation.  Consequently, the BH spin and accretion disk angular
momentum start to be anti-aligned again (\tuniv$\sim$9.8 Gyr). In the
final phase of the evolution of \PSIS, the accretion disk drives the
orientation of the BH spin toward alignment.  While this example may
be an extreme case, it illustrates very well how the long phase of
cosmic inflow of gas from larger scales affects the orientation of the
BH spin.  The detailed study of this galaxy will be presented in a
separate paper (Peirani et al. in prep) which focuses on the formation
and evolution of counter-rotating gas and stellar disks.

The fourth and last example concerns the BH-$549$.  Here again,
  Fig.~\ref{fig_549_map} shows images of the projected distribution
  of stars and gas at different redshifts.  The evolution of its main
  properties is plotted in Fig.~\ref{fig_appendix1} in
  Appendix~\ref{appendix1}.  BH-549 is an interesting case since the
  time evolution of \PSIS indicates that its spin is anti-aligned with
  the galaxy spin during the last $\sim$4.5 Gyr.  We find that 
  the anti-alignment is triggered by the merger with another galaxy at
  \tuniv$\sim$7.2 Gyr that destroyed the galactic gas disk,
   as also demonstrated by \cite{park+19} by studying other \nhs 
   galaxies.  Then, a
  new disk of gas is rebuilt. Due to the specific impact parameter of
  the merger, however, the new gas disk is counter-rotating with
  respect to the original stellar component. This leads to a stable
  configuration in which the BH and galaxy spins are anti-aligned over
  $\sim$4.5 Gyr. 
 It is worth mentioning that
  \cite{capelo_dotti17} found similar trends
  using idealized simulation of galaxy mergers
while highlighting the role of the ram pressure shocks
in the modification of  the angular momentum budget 
of the gas distribution.

Appendix~\ref{appendix1} presents the evolution of other BHs as
  well.  Our analysis of typical cases identified the occurrence of
  three successive phases that an intermediate mass BH (>10$^6$
  M$_\odot$) experiences during its lifetime.  We summarize those
  phases in Fig.~\ref{fig_color_map} that shows the coevolution of
  three BH-galaxy systems using the color code according to
  the value of \COSPSI.  

\begin{figure*}
\begin{center}
\rotatebox{0}{\includegraphics[width=\columnwidth]{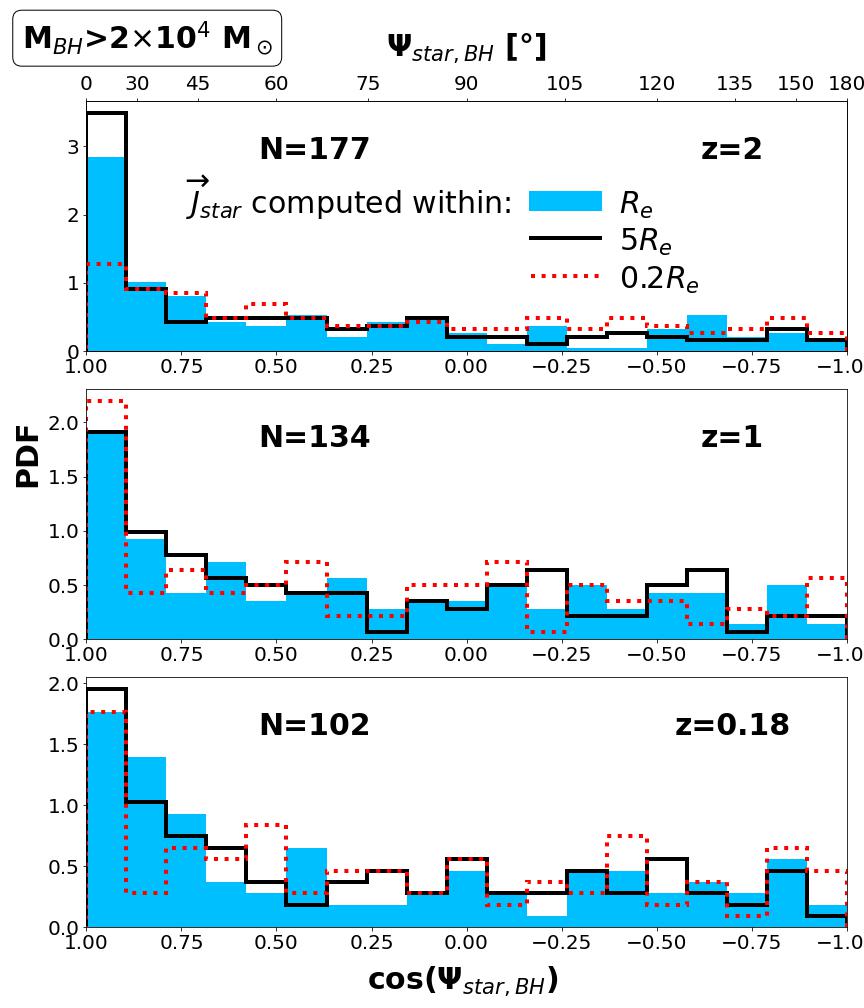}}
\rotatebox{0}{\includegraphics[width=\columnwidth]{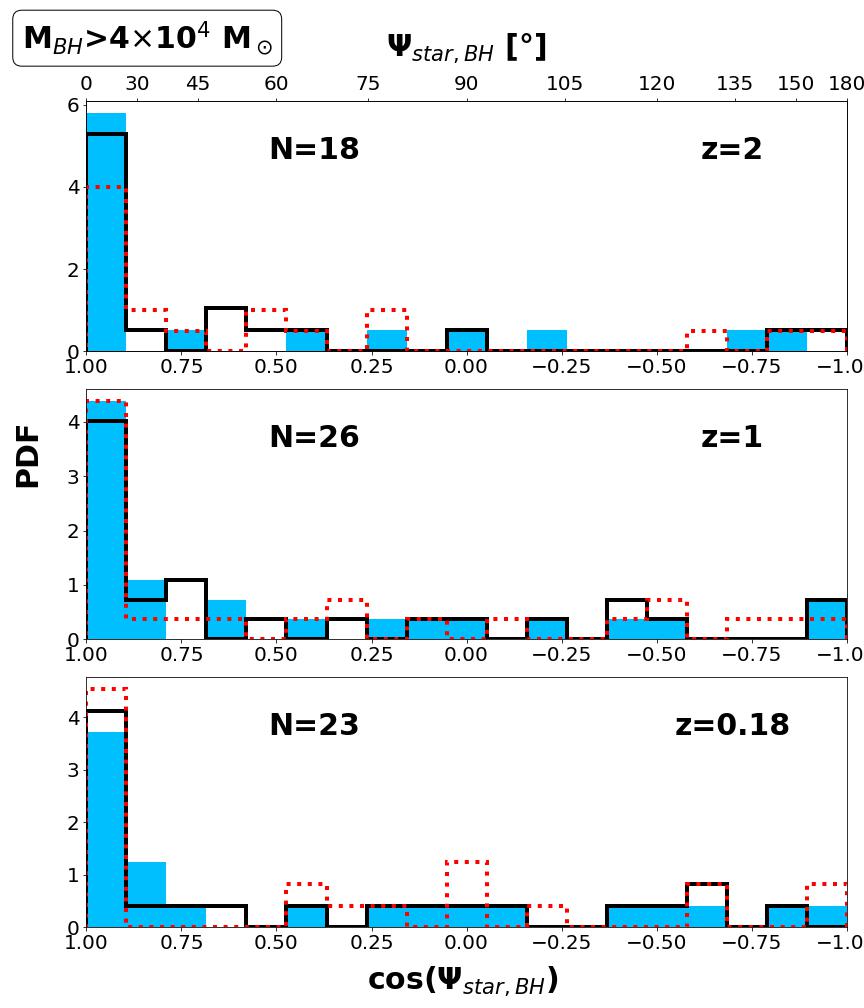}}
\caption{Normalized distributions of \COSPSIS, the angle between the
  BH spin, and the angular momentum of the stellar component of the
  host galaxy. The latter was computed within spheres of radius $R_e$
  (blue histograms), 5$\times R_e$ (black lines), and 0.2$\times R_e$
  (dotted red lines). The results are showns at three different redshifts
  (z$=$2, 1 and 0.18) as well as two lower BH mass limits,
  2$\times$10$^4$ M$_\odot$ (left column) and 4$\times$10$^4$
  M$_\odot$ (right column).  In each panel, the number of primary
  black holes considered to derive the histograms is indicated.
}
\label{fig_pdf}
\end{center}
 \end{figure*}

\begin{figure}
\begin{center}
\rotatebox{0}{\includegraphics[width=\columnwidth]{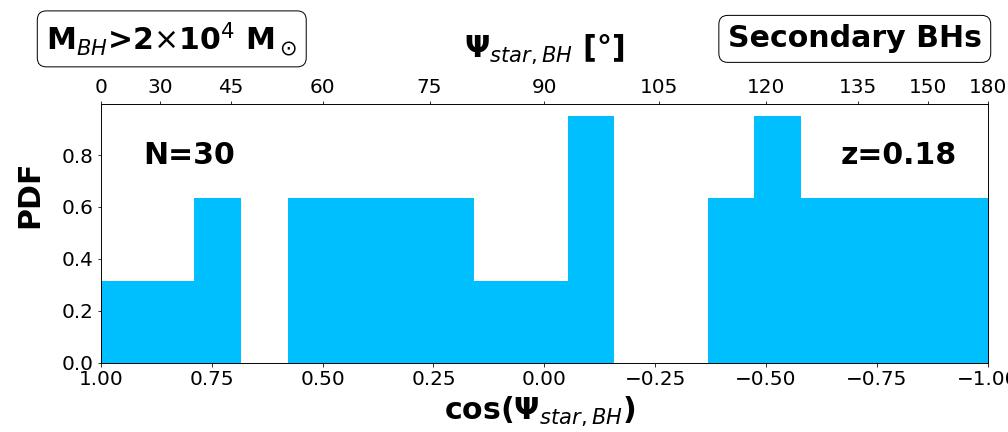}}
\caption{Normalized distribution of \COSPSIS
for secondary BHs at $z=0.18$. The sample consists of 30
BHs with a mass greater than 2$\times$10$^4$ M$_\odot$.
Contrary to primary BHs, the distribution tends to be uniform
suggesting no privileged orientation between the BH spin
and the angular momentum of their host galaxies.}
\label{fig_sec}
\end{center}
 \end{figure}

\subsubsection{Statistical analysis}

We now turn to examining the statistical evolution of \PSIS using the
whole samples of BHs at different redshifts.  In particular, we
examine if the distribution of \COSPSIS exhibits a departure from the
uniform distribution expected from the isotropic BH spin orientation.

Fig.~\ref{fig_pdf} plots the normalized distribution of \COSPSIS for
our sample of primary BHs at three different redshifts. The left
panels are derived from our fiducial samples of
M$_{\mathrm{BH}}>$2$\times$10$^4$M$_\odot$. 
The general tendency of alignment (\COSPSIS close to 1) is
clearly seen between the black hole spin and the angular momentum of
the host galaxy at high ($z=2$), intermediate ($z=1$) and low
redshifts ($z=0.18$).  In section~\ref{sec:dependence}, we study
in detail to what extent the values of \PSIS depend on the system
properties including the BH mass, the host galaxy mass, the BH spin
magnitude, the fraction of mass gained through BH mergers, the
distance between the BH position and the center of the galaxy, and the
morphology of host galaxies.

We confirmed that changing the definition of the radius within which
we estimate the angular momentum of the host galaxies (i.e., $R_e$) has
no significant impact on the distribution. For instance, the same
trends are also seen even if we compute the angular momentum of the
galaxies within 5$R_e$ (black lines) and 0.2$R_e$ (red dotted lines).
It is also the case when we take into account all star particles
selected by {\mbox{{\sc \small Adaptahop}}, as it
was considered in \cite{dubois+14a} and \cite{beckmann+23}.
For clarity, we do not show the results
  in Fig.~\ref{fig_pdf}.
  We note that for the result with 0.2$R_e$ at high
  redshift, the alignment seems to be less pronounced than for those
  with $R_e$ and $5R_e$, but it is likely due to the resolution limit
  required to estimate \JSTARS accurately for regions within $0.2R_e$.

Finally, we also investigate how the results depend on the lower limit
of the BH mass. The right panels of Fig.~\ref{fig_pdf} show the
distributions for BHs with M$_{\mathrm{BH}}$>4$\times$10$^4$M$_\odot$. They present
an even stronger trend of the BH-galaxy spin alignment.  Although our
study is focused on primary BHs, it is also interesting to take a look
at trends obtained
from the sample of secondary BHs with M$_{\mathrm{BH}}$>2$\times$10$^4$M$_\odot$ at
$z=0.18$.  Only 30 BHs satisfy the criteria. Secondary BHs are
indeed mainly off-centered from the galaxy center by definition, and
therefore their mass growth is very modest due to an inefficient gas
accretion.  In this case, the distribution of \COSPSIS 
(Fig.~\ref{fig_sec}) tends to be more uniform (i.e., there is no
preferential orientation between the BH and galaxy spins), in contrast
to the primary BHs.
This may be explained by the fact that off-centered BHs
have generally low accretion rates and low Eddington ratios (see
for instance the evolution of BH-132 in Fig~\ref{fig_appendix1}). 
Consequently, and as mentioned before, the re-alignment
between BH spin and the accretion disk takes 
a long time \citep{lodato+06,lodato+13}. Thus,
 the direction of the BH spin remains virtually unchanged. 
On the contrary, the orientation of the stellar spin 
changes more rapidly and can explain the distribution in
Fig.~\ref{fig_sec}.

A complementary way to characterize the statistics of \PSIS is the
cumulative distribution of \COSPSIS, from which one can easily extract
the percentage of the aligned BH-galaxy systems below a given angle.
The upper panel of Fig.~\ref{fig_cum} plots those at $z=3$, 2, 1, and
0.18.  It appears that the BH-galaxy systems tend to be more aligned
at higher redshifts. About $\sim$44\% of our BH sample have
\PSI$\leq$30$^\circ$ at $z=3$, while this number drops to $\sim$24\% at
$z=0.18$.  Yet, we notice that the trend is reversed between $z=1$ and
$z=0.18$.  To see more clearly the variation of such quantities, we
plot in the lower panel of Fig~\ref{fig_cum}, the time evolution of
the fraction of BH-galaxy pairs that satisfy \PSI$\leq$45$^\circ$ (green
line), \PSI$\leq$30$^\circ$ (blue line) and \PSI$\leq$15$^\circ$ (red line).  Two
different regimes are clearly visible.  First, until $z\sim1.5$, the
fraction of aligned BH-galaxy pairs is decreasing, independently of the choice of the
\PSIS maximum value.  Then it seems to
remain approximately constant until the end of the simulation.
However, we note a slight increase from \tuniv\, close to 11 Gyr. 
We subsequently see in the next section that these variations are strongly
correlated with the BH merger rate.  However, it is worth mentioning
that we cannot exclude that adding an extra refinement level in the
simulation at $z\sim1.5$ and $z\sim0.25$ might cause the spurious
artifacts in the evolution shown in the lower panel of
Fig.~\ref{fig_cum}.  The refinement results indeed in a sudden better
force resolution and an enhanced gas condensation, which may
temporally impact the gas accretion in the BH.

\begin{figure}
\begin{center}
\rotatebox{0}{\includegraphics[width=\columnwidth]{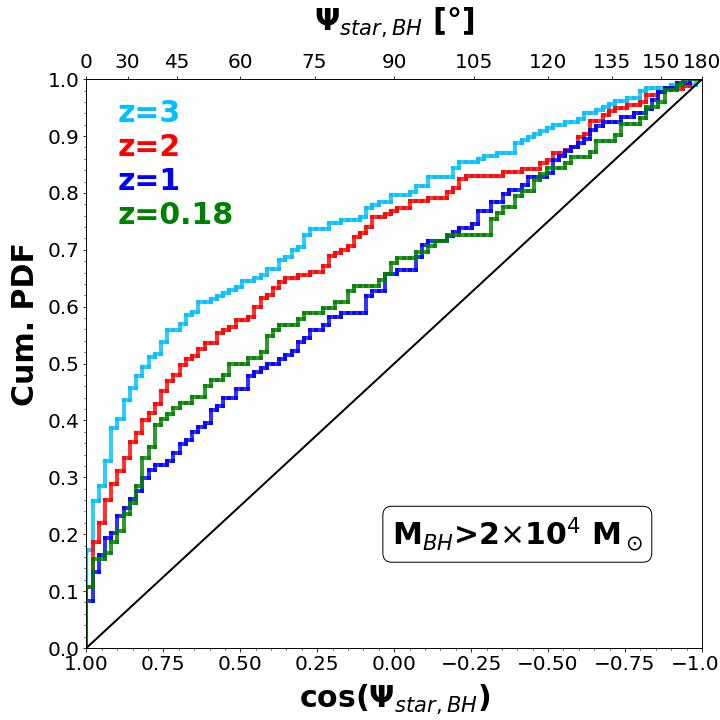}}
\rotatebox{0}{\includegraphics[width=\columnwidth]{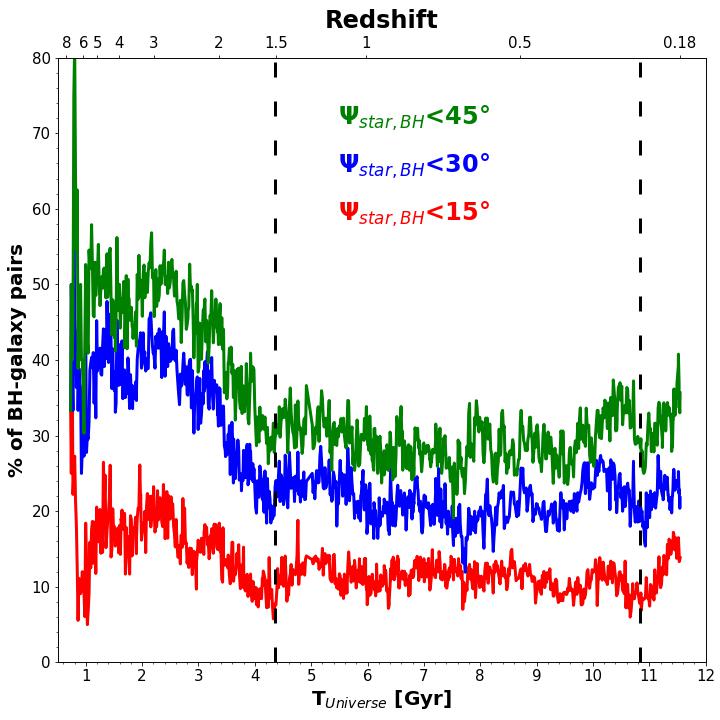}}
\caption{Cumulative distributions of \COSPSIS at
  different redshifts (Upper panel). The black line represents the cumulative
  distribution of a uniform distribution.  In the lower panel, we show
  additionally show the evolution
  of the percentage of BHs or BH-galaxy systems that present an
  alignment angle \PSIS lower than 45$^\circ$ (red line), 30$^\circ$
  (blue line), or 15$^\circ$ (red line). The two vertical dashed lines
  indicate the epochs where an additional level of refinement was
  added in the simulation.  }
\label{fig_cum}
\end{center}
 \end{figure}

\subsubsection{Effect of BH mergers}

The individual evolution of BHs presented in
Figs~\ref{fig_g9685}-\ref{fig_bh1049} and in the
Appendix~\ref{appendix1} suggests that the BH mergers suddenly
decreases the BH spin magnitude as well as the rapid change of its 3-d
orientation. We dedicate this section to study their impact
statistically.

Let us first examine the correlation of \COSPSIS with the BH mass and the BH spin parameter. It is shown in Fig.~\ref{fig_merger1}
for primary BHs with a mass greater than \MASSLIM, where the color of
the symbols indicates the percentage of mass gained through BH-merger
(\fdm).  At z=3, most BHs with M$_{\mathrm{BH}}\sim$ a few 10$^4$M$_\odot$
(i.e., have doubled or tripled their seed mass) have in general $|a|$
close to unity; they are still in
the process of spinning up through early gas accretion as suggested by
\fdm\, close to 0\%.  This behavior was already seen in the individual
evolution of black hole right after their birth in
Figs.~\ref{fig_g9685} and \ref{fig_bh1049} 
\citep[see also Fig.~23 of ][]{nhz}. 
During this phase, their spin is more likely
to be aligned with the angular momentum of the stellar component.  We
also note the existence of a population of BHs with very low \fdm\,
and high \PSIS values. This population of BHs most likely experienced
the phase of chaotic gas accretion seen again in the individual
evolution of BHs.  We also note that other BHs have already undergone
major mergers (red filled circles), which tend to decrease the
magnitude of their spin while increasing the scatter in the
distribution of \COSPSI.  At low redshifts, the same trend is observed
in the \COSPSI-$|a|$ diagram; BH mergers tend to decrease the
magnitude of the BH spin.  Regarding the correlation between \COSPSIS
and \fdm\, no clear trend is visible. In fact, even BHs with high
\fdm\, can have their spin well-aligned with their host. This is
because their spin can be realigned rapidly as suggested by the
evolution of BH-166 in Fig.~\ref{fig_bh166} and several BHs in
Appendix~\ref{appendix1}.  This is the case for the two most massive
black holes of our sample, BH-166 and BH-455, which have a high \fdm\,
value (27.3\% and 23.8\% respectively) while having low values of
\PSIS (7.7$^\circ$ and 4.9$^\circ$ respectively).

\begin{figure*}
\begin{center}
\rotatebox{0}{\includegraphics[width=\columnwidth]{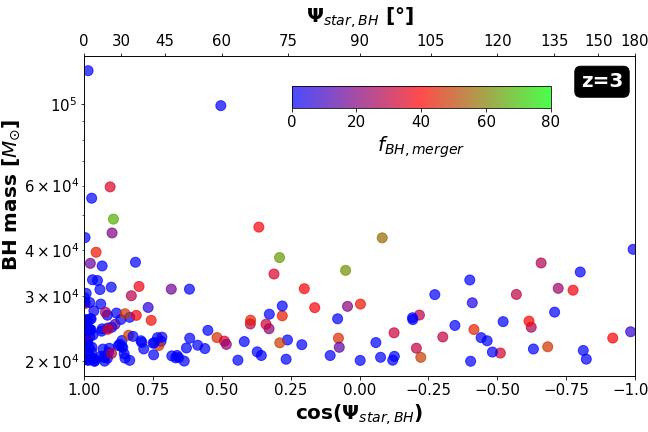}}
\rotatebox{0}{\includegraphics[width=\columnwidth]{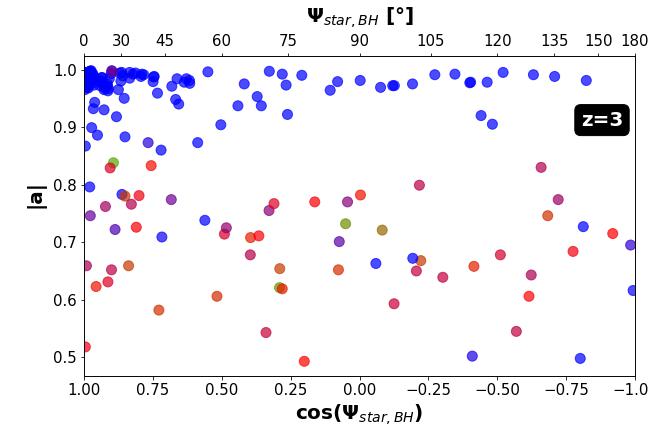}}
\rotatebox{0}{\includegraphics[width=\columnwidth]{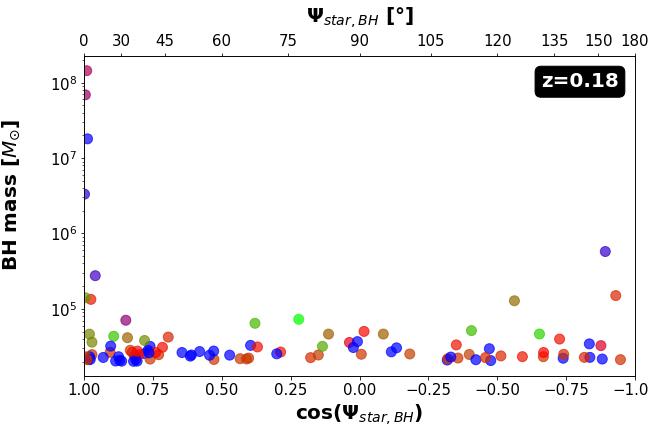}}
\rotatebox{0}{\includegraphics[width=\columnwidth]{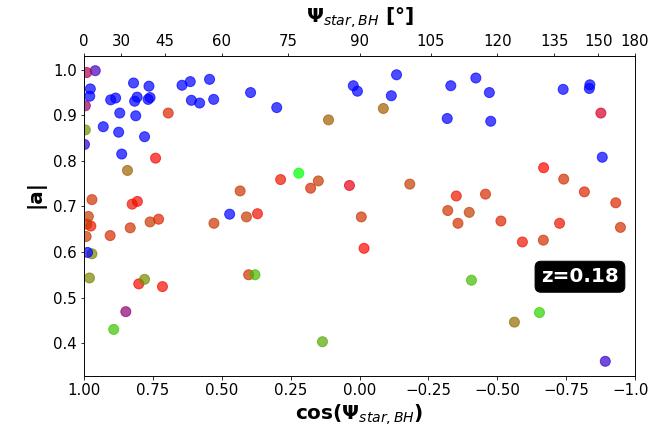}}
\caption{Correlation of \COSPSIS against the BH mass (for the
  primary BH with a mass greater than \MASSLIM) (left panels) and the
  BH spin magnitude (right panels), with color coding according to the
  percentage of mass gain through BH-merger \fdm\, at each specific redshift.
  At $z=3$, most BHs
  whose mass is close to their seed mass ($\sim$[1-3]$\times$10$^4$
  M$_\odot$) are still in the process of spinning up through early gas
  accretion (i.e., blue filled circles).  Their dimensionless spin
  parameters are close to unity, and tend to be aligned with the
  stellar component angular momentum. Other BHs have already undergone
  major mergers (red and green filled circles) which tend to decrease the
  magnitude of their spin while increasing the scatter in the
  distribution of \COSPSI. At low redshift, the same trend is observed in
  the \COSPSI-$|a|$ diagram.
  The corresponding correlations between  $|a|$ and the BH mass
  can be directly seen in Fig.~23 of \cite{nhz}.
}
\label{fig_merger1}
\end{center}
 \end{figure*}

We also study the correlation of \fdm\, with
\COSPSIS in the first column of Fig.~\ref{fig_scatter2}, at $z=2, 1$ and $0.18$.  A
visual inspection indicates that BHs which have experienced a few
merger episodes in their life, for instance those with \fdm< 10\%
(blue filled circles), are more likely to have their spin aligned with
the angular momentum of their host galaxy. This trend is clearly
visible at every redshift.  On the contrary, BHs that have undergone
more merger episodes and characterized by high \fdm\, values
(e.g., \fdm>10\%, red filled circles), have their spins more uniformly
distributed at $z=2$ and $z=1$.  This is not the case at $z=0.18$
because, as explained in the next paragraph, BH spins tend to be
realigned to the galaxy spins since the subsequent BH mergers occur
rarely.

Finally, Fig.~\ref{fig_merger_rate} plots the BH merger rate of our
primary BHs in the volume of the studied simulation. We note first an
increase of the number of BH mergers prior to $z=2$, which is
associated with a period of BH formation. Between $z=2$ and $z=1$, on
the other hand, the number of BH mergers decreases, and becomes nearly
constant after $z=1$.  Such trends are clearly correlated with the
global evolution presented in the lower panel of
Fig.~\ref{fig_cum}. At high redshift, the number of BHs that are
aligned below a threshold angle decreases due to the high BH merger
rate.  At lower redshifts, BH mergers become rarer and have less
impact on the variations of \PSIS. Thus, the BH spin becomes
progressively realigned with the galaxy spin.  This also explains the
trend in the left panels of fig.~\ref{fig_scatter2} at $z=2$ and
$z=1$: BHs that have undergone more merger episodes (red filled
circles) have generally exhibited the spin misalignment relative to
the galaxy spin.  Between $z=1$ and $z=0.18$, as BH mergers become
less frequent, their impact of the BH-galaxy spin alignment is
reduced. Thus, during the last 5 Gyr, BH spins become realigned with their
host galaxy spin.  This explains why at z=$0.18$ even merger-dominated
BHs can have their spin aligned with \JSTAR.  These results are in
good agreement with \cite{beckmann+23}, who found from \hagn\, that BHs
dominated by non-merger growth (\fdm<10\%) are more likely to be
aligned to their galaxy spins than BHs dominated by merger growth.

\begin{figure}
\begin{center}
\rotatebox{0}{\includegraphics[width=\columnwidth]{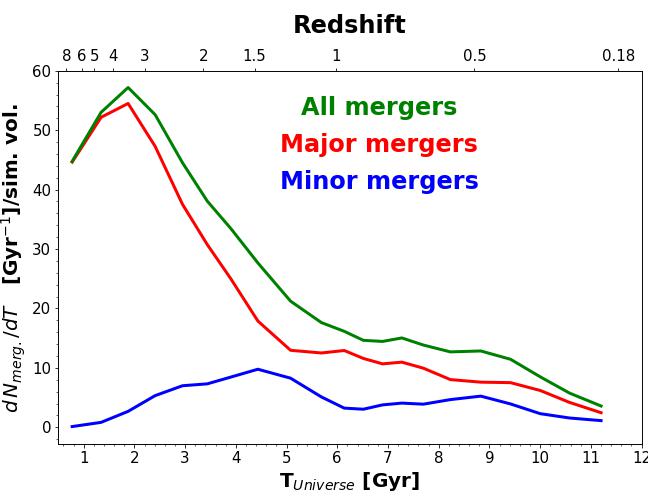}}
\caption{Evolution of the merger rates of primary black holes with
  a mass greater than 2$\times$10$^4$ M$_\odot$ in the simulated
  volume (green line). The red and blue lines correspond respectively to major and
  minor mergers delimited by a 1:4 mass ratio. We have slightly
  smoothed the curves to emphasis the general trends.  }
\label{fig_merger_rate}
\end{center}
 \end{figure}


\subsection{Dependence on BH and galaxy properties} 
\label{sec:dependence}

Finally, we investigate the potential dependence of \COSPSIS on the BH and
galaxy properties. Let's consider first the correlation of \COSPSIS and the BH mass. The
normalized distribution of \COSPSIS for our samples of primary BHs are
shown in the left panels of Fig.~\ref{fig_scatter1}. For clarity, we
have divided our sample into two groups: BHs with mass lower (blue
filled circles) or greater (red filled circles) than $10^5$ M$_\odot$.
As already found in the last sections, BHs with
M$_\mathrm{{BH}}$>10$^5$ M$_\odot$ preferentially show the spin
alignment with \JSTAR.  This result is consistent with
\cite{dubois+14a}.  However, they also found that for more massive BHs
(M$_\mathrm{{BH}}$>10$^8$ M$_\odot$), spins seem to be more randomly oriented
with respect to their host galaxy angular momentum.
Their finding was not confirmed in the present analysis, mainly
because our sample lacks in very massive black holes due to the
limited volume size.
The spins of less massive BHs are more likely to be aligned at high
$z$ since most of them are still in the process of spinning up through
an early efficient gas accretion.  This effect is less visible for
intermediate mass BHs at low redshifts, as already indicated in
Fig.~\ref{fig2}.

Second, we examine the correlation between \COSPSIS and the host
galaxy mass; see middle panels of Fig.~\ref{fig_scatter1} for two
separated samples of galaxies selected with a mass lower (blue circles)
or greater (red circles) than $10^9$ M$_\odot$.  In general, 
the two subsamples present a quite similar
trend that BH-galaxy spins are more likely to be aligned in both
samples. However, as already noticed in Fig.~\ref{fig2}, the trend
is more prominent for low-mass galaxies at high redshifts; see
the blue histogram at $z=2$.

Finally, the right panels of Fig.~\ref{fig_scatter1} plot the
correlations between \COSPSIS and the dimensionless BH spin parameter.
At high redshift ($z>1$), BHs with a higher spin magnitude ($|a|>0.8$)
tend to have a spin more aligned with the galaxy spin than those with
$|a|<0.8$.  This is consistent since BHs spin up rapidly due to an
efficient gas accretion \citep{volonteri+13}. In this case, the
angular momentum transferred to the black hole at the accretion phase
is sufficient to keep the BH spin mostly aligned, in agreement with
\cite{bustamante+19}.  At low redshifts, the trends are even more
pronounced. This is due to the fact that minor/major mergers become
less frequent. Consequently, BH spin magnitudes tend to increase
through a coherent gas accretion, which efficiently aligns
their spin to the angular momentum vector of the stellar
component.

Additionally, we consider the effect of retrograde gas accretion
disks.  Fig.~\ref{fig_a_neg} shows the correlation of \COSPSIS and
$a$; the positive and negative signs of $a$ indicate the prograde
(red) and retrograde or counter-rotating (blue) gas disks,
respectively.  The two histograms in the upper panel indicate counter-rotating gas disks generally
lead to BH-galaxy spins less aligned.

Middle panels of Fig.~\ref{fig_scatter2} plot the distances between
the BH positions and the center of their host galaxy, \dsbh, against
\COSPSIS.  Those BHs located closer to the center of their host galaxy
(<1 kpc) have generally their spin well-aligned with \JSTAR. On the
contrary, for off-centered BHs, the values of \COSPSIS tend to be more
uniformly distributed. This trend was already noticed when studying a
sample of secondary BHs that are off-centered according to our
definition.


Finally, we look into the dependence of \COSPSIS on the morphology type
of the host galaxies.  For each galaxy, we have computed the radial,
tangential and vertical velocity components of stellar particles.  In
doing so, we choose the z-axis of the cylindrical coordinate as the
total angular momentum vector direction of the stellar
component. Then, we estimate the rotation velocity $V$ of the galaxy
by averaging the tangential velocity component. We also compute the
velocity dispersion $\sigma$ from the dispersion of the radial
$\sigma_r$ , the tangential $\sigma_\theta$ and the vertical velocity
$\sigma_z$ components with respect to their averaged values; $\sigma^2
= (\sigma_r^2 + \sigma_\theta^2 + \sigma_z^2 )/3$.  In what follows,
we use the quantity $V/\sigma$ as a proxy of galaxy morphology.

The right panels in Fig.~\ref{fig_scatter2} show the correlation between
\COSPSIS and $V/\sigma$. In those panels, we select more
rotation-dominated galaxies with $V/\sigma > 0.6$, and more
dispersion-dominated galaxies with $V/\sigma <0.6$.  In general, disk
galaxies (i.e., objects with high $V/\sigma$ values) clearly have their
spin well-aligned with that of the associated BH spin.
The trend is however less pronounced for more spheroidal galaxies, especially at high redshifts.  
 This can be explained by the fact that 
 BH spins tend to align with the angular momentum
of the accreted material, which is in general
related to the large scale kinematics of the galaxy.
To this regard, \cite{sesana+14}  found that 
different galaxy morphologies result in different 
spin distributions, in particular that BHs hosted in ellipticals tend to
have lower spins than those hosted in spirals.
Another possible explanation for this discrepancy is
that mergers contribute more significantly to the build-up of such
galaxies at high redshifts.
\citep[e.g.,][]{Boylan-Kolchin+06,trujillo+10,volonteri+13_morpho}.
It could also be a selection effect since galaxies with low $V/\sigma$ values tend to be less massive and characterized  by a more turbulent gas. Consequently, BHs lie in a gas environment owing 
a more random angular momentum which affects their spin direction.
Yet, at z=0.18, the
trend looks quite similar for the two subsamples.  This is consistent
with \cite{beckmann+23} who have derived the distribution of \COSPSIS
from the \hagn\, simulation at z=0 by classifying the galaxy morphology
according to the bulge-to-total ratio (B/T).  They found that
bulgeless systems (with B/T<0.1) exhibit no statistical difference in
the BH-galaxy spin alignment relative to the rest of the sample
(B/T>0.1).


\begin{figure*}
\begin{center}
\rotatebox{0}{\includegraphics[width=6cm]{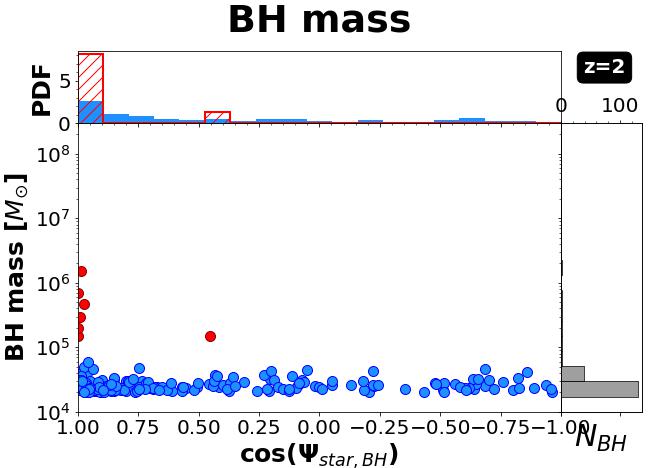}}
\rotatebox{0}{\includegraphics[width=6cm]{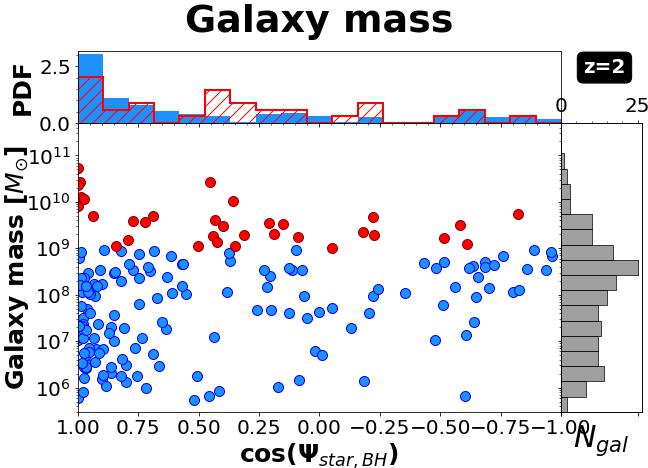}}
\rotatebox{0}{\includegraphics[width=6cm]{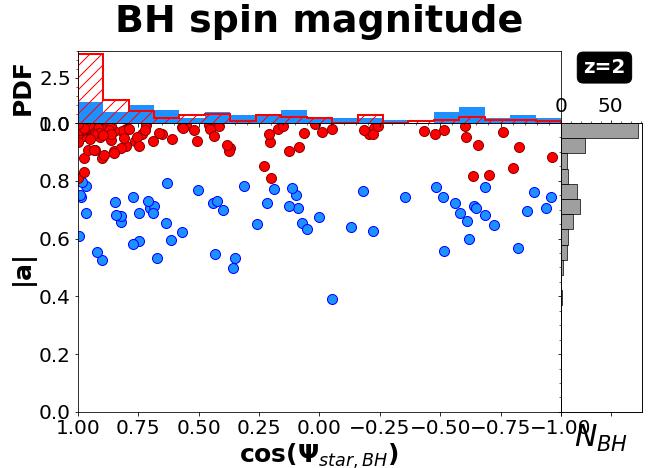}}
\rotatebox{0}{\includegraphics[width=6cm]{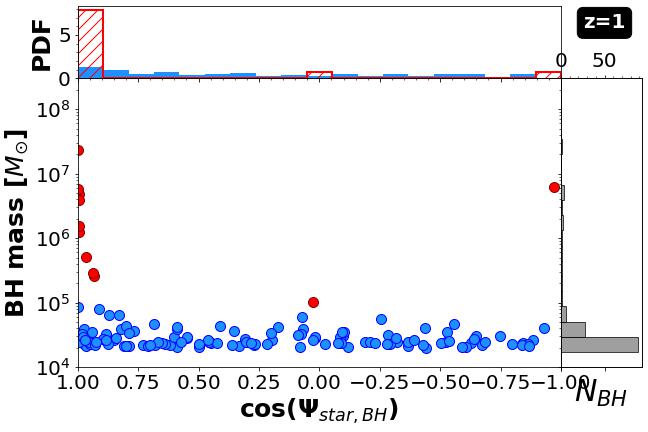}}
\rotatebox{0}{\includegraphics[width=6cm]{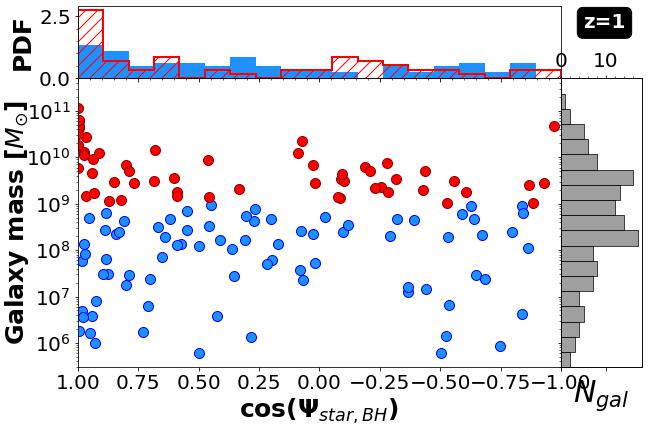}}
\rotatebox{0}{\includegraphics[width=6cm]{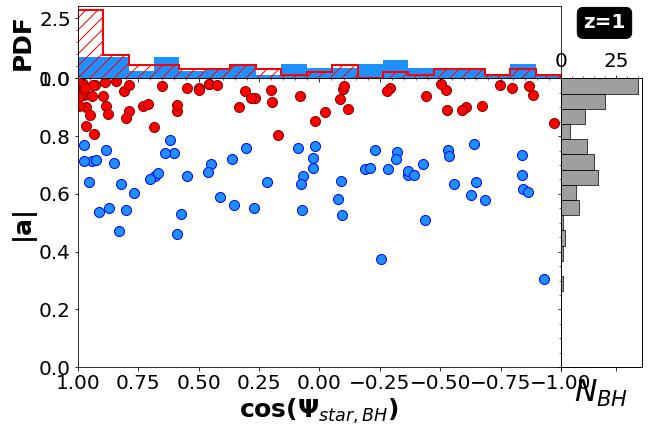}}
\rotatebox{0}{\includegraphics[width=6cm]{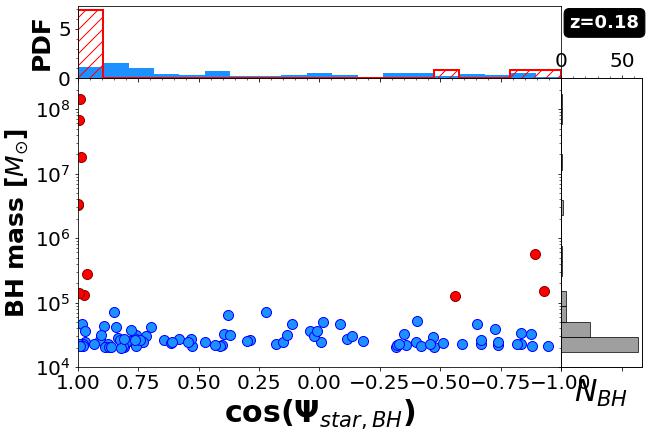}}
\rotatebox{0}{\includegraphics[width=6cm]{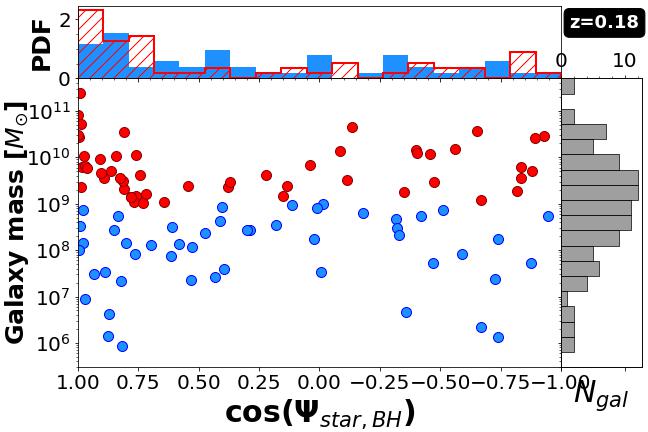}}
\rotatebox{0}{\includegraphics[width=6cm]{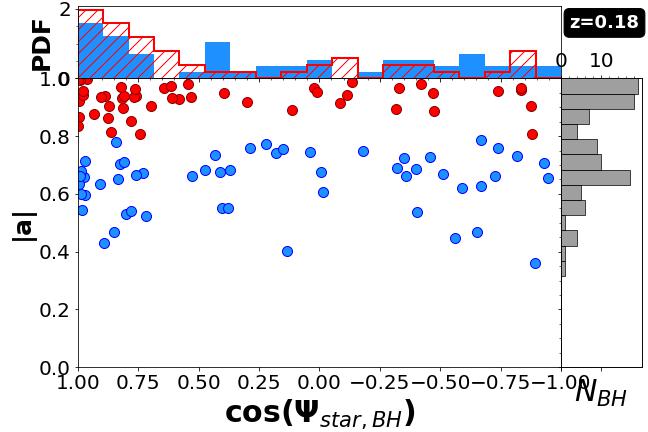}}
\caption{Variations of \COSPSIS from our sample of primary BHs
  with a mass M$_{\mathrm{BH}} > 2\times$10$^4$ M$_\odot$ with respect to the BH
  mass (first column), the host galaxy mass (second column), and BH spin (third column) at three different redshifts ($z=$2, 1, and
  0.18).  In each panel, we divided the BH sample into two
  subsamples, characterized by the blue and red colors, and show the
  corresponding distributions of \PSIS in the upper part of the
  panels.  Here, the angular momentum of each associated galaxy is
  calculated within spheres of radius $R_e$.  }
\label{fig_scatter1}
\end{center}
 \end{figure*}

\begin{figure}
\begin{center}
\rotatebox{0}{\includegraphics[width=\columnwidth]{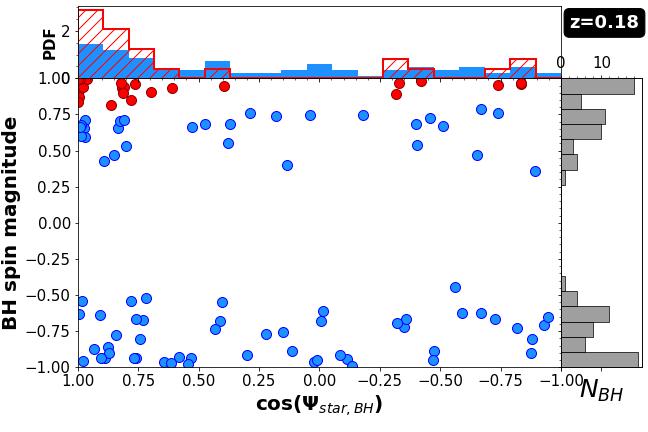}}
\caption{Variations of  \COSPSIS 
with respect to the BH spin magnitude ($a$) at $z=0.18$. Negative values
of $a$ is associated with a retrograde gas accretion.
 }
\label{fig_a_neg}
\end{center}
 \end{figure}

\begin{figure*}
\begin{center}
\rotatebox{0}{\includegraphics[width=6.cm]{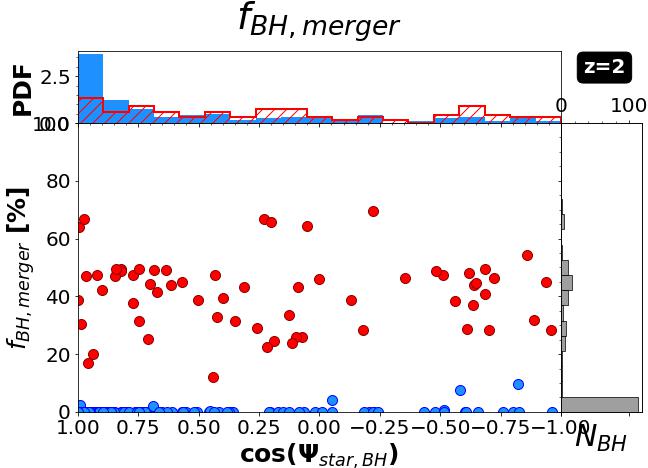}}
\rotatebox{0}{\includegraphics[width=6.cm]{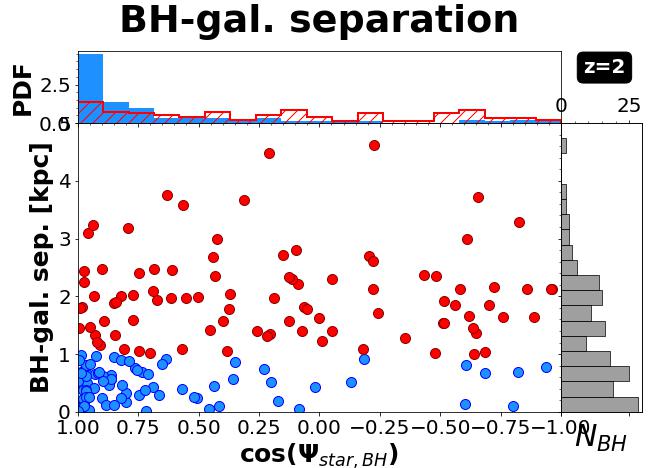}}
\rotatebox{0}{\includegraphics[width=6.cm]{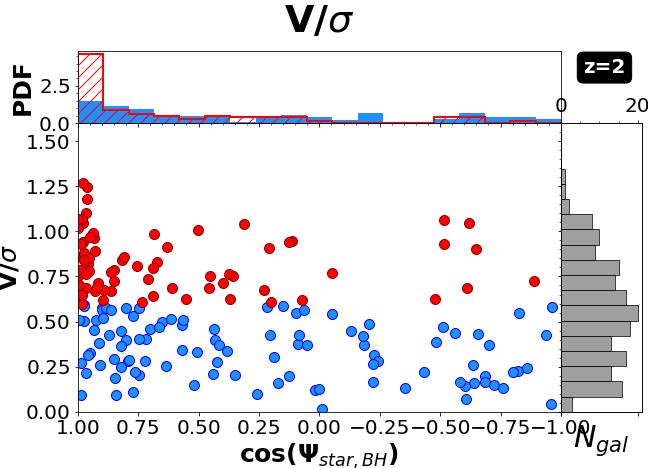}}
\rotatebox{0}{\includegraphics[width=6cm]{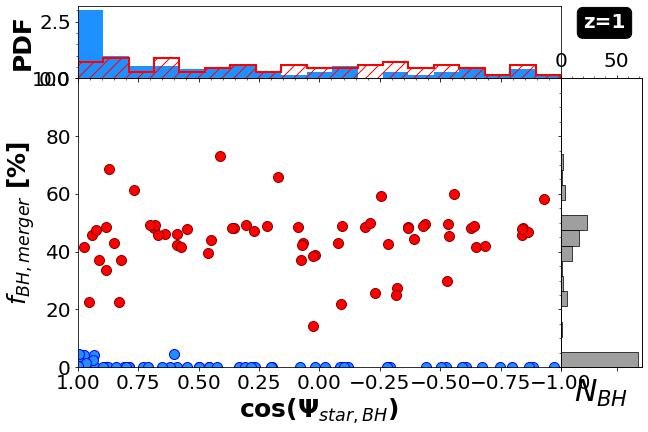}}
\rotatebox{0}{\includegraphics[width=6cm]{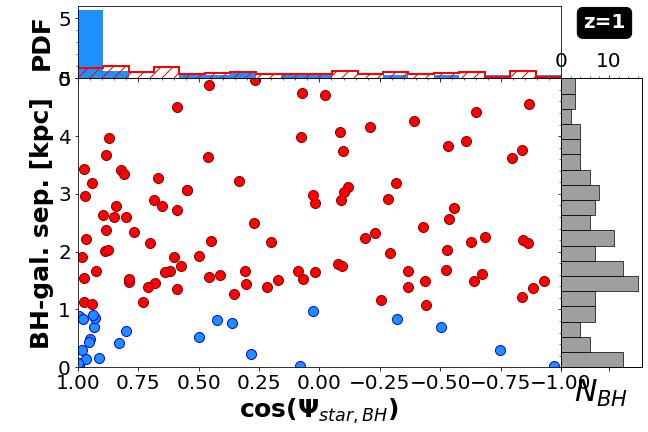}}
\rotatebox{0}{\includegraphics[width=6cm]{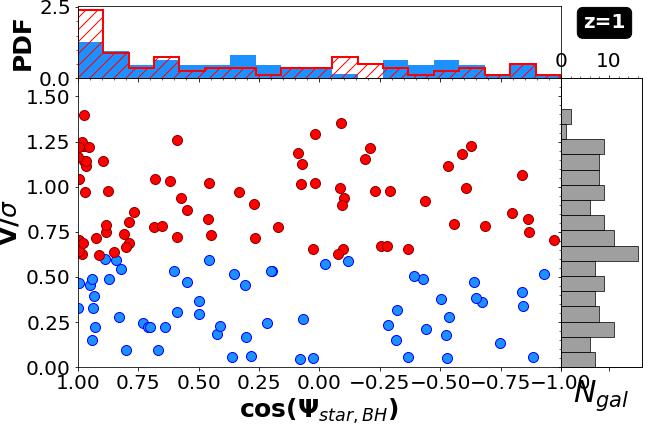}}
\rotatebox{0}{\includegraphics[width=6cm]{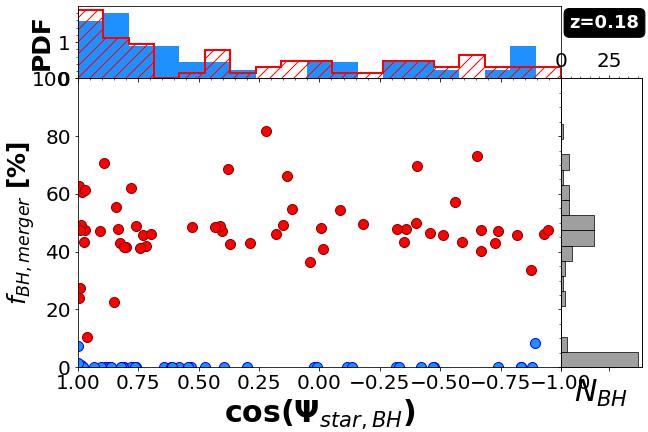}}
\rotatebox{0}{\includegraphics[width=6cm]{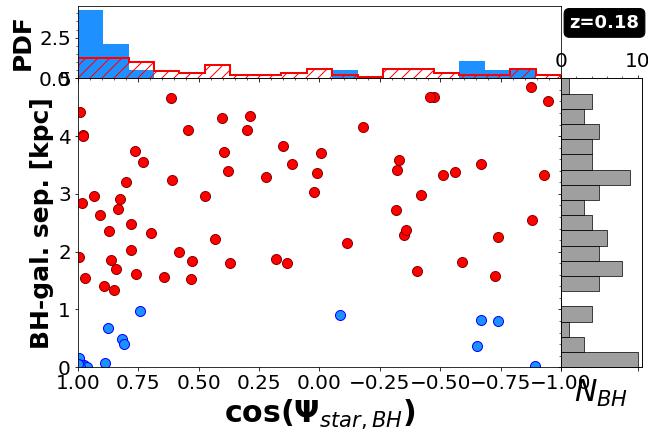}}
\rotatebox{0}{\includegraphics[width=6cm]{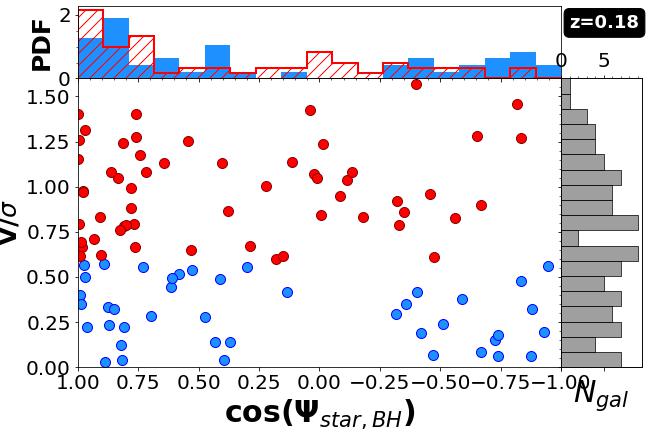}}
\caption{Same as Fig.~\ref{fig_scatter1} but for the variations of \COSPSIS
with respect to the percentage of BH mass gained through BH mergers
(\fdm\,, first column),
the distance separation between the BH 
and the center of its host galaxy (second column),
and $V/\sigma$ as a proxy for galaxy morphology (third column). 
 }
\label{fig_scatter2}
\end{center}
 \end{figure*}

\section{Projected angles statistics}
\label{sec:progectedanalysis}

From an observational point of view, the accurate measurement of the ``true''
3-d angle \PSIS  may prove to be a difficult task
since the stellar component as well as spin vectors are ``seen'' in 2-d projection on the sky.
Instead, it might be relevant to focus on  
the so-called 2-d projected angles $\lambda$ \citep[e.g.,][]{Ohta2005,Benomar2014b}, which is the angle 
between the projected BH spin and the galaxy spin on the sky as schematically
shown in Fig.~\ref{fig_scheme2}. As for $\Psi$, $\lambda$
lies between 0 and 180 degrees.

We therefore would like to present relevant statistics of $\lambda$
 which can be directly
derived from our 3-d statistical analysis.
This could potentially help the interpretation of future observations
of BH-galaxy spin alignment through the measurement of $\lambda$,
for instance by combining the projected velocity field (to estimate the direction
of \JSTAR) and the direction of jet to infer the direction of the BH spin.

We recall that for a 3-d isotropic distribution of BH spins, the distribution
of \COSPSIS is uniform. In 2-d, it is the distribution of $\lambda$ which is
uniform and for this reason, we study in the following
the probability density function \PLP, where $i_o$ is the
galaxy disk inclination relative to the observer.

\begin{figure}
\begin{center}
\rotatebox{0}{\includegraphics[width=\columnwidth]{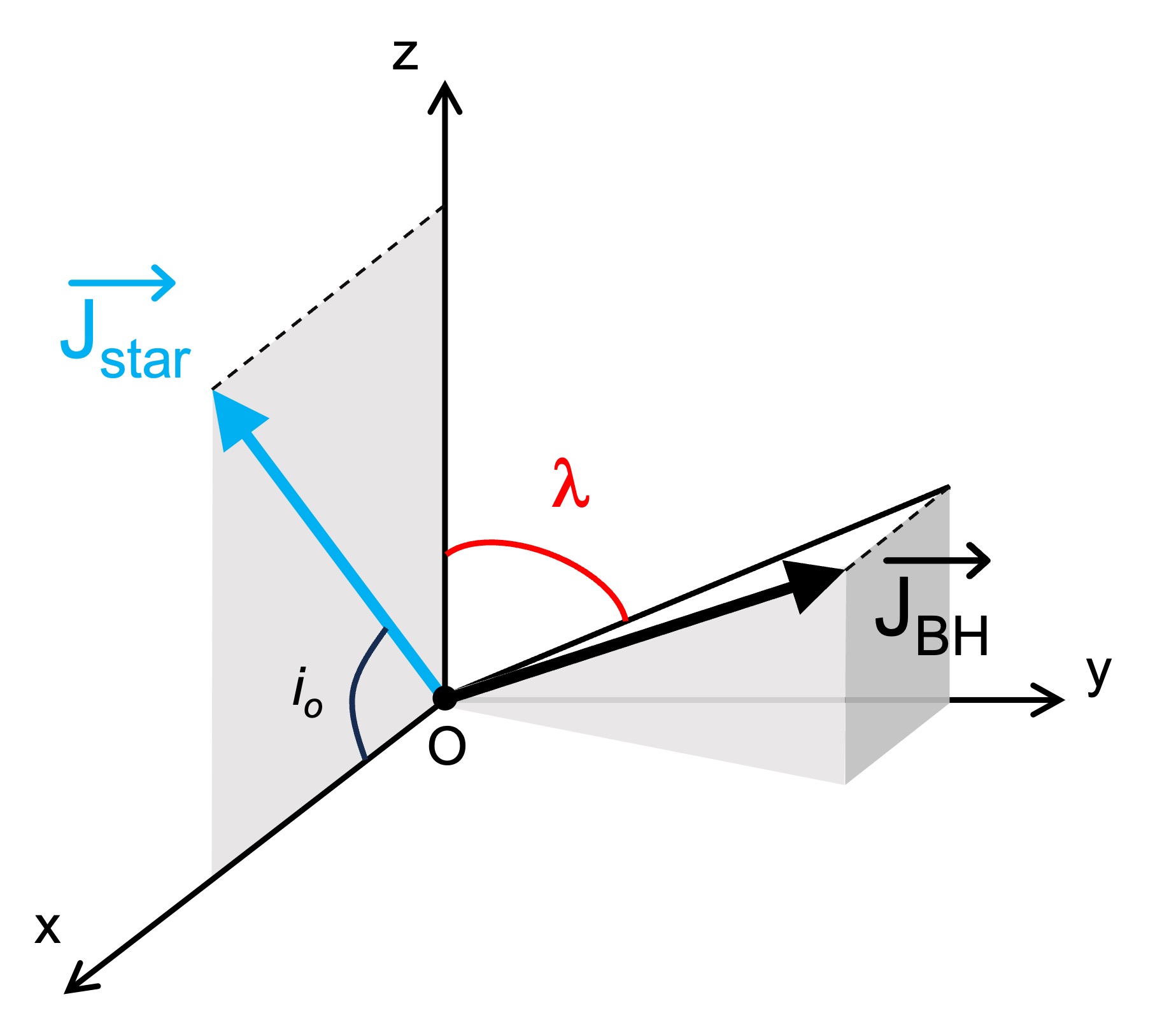}}
\caption{Schematic view of the projected angle. The $x$-axis is 
pointing toward an observer. The angle $i_o$ 
is the inclination which defines the angle 
through which the galactic plan is seen by the observer. 
The projected angle $\lambda$ of the 3-d angle \PSIS is contained
in the $y$-$z$ plane (with $x$=0).
}
\label{fig_scheme2}
\end{center}
 \end{figure}

\subsection{3-d and 2-d projected angle relation}

We use a simple Monte-Carlo method to derive the
projected angle from a given BH-galaxy pair of our samples
with a random orientation in space as well as a random $i_o$.
We describe the main steps here.
First, for a given studied BH-galaxy pair, 
we have rotated both the angular momentum
of the stellar component (galaxy spin, \JSTAR) and the associated BH spin
from their original 3-d orientation in space in the simulation
so that \JSTARS  coincides now exactly with 
the $z$-axis of the frame presented in Fig.~\ref{fig_scheme2}. Then we have
randomly drawn 
an angle between 0 and 2$\pi$ and rotate accordingly the BH spin with respect to 
the $z$-axis to generate a new configuration.
This operation does not modify the original value of the 3-d angle \PSI.
The final step consist in applying an orbital inclination $i_o$.
To do so, we randomly draw an angle between 0 and $\pi/2$ and
rotate the BH spin with respect to the $y-$axis.
The projected angle is then the angle between the $z$-axis and the
projection of the 3-d BH spin onto the x=0 plane.

The relation between the 3-d ($\Psi$) and 2-d projected ($\lambda$) 
angles has already been 
studied in works related to the stellar spin-orbit misalignment
angle in planetary systems \citep{fabrycky+09,crida+14}.
But in the majority of cases, $i_o$ is close to $\pi/2$ 
since the observer is almost exactly in the orbital plane
of the transiting planets. In that case,
a simple analytical expression of the 
probability density function of $\lambda$ for fixed $\Psi$,
\PLPpi,
can be derived: equation (4) in \cite{crida+14} or equation (19) in
\cite{fabrycky+09}.

To test the validity of our sampling,  
we compute \PLPpi\,
for two BH-galaxy pairs from our sample at $z=0.18$ for which
\PSI=7.7$^\circ$ (BH-166) and \PSI=153.2$^\circ$ (BH-549), respectively.
For each of them, we have generated 10\;000 random realizations using the steps described
above,
and computed the corresponding projected angles.
The probability density functions are shown in Fig.~\ref{fig_lambda_prob}.
The good agreement with the analytical expressions derived by \cite{crida+14}
ensures that our sampling is reliable. Moreover, it is instructive
to see that
for $i_o=\pi/2$, the projected angle can take many values ranging from 
[0-$\Psi$] (if $\Psi$<$\pi/2$) or from [$\Psi$-$\pi$] (if $\Psi$>$\pi/2$).
We have also checked that our sampling is in good agreement with analytical
solutions for more complex cases ($i_o\ne\pi/2$). 
For instance we have 
correctly reproduced the probability functions P($\lambda$\,|$\Psi$,\,$i_o$=$80^\circ$) presented in Fig.~3 of \cite{fabrycky+09}.

\begin{figure}
\begin{center}
\rotatebox{0}{\includegraphics[width=\columnwidth]{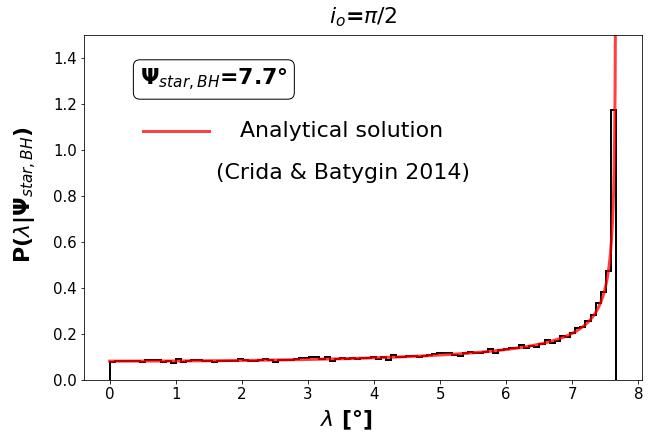}}
\rotatebox{0}{\includegraphics[width=\columnwidth]{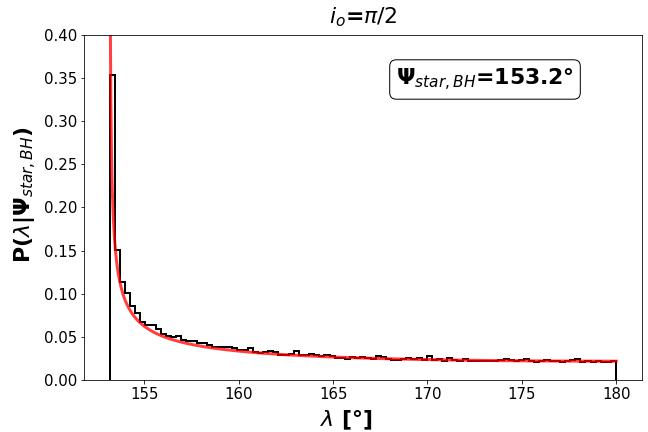}}
\caption{Probability density function \PLPpi\, 
of the 2-d projected angle ($\lambda$) for two
fixed values of \PSI, 7.7$^\circ$ (top panel) and 153.2$^\circ$ (lower panel).
In the two cases, we have considered an orbital inclination
of $i_o=\pi/2$. The histograms are the results obtained from
a Monte Carlo method (using 10\;000 random lines-of-sight)
while the red lines are analytical solutions.
 }
\label{fig_lambda_prob}
\end{center}
 \end{figure}

\subsection{2-d projected angles statistics}

Now, since \COSPSIS and therefore \PSIS have their own distribution at $z=0.18$
(the lower left panel of Fig.~\ref{fig_pdf} for \COSPSI), what
is the corresponding distribution of $\lambda$?

To answer this question, we have applied the Monte-Carlo method 
to the 102 BH-galaxy pairs of our sample at $z=0.18$.
For each original BH-galaxy pair, we have generated 500 random configurations and
computed the corresponding projected angles.
The resulting distribution of $\lambda$ is shown in the upper left
panel of Fig.~\ref{fig_mis1}. We also show the spline interpolated function to 
describe it.
The tendency of spin alignment 
seen in the 3-d analysis is still observable
in the 2-d case. 
In the lower left panel, we show the different distributions obtained 
using spline functions for different values of $i_o$
(every 10$^\circ$). Since observational measurements might be preferentially obtained for high
$i_o$ values (i.e., close to $\pi/2$), this result could reveal some selection effects.
We notice that the tendency of alignment is less and less clear as $i_o$ decreases
(more face-on views).  
Note that we have assumed in our model a uniform distribution of $i_o$.
However, it is now well known
that galaxies do not form everywhere but in filaments and nodes
implying some preferred directions for the orientation of the spins 
within the cosmic web, as suggested by
numerical simulations \citep[e.g.,][]{codis+18, Kraljic+20, zhang+23}
or observational analysis \citep[e.g.,][]{welker+20, Kraljic+21, desai+22}.
This effect could be taken into account for even more realistic modeling. 

\begin{figure*}
\begin{center}
\rotatebox{0}{\includegraphics[width=\columnwidth]{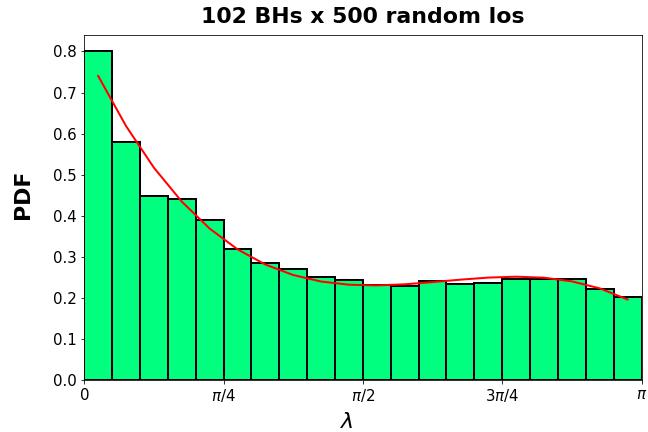}}
\rotatebox{0}{\includegraphics[width=\columnwidth]{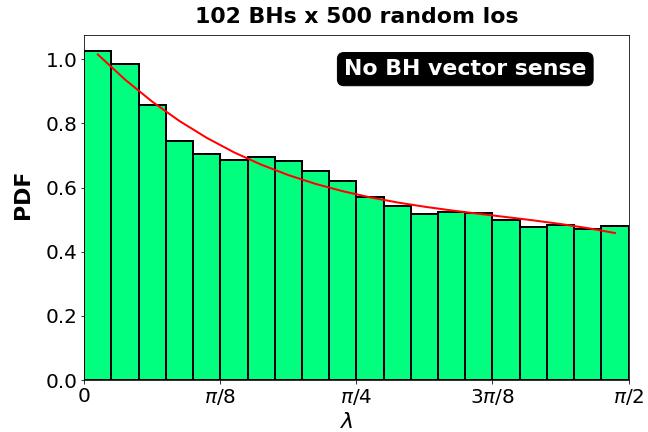}}
\rotatebox{0}{\includegraphics[width=\columnwidth]{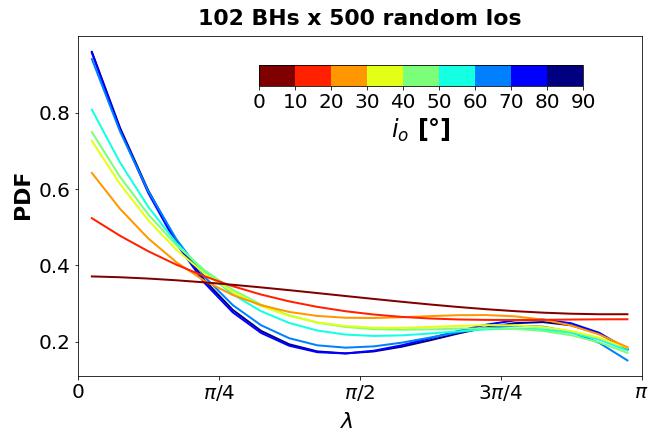}}
\rotatebox{0}{\includegraphics[width=\columnwidth]{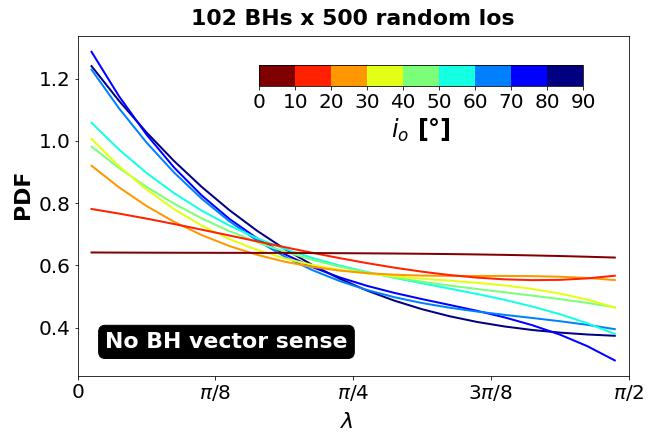}}
\caption{
{\it Left column}: Probability density function \PLP\, 
of the 2-d projected angle ($\lambda$) obtained from the distribution
of \PSIS at $z=0.18$. We also use a spline function to 
describe the obtained PDF.
In the lower left panel, we show the different distributions obtained 
through spline function description for different values of $i_o$
(every 10$^\circ$). {\it Right column}: 
We generated the plots using the same statistics but without specifying the sense of the BH vector to maintain consistency with observations.
 }
\label{fig_mis1}
\end{center}
 \end{figure*}

If the orientation of jets is used as a proxy to the 
direction of the projection of any
BH spin vector, the sense of its rotation actually cannot be known. 
Thus, to be more consistent
with future observational measurements based on jet orientation,  
we plot in the right side of  Fig.~\ref{fig_mis1} the same statistics
as the left part by redefining $\lambda$ in the range 
between 0 and $\pi/2$.
In that case, among the two 
possible projected angles $\lambda$ and $\pi-\lambda$,
we always choose the smallest angle
which ranges from 0 to $\pi/2$. We note that the tendency of
spin alignment is still present but less pronounced than when taking
into account the sense of the BH spins.
Note also that, for a higher consistency, only BHs with a low Eddington ratio
at the studied redshift (or in an redshift interval) should be considered.
For instance, at z=0.18, there are only 5 and 8 BHs that have a Eddington ratio greater that 0.01 and 0.001 respectively. We have checked that the
main trends displayed in the right part of Fig.~\ref{fig_mis1} remain
similar even if we remove these BHs from the sample. 
Additionally, we have considered the whole sample of BHs (102 in total) within a small redshift interval (for instance 0.18< z<0.28 which corresponds to the last 1 Gyr), and focused on those BHs when their Eddington ratio was low (i.e., radio mode). Even in those cases, we made sure that our trends remain the same.
Also, it is worth mentioning that the orientation 
of radio jets might not be necessarily aligned with BH spins because 
the propagation of jets from the accretion disk up to 
galactic scale can be strongly perturbed by the multiphase structure 
of the interstellar medium \citep[e.g.,][]{mukherjee+18, cielo+18, M87},
or because of the jet precession \citep{duun+06, krause+19,ubertosi+24},
or also due to relative motions between the jets and the hot atmosphere in intra-galaxy clusters \citep[e.g.,][and references therein]{heinz+06,morsony+10,morsony+13,odea+23}. All of these effects might affect the observed distribution of $\lambda$.

Additional statistics on the 2-d projected angles according to
the galaxy morphology are derived in Appendix~\ref{appendix2}.

\section{Conclusion and discussion}
\label{sec:conclusions}

The BH spin is a key parameter in black hole physics as well as
in the formation process of the host galaxies. 
Characterizing the evolution of its amplitude and orientation through
cosmic time helps distinguish among different scenarios for BH-galaxy
coevolution.
Using the \nhs and \galacticas simulations, 
we have investigated the long-term evolution of the 3-d angle between
BH spins and the angular momentum vectors of their host galaxies. These simulations have sufficient resolution to capture the injection scale of gas turbulence, and has a multiphase interstellar medium.
Catalogs of BH-galaxy pairs have been produced by selecting
primary BHs with mass greater than \MASSLIM\, and the angular momentum
of each galaxy has been estimated using all star particles within one effective radius.
Our main conclusions can be summarized as follows:

{\bf 1.}
The analysis of typical individual evolution of the most massive BHs 
of our catalogs 
suggests the occurrence of three phases
in the evolution of \PSIS:

\noindent
$\bullet$ Right after being seeded with 
an initial mass of 10$^4$ M$_\odot$ and no spin, most of
BHs are rapidly spun up to a high or a maximum spin value as
they double or triple their mass 
through gas accretion (Fig.~\ref{fig_merger1}),
in agreement with \cite{nhz}, \cite{bustamante+19} and \cite{sala+23}.
During this relatively short period, 
BH spins are likely to be well aligned with the spin of their
host galaxies. And as a consequence, low-mass BHs at high redshift 
present in general low values of \COSPSIS (Figs.~\ref{fig2}, \ref{fig_color_map}, \ref{fig_merger1});

\noindent
$\bullet$ A second phase (M$_{\mathrm{BH}}$$\lesssim$10$^5$M$_\odot$) is dominated by a rather
chaotic and inefficient gas accretion,
reflecting the patchy morphology of forming proto-galaxies.
In this phase, \PSIS presents erratic variations  
mainly driven by the rapid change of the stellar angular momentum
orientation, rather than that of the BH spin direction (see the evolution 
of $\theta_{\rm BH}$ and $\theta_{\rm star}$ in
the individual evolution shown in Figs.~\ref{fig_g9685}-\ref{fig_bh1049} and Figs.~\ref{fig_appendix1},\ref{fig_appendix2}).
This phase is indeed characterized by low accretion rates and low
Eddington ratios, which favor a very long alignment time 
between BH spin and the accreted material \citep{lodato+06,lodato+13}.

\noindent
$\bullet$ The last phase follows, in which the gas accretion
becomes much more coherent and in which BHs are generally well settled in the center of their host (10$^5$M$_\odot\lesssim$ M$_{\mathrm{BH}}$). From the study of the most massive BHs of
our samples, the transition happens generally when
the host galaxy has a mass ranging from 5$\times$10$^9$ to
5$\times$10$^{10}$ M$_\odot$.
The BH spins are likely to be well aligned
with the angular momentum of the galaxy, 
and this configuration can last over Gyr time periods, even though
BH merger episodes can temporally misalign their spin.

{\bf 2.}
The majority of the most massive BHs of our samples (M$_{\mathrm{BH}}$>10$^5$M$_\odot$) 
have BH spins highly aligned with \JSTAR, even at low redshift $z=0.18$. 
This result is consistent with those
from \cite{dubois+14a}. However, in that paper they also found that for
more massive BHs (M$_{\mathrm{BH}}$>10$^8$M$_\odot$), spins are more randomly
oriented with respect to their host galaxy angular momentum
and tend toward \COSPSI=0. In these galaxies, BHs indeed accrete, at very low levels, from the turbulent hot intra-cluster gas 
that bears no connection to the angular momentum of the stellar component of the galaxies.
Such trend could not be confirmed
in the present analysis since our sample of very massive black
holes is too limited.

{\bf 3.} 
Statistically, the distribution of \COSPSIS at a given redshift indicates that
BH spins are more likely to be aligned with their host galaxy
spin, in agreement with \cite{beckmann+23} and with a recent
observational analysis \citep{zheng+24}.
This tendency to alignment is more pronounced at high redshift ($z=3-2$),
while it decreases until $z\sim$1.5 due to BH-merger events,
and becomes nearly constant. We also note that at late time, 
BH spin starts again to be more aligned with the galaxy spin
since the BH mergers become less frequent.

{\bf 4.}
BH merger events is one of the main drivers of 
3-d angle misalignment. Our analysis suggests that 
BHs which have undergone more merger episodes in their history, and
characterized by high \fdm\, values, are more likely
to have their spin misaligned than those dominated by
a smooth gas accretion (Fig.~\ref{fig_merger1}  
and first column in Fig.~\ref{fig_scatter2}).
This is expected since the orbital angular momentum
is redistributed in BH mergers.
Our investigation of individual BH history also suggests
that after a BH-merger episode, the BH spin tends to realign to the galaxy spin
in a relatively short period ($\Delta t\sim$100-200 Myr). 
A similar timescale is suggested in Fig.~9 of \cite{bustamante+19}
regarding the average variations of the \PSIGAS\, after a merger. 

{\bf 5.}
Long phase of retrograde accretion of gas induced by either
cosmic inflows or galaxy mergers can produce strong effects,
by anti-aligning the BH-spin in a long period, as suggested by the evolution of 
BH-$1049$ and BH-$549$.
From a statistical point of view,
counter-rotating gas disk with respect to the stellar component are more likely to lead to BH-galaxy spin misalignment (Fig.~\ref{fig_a_neg}).
\cite{dubois+14a} found that about 
10-30\% of BHs have spins counter-aligned with their
host galaxy. This occurs in galaxies where
the central regions have been deprived of cold gas 
or because BH coalescences have flipped the spin
direction. \cite{bustamante+19} confirmed this statement
and found that 
20\% of BH merger events may lead to momentarily counter-rotating
gas disk. 
The individual evolution of BH-$1049$ and BH-$549$ present different scenarios 
as the retrograde gas accretion onto the BH
is not induced by a punctual effect (such as a BH merger). 
Instead, it is due to the continuous accretion supplied by 
the galactic gas disk.
For this reason, a BH-galaxy spin anti-alignment configuration 
could persist for an extended period, 0.5 Gyr and >4 Gyrs for BH-$1049$ and BH-$547$ respectively.
The present work therefore predicts that galaxies presenting strong
star-gas spin angle misalignment 
\citep[e.g., from the ManGA survey:][]{beom+22, xu+22,katkov+23}
are likely to host BHs owning to a clear misaligned or anti-alignment
with respect to the stellar component.
We will present in a forthcoming paper, a detailed analysis about 
the formation of counter-rotating gas/stellar disks 
(Peirani et al. in prep).


{\bf 6.}
The analysis of the 2-d misalignment angle ($\lambda$) revealed that
the distribution tends to keep record of the alignment tendency seen
in the 3-dimensional predictions.
It might then be possible to identify such signal from
observational analysis measuring 2-d misalignment angles
(Lin et al. in prep).

\medskip
This paper is complementary to Beckmann et al. (in prep) which focuses
on the cosmic evolution of BH spin {\it magnitude} using \nhs and the same BHs catalogs.
They also found that the evolution of BH spins follow three distinct phases
while highlighting the role of BH mergers during the second phase 
which tend to increase the scattering of |a|.
This is consistent as the magnitude and orientation of spins are 
connected:
high/low values of spin magnitude are generally associated with   
BH-galaxy spin alignment/misalignment, as suggested by
Figs.~\ref{fig_merger1} and \ref{fig_scatter1}.

The theoretical predictions presented in this paper heavily rely on sub-grid models, which may not completely follow all relevant physical processes, even in the currently best simulations. In particular, sub-grid modeling of
   BH physics should be interpreted with caution.
For instance, the gas accretion disk falling onto the BH is not
spatially resolved in our simulation. 
Thus, we have assumed that the
gas accretion disk around the BH retains the same angular momentum of
gas component estimated at $4\Delta x\sim$136 pc
from the location of each BH.
While this scale is much larger than the physical scale of the
accretion disk, it is expected to be a reasonably good
approximation to assume that its angular momentum, especially its orientation, is conserved
during the accretion process of the gas toward the disk.
This is consistent with the result of \cite{maio+13} using high
resolution simulations down to 1 pc, and even in the presence of strong star
formation feedback.
Yet, we note that there are different suggestions in the
literature. \cite{levine+10} found that the angular momentum vector on
scales $\sim$100 pc may vary substantially from the direction of
angular momentum on kiloparsec scales between z=4 and z=3.  This is
mainly due to the interaction of clump of gas moving toward the center
of the disk, which results in dramatic change of the nuclear gas
angular momentum.  \cite{hopkins+12} also predicted a weak correlation
between the nuclear axis and the large-scale disk axis, using
high-resolution simulations of gas inflows from galaxy to parsec
scales around AGN.  Sudden misalignments may be caused either by
single massive clumps falling into the center slightly off-axis, or
due to gravitational instabilities.  The two latter works, however, focused on gas rich galaxies at high redshifts where the gas turbulence may still play a strong role in the gas angular momentum alignment. 

Another important assumption in the present work is that two BHs
instantly merge when they come close less than $4\Delta x$. In reality,
however, they would first form a binary for a while.  If there is a
sufficient amount of gas supply, the BH spins may evolve during the merger process.  If the circumbinary disk is misaligned with respect to
the BH binary orbital plane, it leads to tearing or warping of the disk structure,
which may affect the spin alignment (or not) between the two BHs
\citep[see, for
  instance,][]{nixon+11,nixon+13,gerosa+15,moody+19,nealon+22,bourne+23}.
However, our analysis suggests that BH spins tend to rapidly align
after the BH merger especially during long and coherent phases of gas
accretion. Thus, we expect that these missing processes should not
significantly affect our predictions, at least statistically.

We should also note that we have not included any analytically model
for the unresolved dynamical friction experienced by BHs from stars
and DM particles.  This has been shown to play a crucial role in
sinking BHs closer to the galaxy center, and also in enhancing the BH
merger rate \citep{pfister+19,chen+22, ma+23}.  The dynamical friction
from gas is expected, however, to have a lower impact especially at
increased resolution due to instabilities in the wake
\citep{beckmann+18} and the turbulent nature of the gas
\citep{lescaudron+23}.  The absence of dynamical friction in our model
might at some point delay the alignment of the BH spin since off-centered
BHs are more likely to have misaligned spins. On the contrary, a
higher BH merger rate should statistically reduce the tendency of
alignment. Thus, more quantitative and reliable predictions need the refinement of the sub-grid physics.

Finally, we have assumed that all BH at birth have an initial mass of
$10^4$ M${_\odot}$ with no spin.  Depending on the initial BH
spin, however, accretion onto the BH after birth can alter the spin magnitude
and/or torque the BH spin alignment \citep{mckernan_et_ford23}.
  Furthermore, \citet{sala+23} recently employed a very similar
 numerical scheme to compute the BH spin evolution in idealized or
 cosmological simulations. They found the rapid increase of the spin
 parameter for $5\times 10^5 $M$_\odot <M_{\rm BH}<10^6 $M$_\odot$
 (see their Figure 13), which is consistent with ours qualitatively
 \citep[or see Fig.~23 of ][for a more direct comparison]{nhz}.  Since
 the mass range is very close to their initial BH seed mass of
 $5\times 10^5 $M$_\odot$, the result may be suspected to be biased
 due to the mass resolution.  In fact, our BH seed mass is $10^4
 $M$_\odot$, 50 times smaller than theirs, and we found the similar
 trend around $10^4 $M$_\odot<$M$_{\mathrm{BH}}<2\times 10^4 $
 M$_\odot$. Even though we selected BHs of $M_{\mathrm{BH}}>2\times
 10^4$ M$_\odot$, it is likely that they still suffer from the
 numerical artefact to some extent.  A similar statement can be drawn
 from Fig.~7 of \cite{bustamante+19} as they used a seed mass close to
 $\sim$10$^6$ \msun.  This is why we examined the behavior of four
 examples in detail (BH-G9685, BH-166, BH-1049 and BH-541), which are
 supposed to be free from such possible numerical effects, at least in
 the late epochs ($z <2$). In any case, our primary focus is the
 orientation of the BH spin and galaxy orbit, and thus the present
 paper is very complementary to those previous works that are more
 focused on the cosmic evolution of the amplitude of the BH spin
 \citep[e.g.,][]{dubois+14a, nhz, bustamante+19, sala+23}.

\subsection*{Acknowledgements}
We warmly thank the referee for an insightful review that considerably improved the quality of the original manuscript.
S.P. acknowledges the support from the JSPS (Japan Society for the
Promotion of Science) long-term invitation program, and is grateful for the
hospitality at Department of Physics, the University of Tokyo.
This research is partly supported by the JSPS KAKENHI grant Nos.
JP18H01247 and  23H01212 (Y.S.).
S.K.Y. acknowledges support from the Korean National Research Foundation (2020R1A2C3003769, 2022R1A6A1A03053472). This work was granted access to the HPC resources of CINES under the allocations c2016047637, A0020407637, and A0070402192 by Genci, KSC-2017-G2-0003 by KISTI, and as a “Grand Challenge” project granted by GENCI on the AMD Rome extension of the Joliot Curie supercomputer at TGCC. The large data transfer was supported by KREONET which is managed and operated by KISTI.
Y-T.L. is grateful for support from the National Science and Technology
Council of Taiwan under grants MOST 111-2112-M-001-043 and NSTC
112-2112-M-001-061.
C.P. is partially supported by the grant
\href{https://www.secular-evolution.org}{Segal} ANR-19-CE31-0017 of the French Agence Nationale de la Recherche and by the National Science Foundation under Grant No. NSF PHY-1748958.
This work was carried within the framework of the
Horizon project (\href{http://www.projet-horizon.fr}{http://www.projet-horizon.fr}).
Most of the numerical modeling presented here was done on the Horizon cluster at Institut d'Astrophysique de Paris (IAP).

\bibliographystyle{aa}
\bibliography{bhspin}


\appendix

\section{Other individual evolutions}
\label{appendix1}
We present in this appendix additional individual evolution of BHs 
summarized in Table~\ref{tab1}.
Most of them are displaying a typical \PSIS evolution with different
regimes described in section~\ref{sec:individual}.
This is the case for the BH-$146$, BH-$348$ and BH-$541$ owing mass
greater that 10$^7$M$_\odot$ but are contaminated with less than
0.1\% of their mass with low resolution DM particles. We reasonably believe that
this should not affect the results, in particular the estimation
of \JSTAR, though such objects are discarded from the statistical 
study conducted in  section~\ref{sec:3Danalysis}.
On the contrary, BH-$549$ and BH-$132$ show different evolutions.
BH-$549$ is
displaying a long phase where its spin is anti-aligned with the stellar
angular momentum (see Fig~\ref{fig_549_map} and the end
section~\ref{sec:individual} for more details).
As far as BH-$132$ is concerned, it is
off-centered from the galaxy center. The efficiency of both the gas accretion
and the Eddington ration are
generally low and the BH spin is most of the time mis-aligned with \JSTAR.

\begin{figure*}
\begin{center}

\rotatebox{0}{\includegraphics[width=\columnwidth]{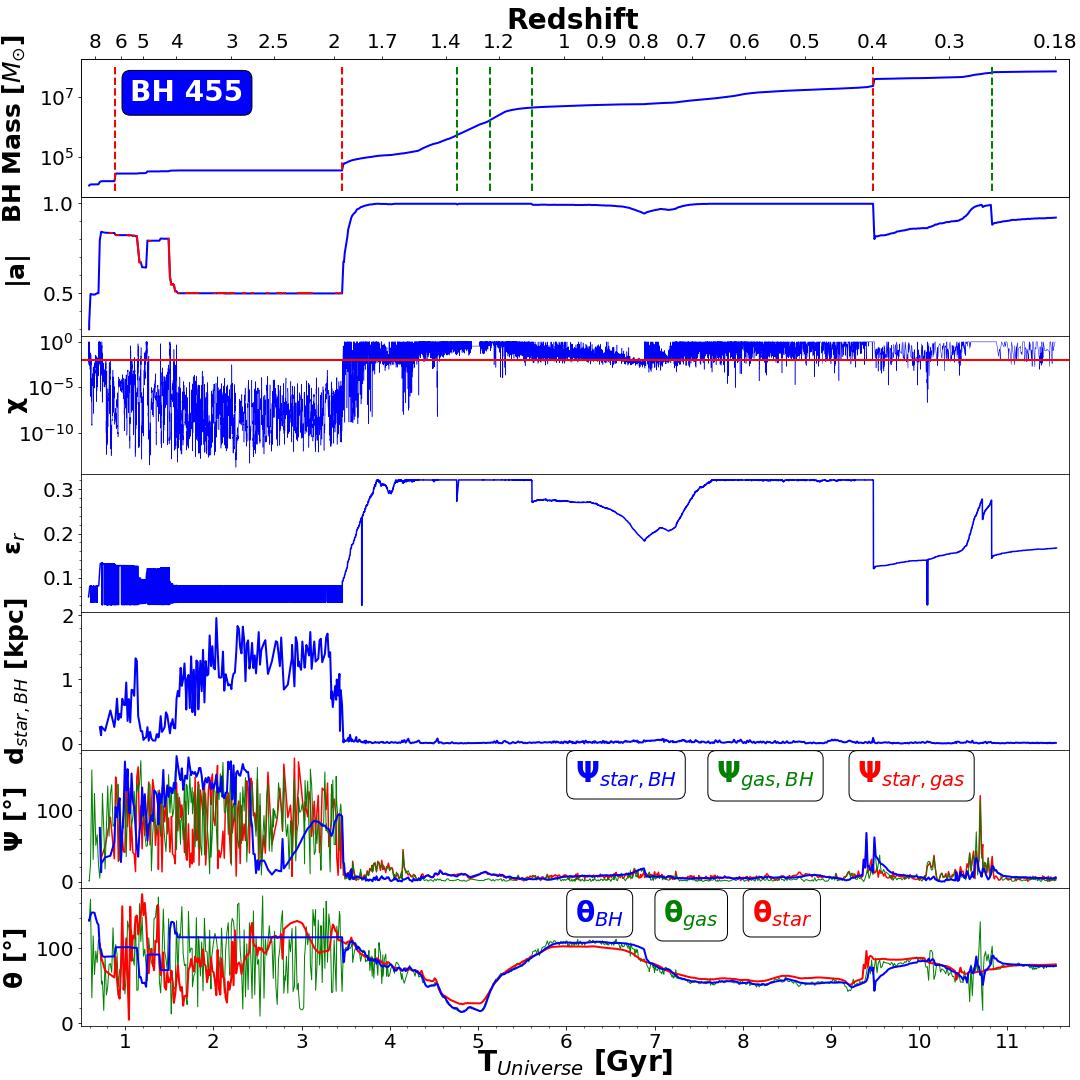}}
\rotatebox{0}{\includegraphics[width=\columnwidth]{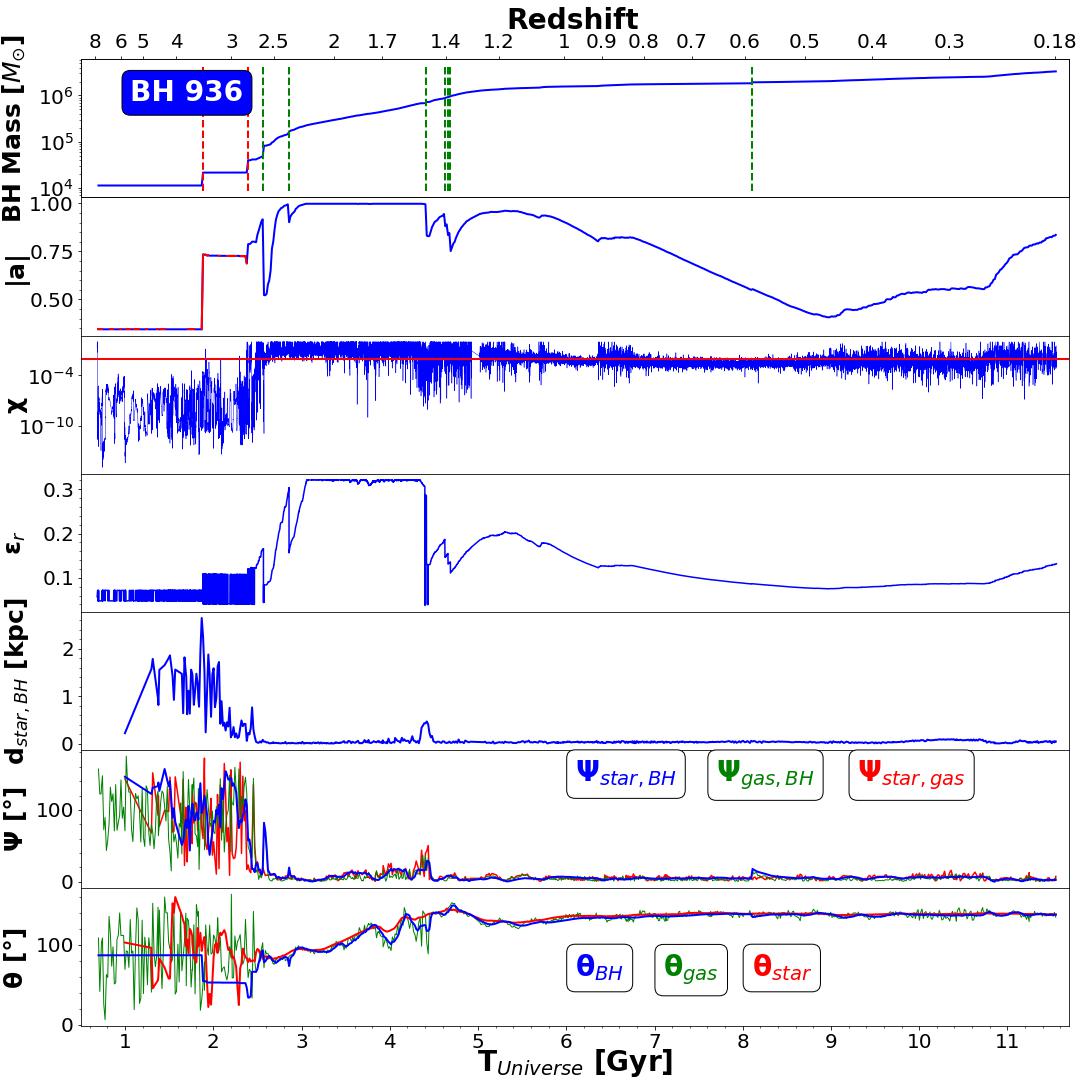}}

\rotatebox{0}{\includegraphics[width=\columnwidth]{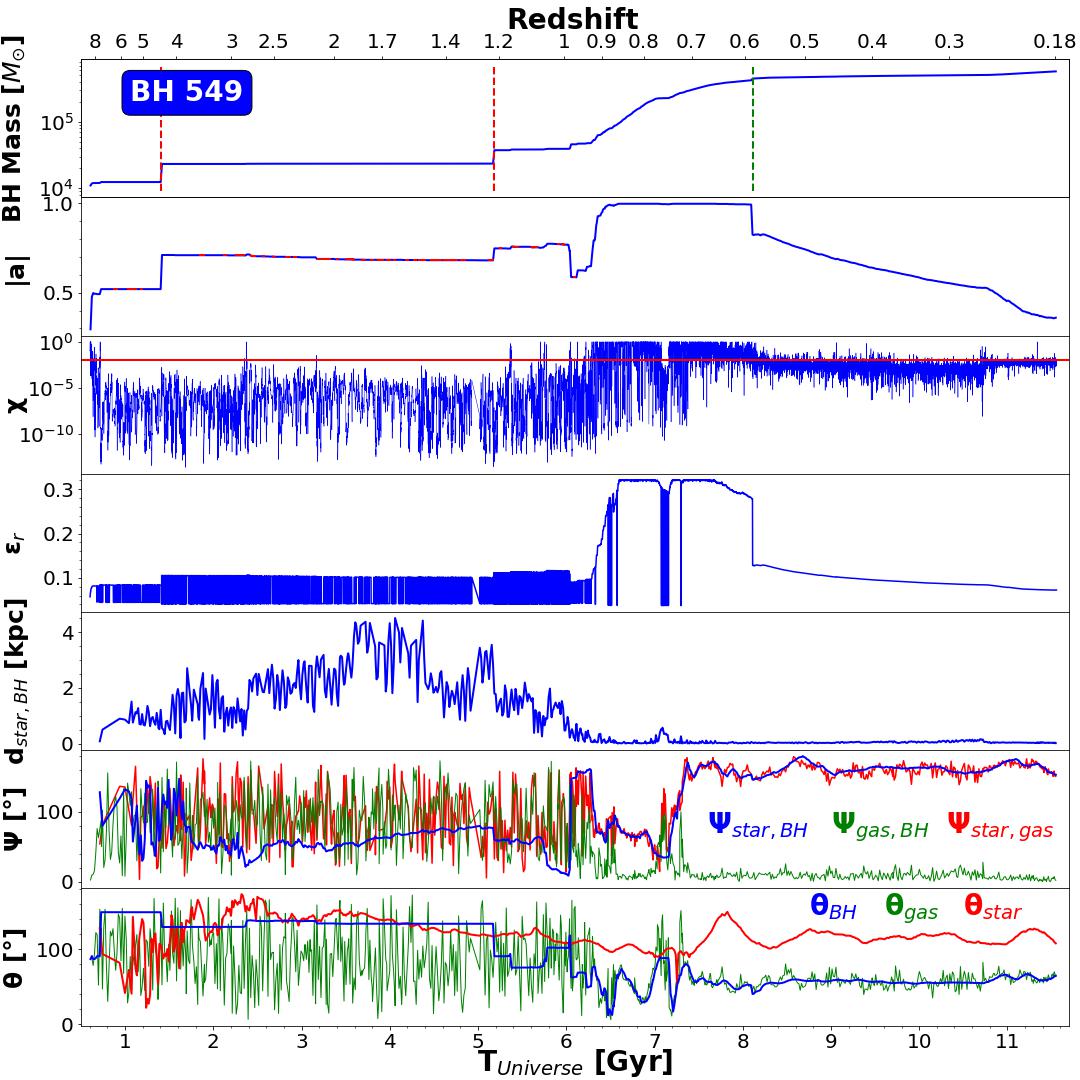}}
\rotatebox{0}{\includegraphics[width=\columnwidth]{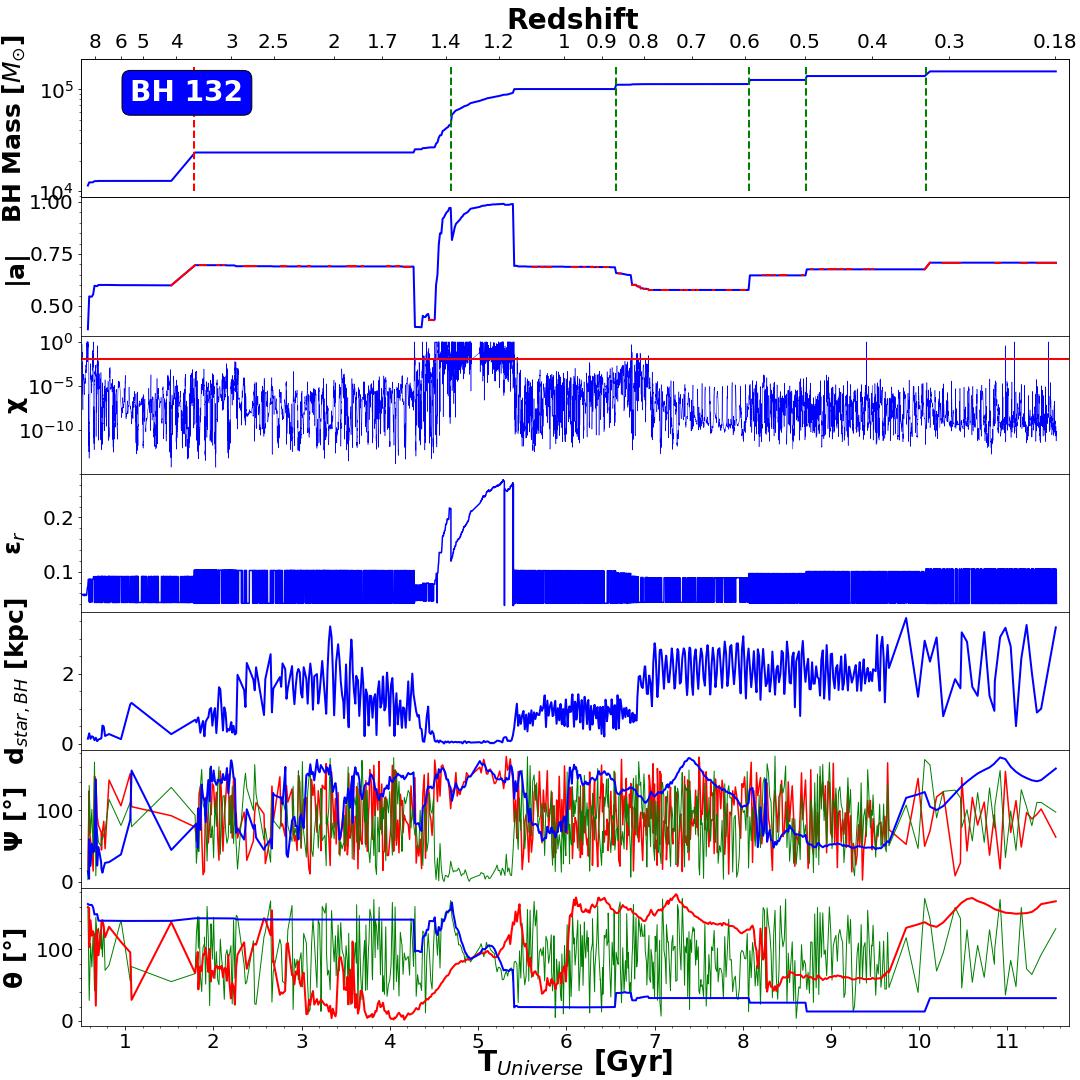}}
\caption{Cosmic evolution of BH-$455$ (upper left panel), BH-936 (upper
  right), BH-$549$ (lower left panel), and BH-$132$ (lower right
  panel).  Same as Fig.~\ref{fig_g9685} but for  BH-$455$ and BH-936 display
  typical \PSIS evolution.  BH-$549$ is displaying a long phase where
  its spin is anti-aligned with the stellar angular momentum. In this
  example, the gas accretion disk is in counter-rotation with respect to the
  stellar component. As far as BH-132 is concerned, it is off-centered
  from the galaxy center. In this latter case, the efficiency of gas accretion and Eddington ratio are generally
  low and BH spin is mis-aligned with \JSTAR.}
\label{fig_appendix1}
\end{center}
 \end{figure*}

\begin{figure*}
\begin{center}

\rotatebox{0}{\includegraphics[width=\columnwidth]{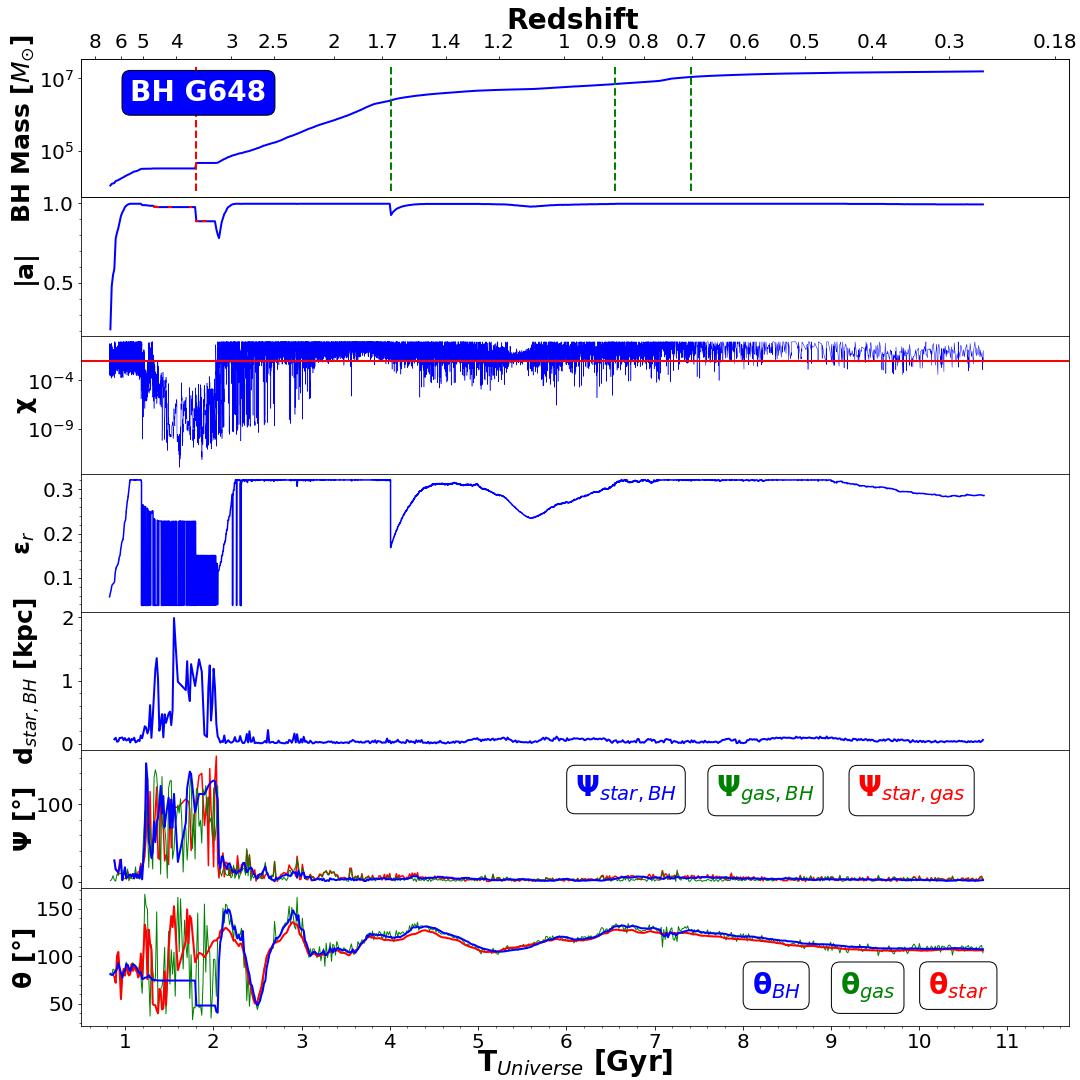}}
\rotatebox{0}{\includegraphics[width=\columnwidth]{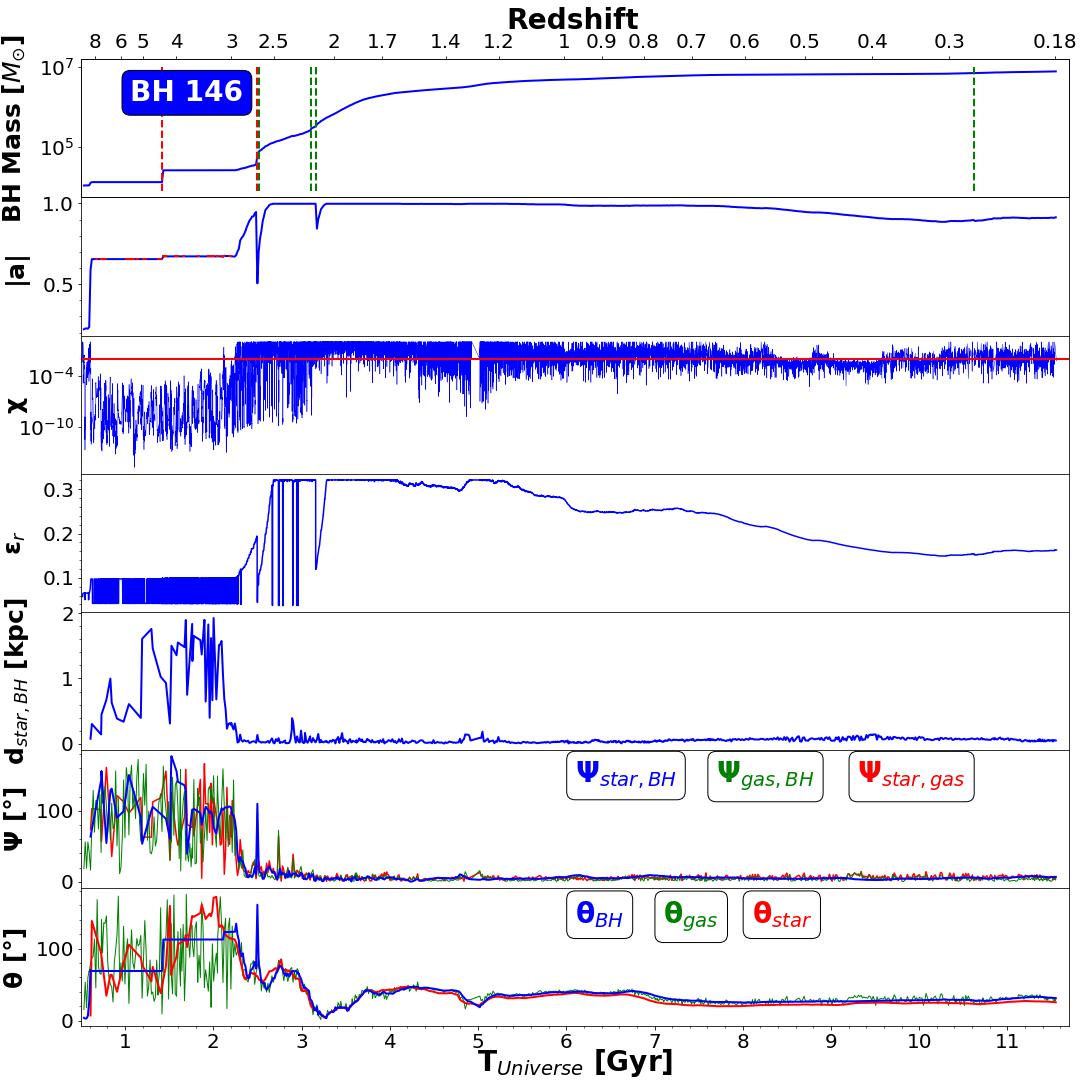}}

\rotatebox{0}{\includegraphics[width=\columnwidth]{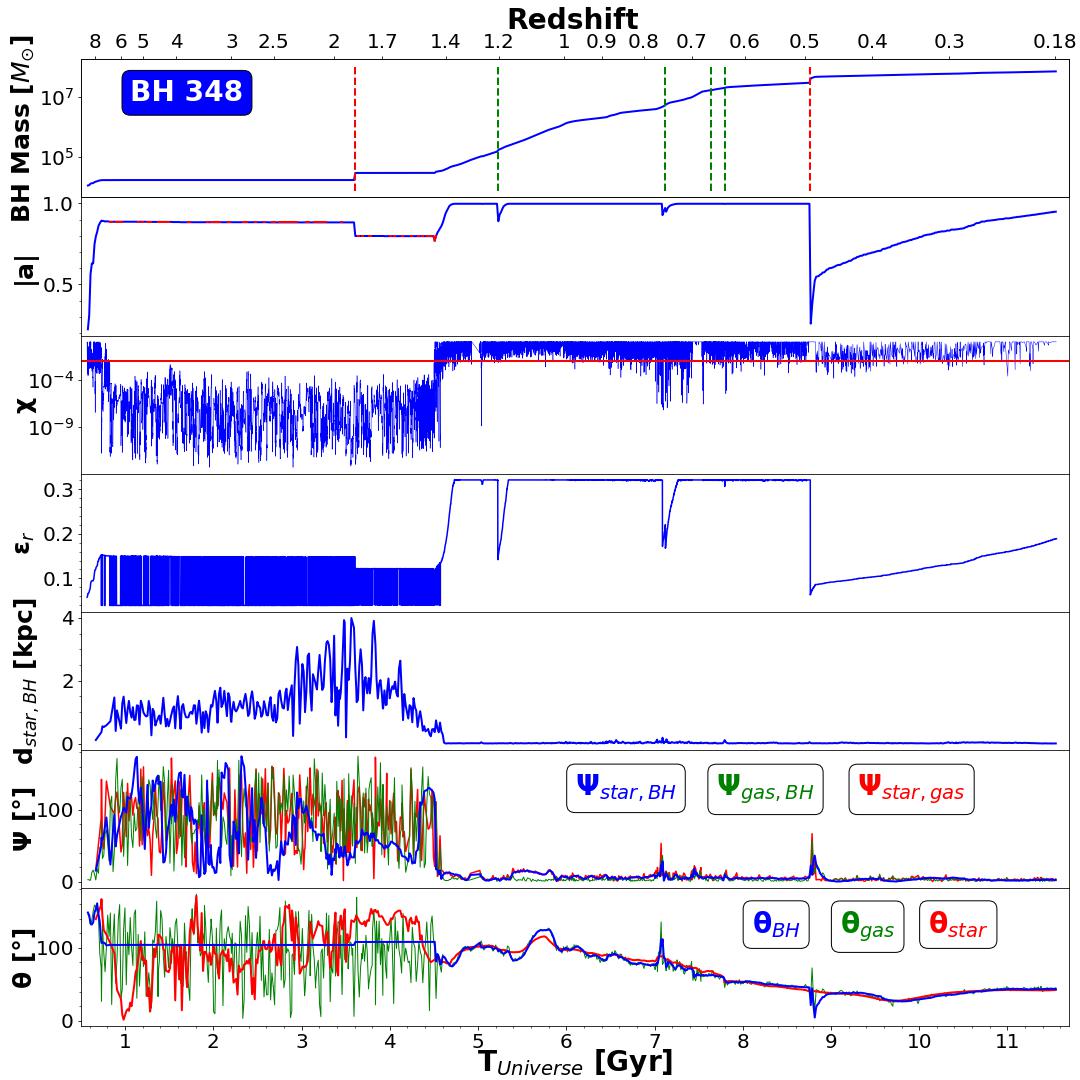}}
\rotatebox{0}{\includegraphics[width=\columnwidth]{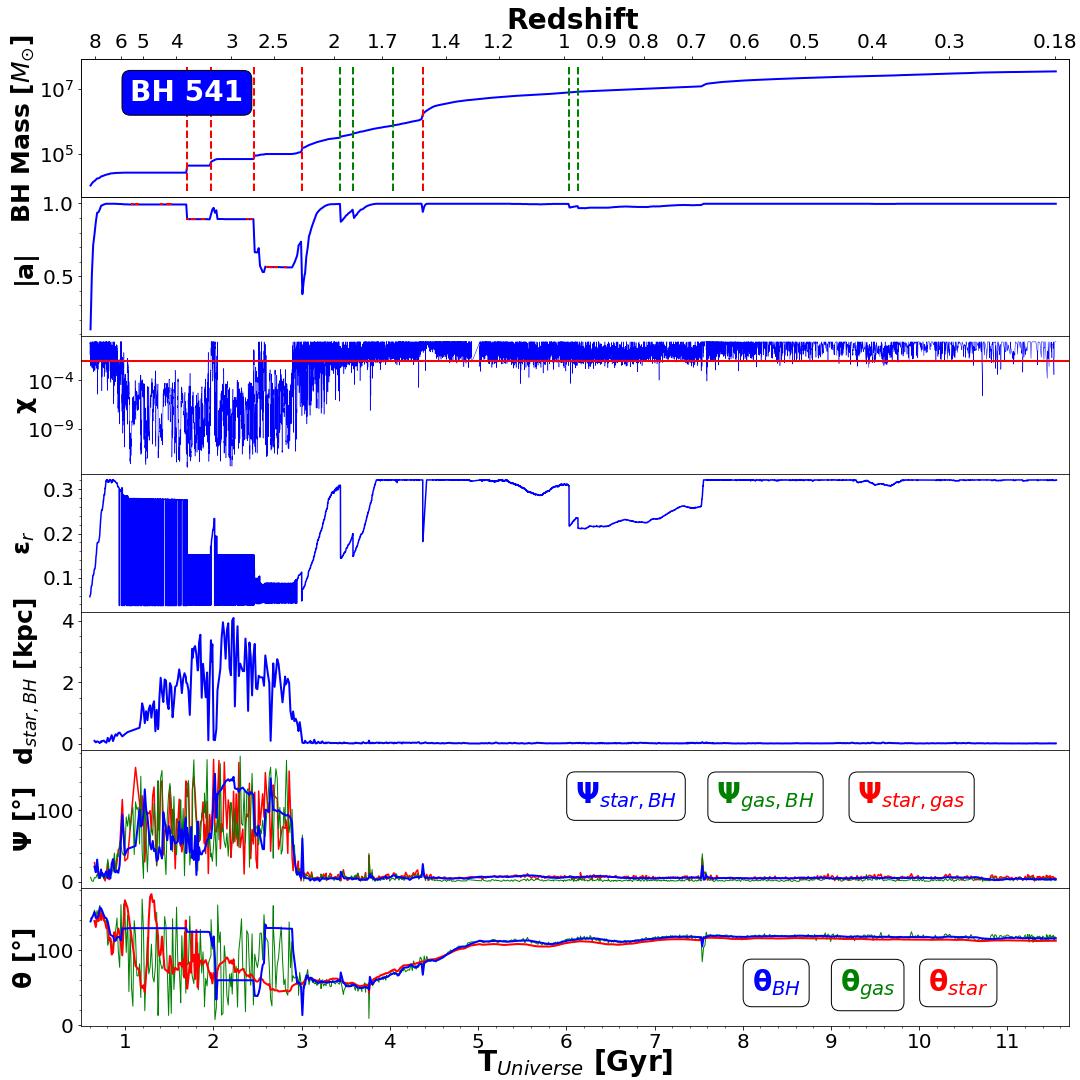}}
\caption{Cosmic evolution of BH-G$648$ (upper left panel),
BH-146 (upper right), BH-$348$ (lower left panel), and BH-$541$ (lower right panel).
BH-G$648$ is a black hole from a \galacticas zoom. The other BHs are extracted
from \nhs and are hosted by DM halos that have less than 0.1\% of their
mass composed by low resolution particles (contamination).
}
\label{fig_appendix2}
\end{center}
 \end{figure*}

\begin{figure*}
\begin{center}
\rotatebox{0}{\includegraphics[width=9.1cm]{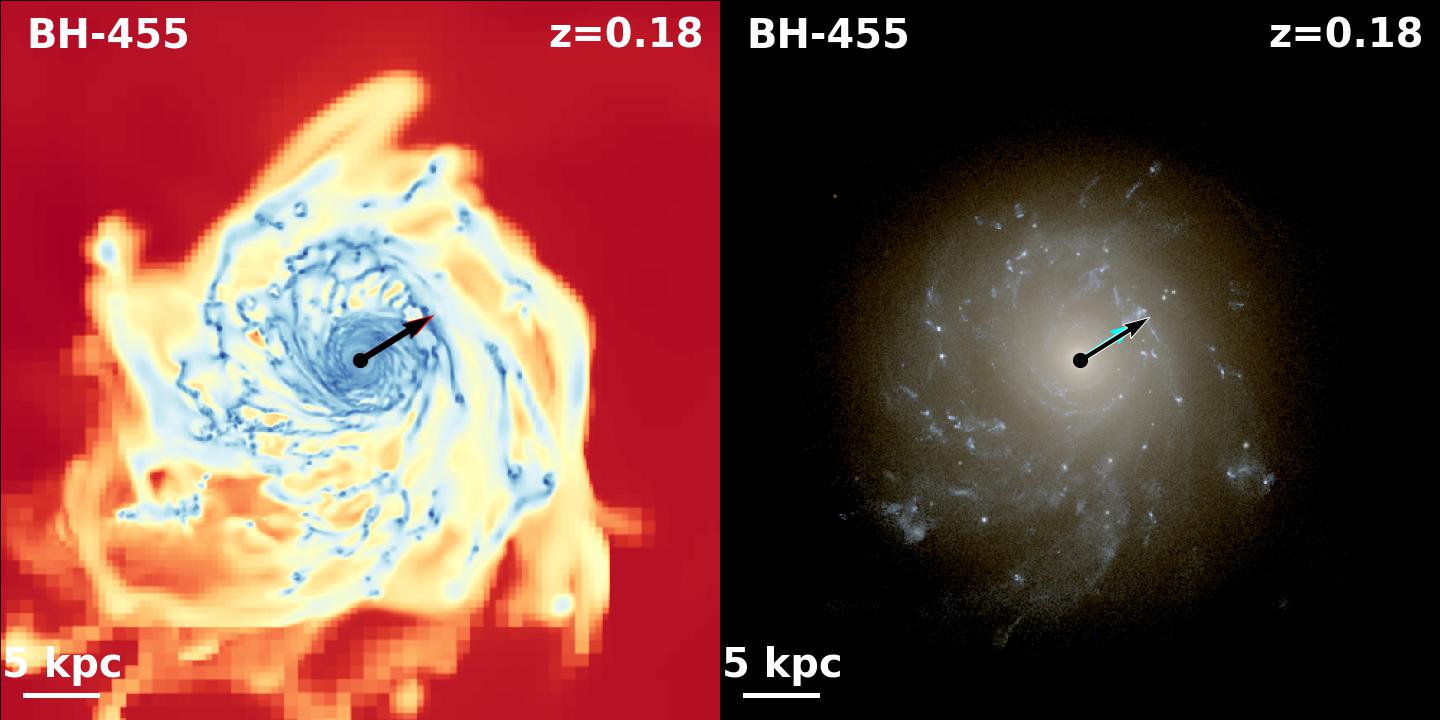}}
\rotatebox{0}{\includegraphics[width=9.1cm]{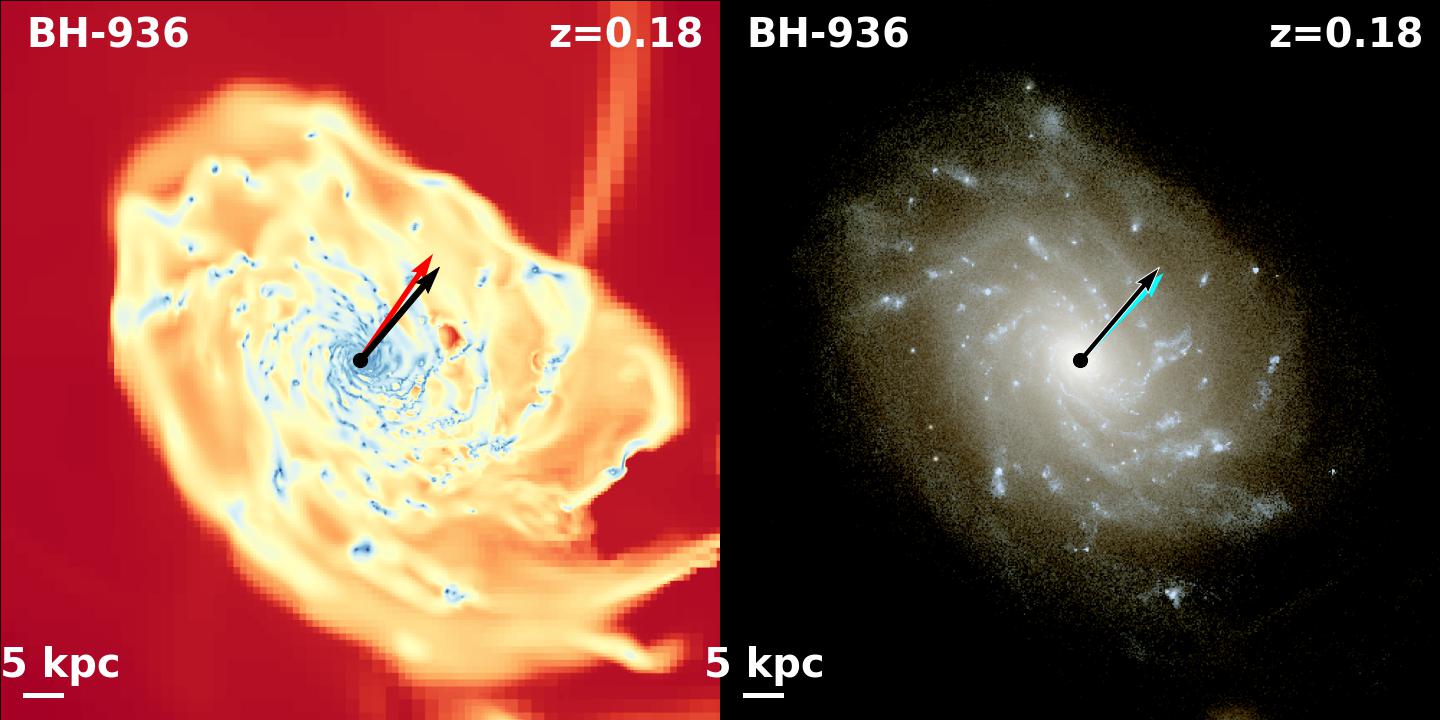}}
\rotatebox{0}{\includegraphics[width=9.1cm]{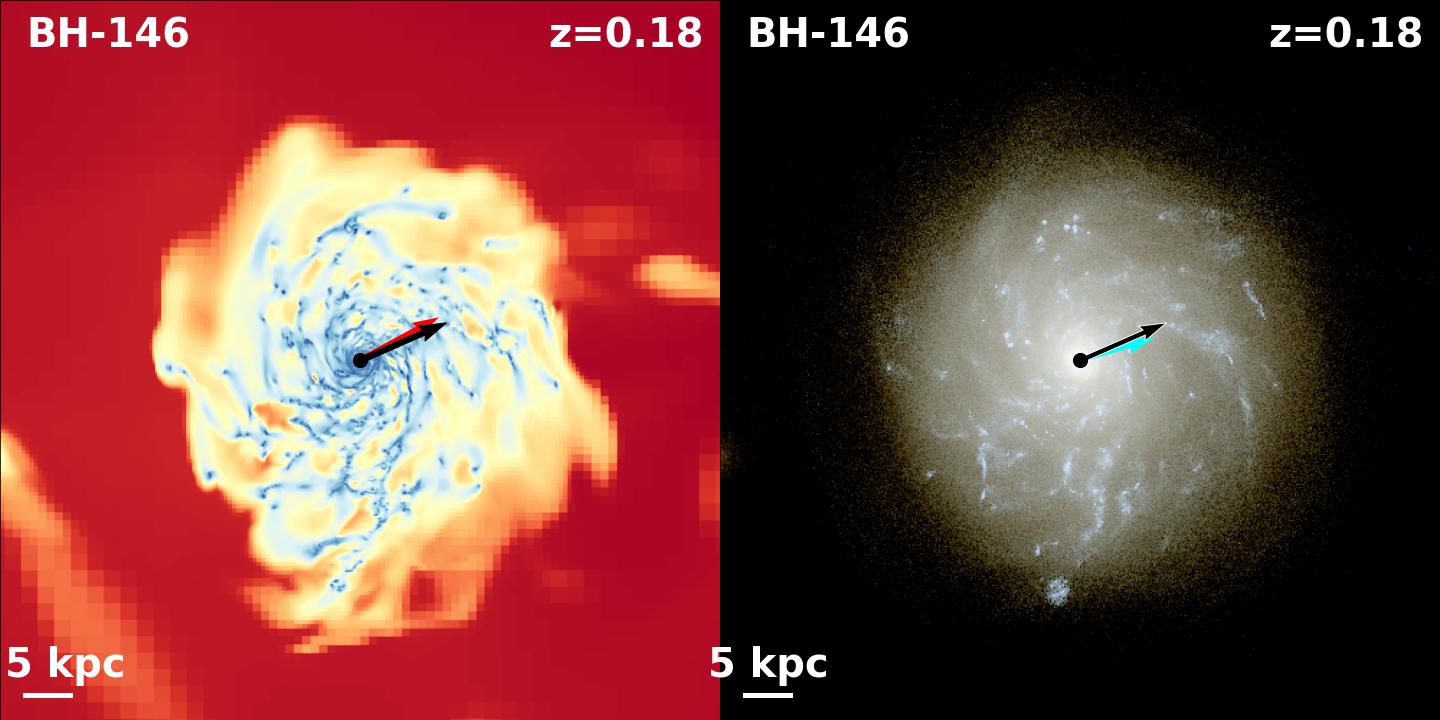}}
\rotatebox{0}{\includegraphics[width=9.1cm]{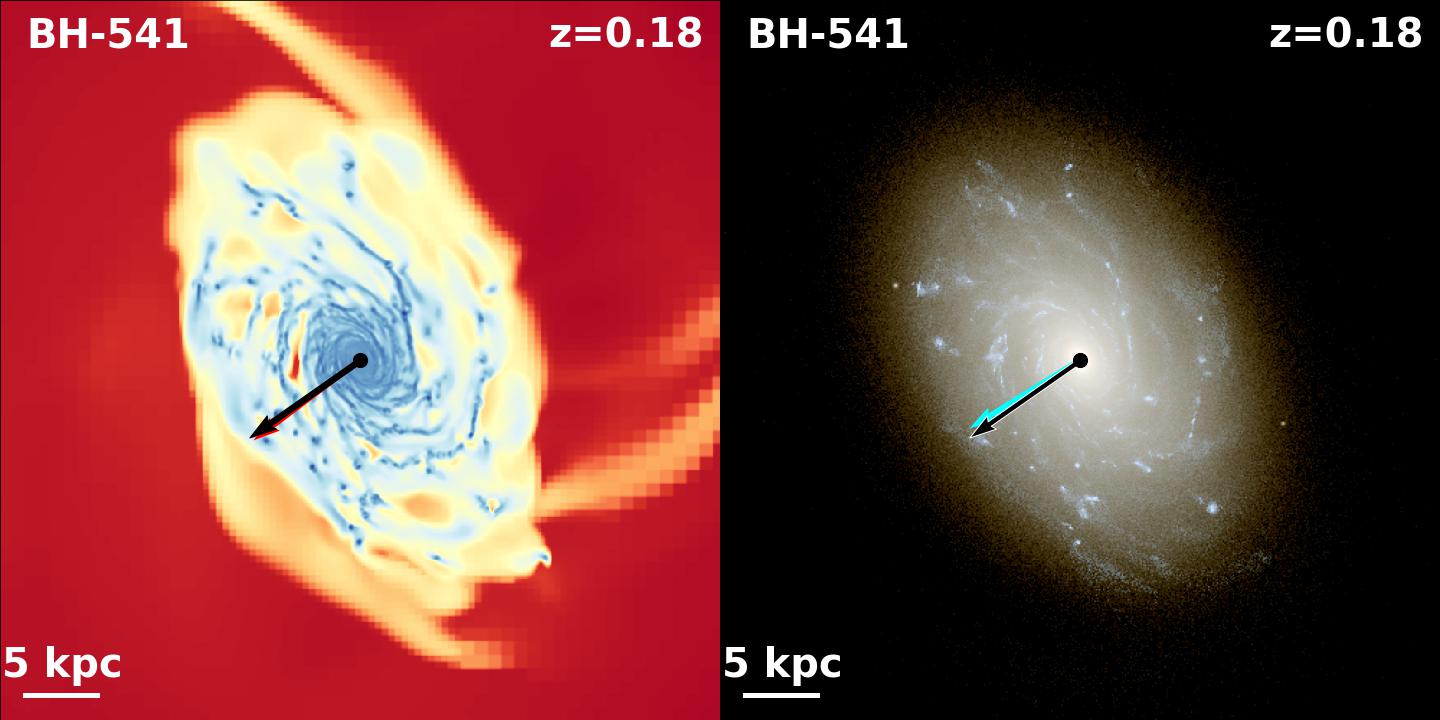}}
\rotatebox{0}{\includegraphics[width=9.1cm]{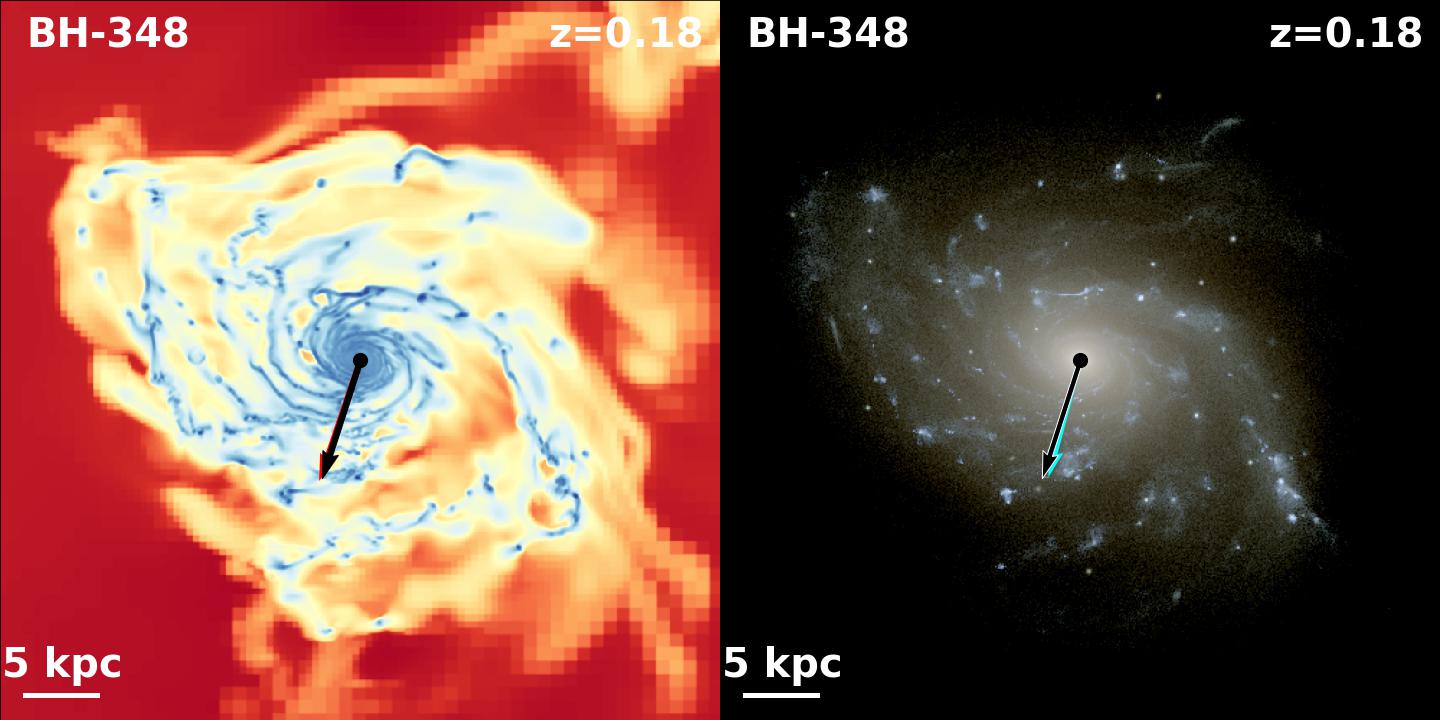}}
\rotatebox{0}{\includegraphics[width=9.1cm]{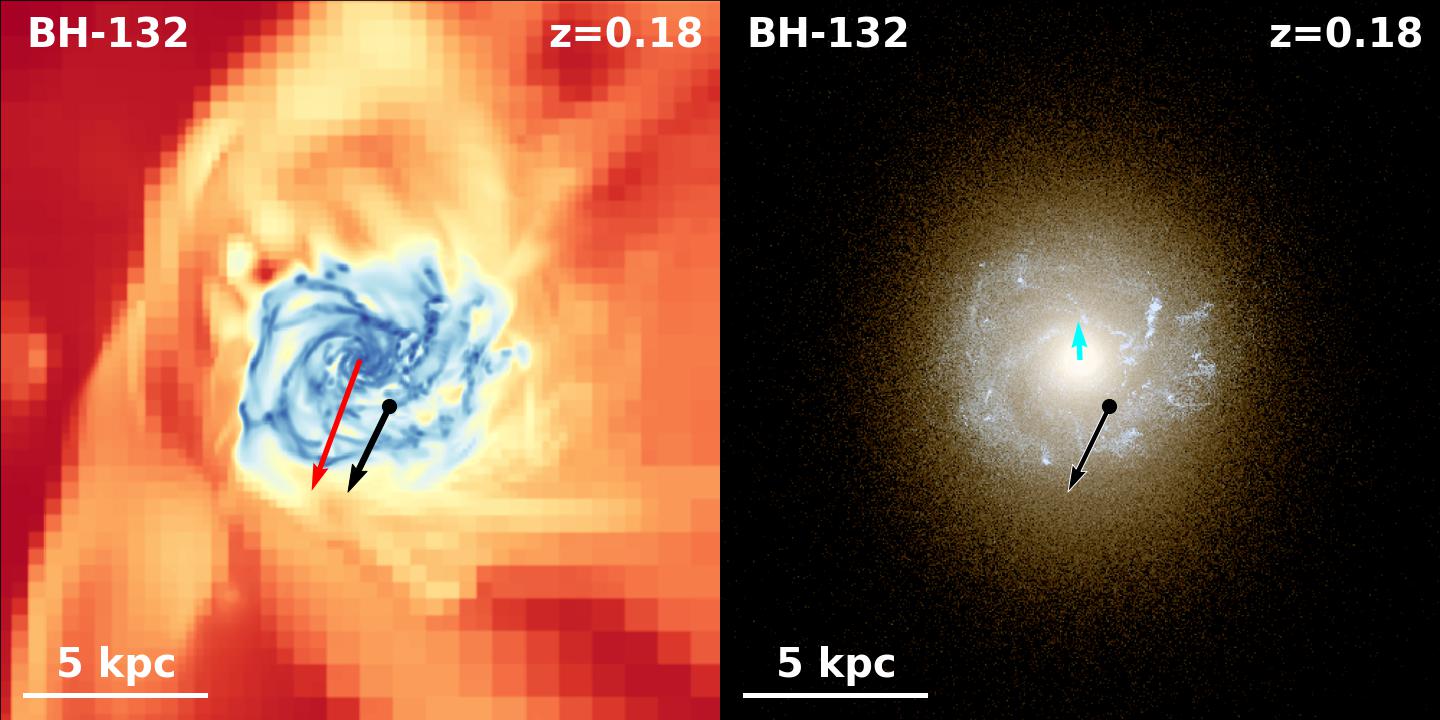}}
\caption{Projected
distribution of gas and starts (u-g-r band images) at z=0.18 of the
different BH-galaxy pairs studied in this Appendix.
}
\label{fig_appendix3}
\end{center}
 \end{figure*}

\section{2-d projected angles statistics and galaxy morphologies}
\label{appendix2}

In order to help the comparison with observational analysis,
we derive some statistics on the 2-d projected angles
depending on the galaxy morphology. 
We select from our BH-galaxy sample at $z=0.18$, 
either galaxies with $V/\sigma$<0.4 (i.e., spheroidal-dominated galaxies, see the definition
at the end of section~\ref{sec:dependence})  or $V/\sigma$>0.6 (i.e., disk-dominated galaxies). 

We then repeat our Monte-Carlo method and derive the
new distributions of $\lambda$
with and without
specifying the sense of the BH vector
in Figs.~\ref{fig_mis2} and  ~\ref{fig_mis3}.
Nevertheless, the results are quite similar
to those of presented Fig.~\ref{fig_mis1} without any 
constraint on galaxy morphology.



\begin{figure*}
\begin{center}
\rotatebox{0}{\includegraphics[width=\columnwidth]{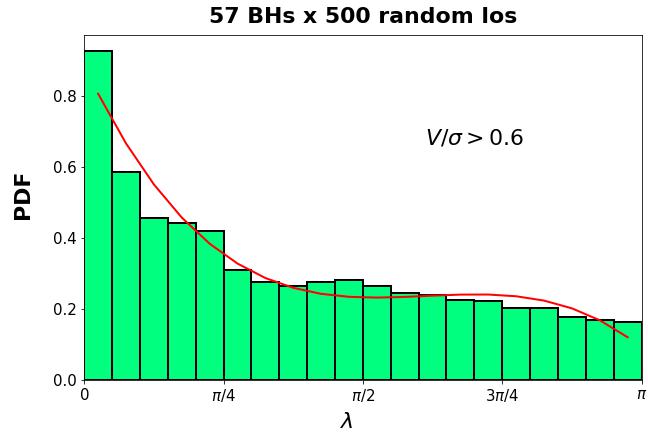}}
\rotatebox{0}{\includegraphics[width=\columnwidth]{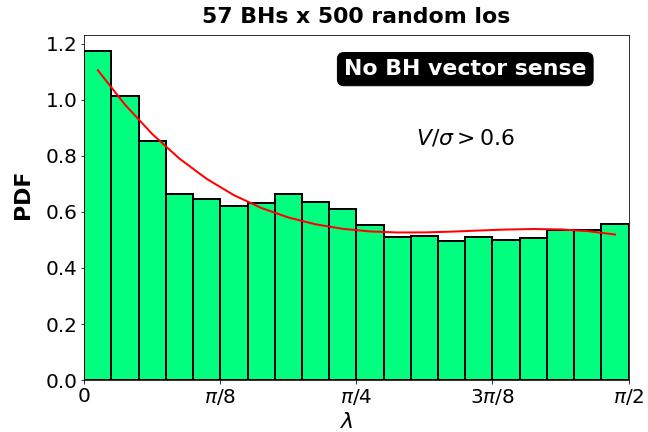}}
\rotatebox{0}{\includegraphics[width=\columnwidth]{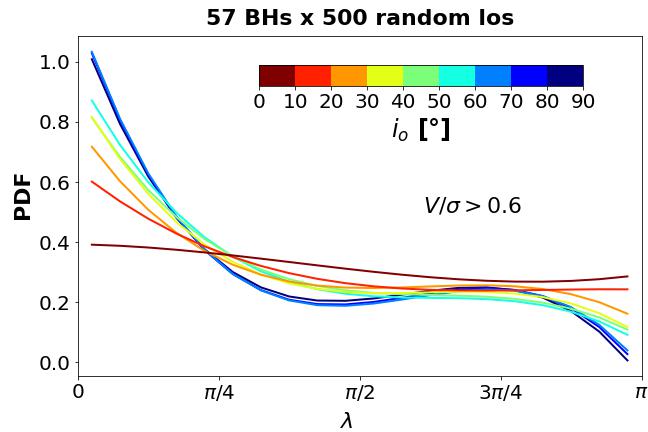}}
\rotatebox{0}{\includegraphics[width=\columnwidth]{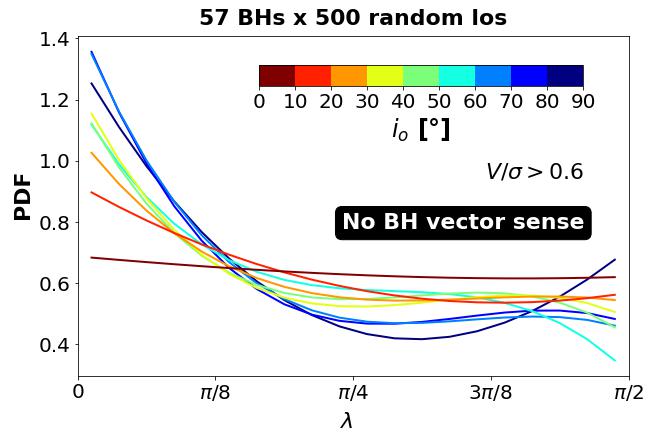}}
\caption{Same as Fig.~\ref{fig_mis1}, but selecting galaxies with 
$V/\sigma$>0.6, i.e.,
preferentially disk-dominated galaxies (and discard spheroidal-dominated galaxies).
 }
\label{fig_mis2}
\end{center}
 \end{figure*}

\begin{figure*}
\begin{center}
\rotatebox{0}{\includegraphics[width=\columnwidth]{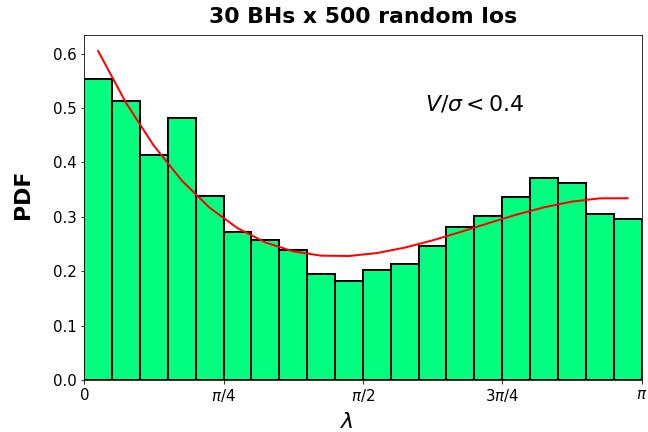}}
\rotatebox{0}{\includegraphics[width=\columnwidth]{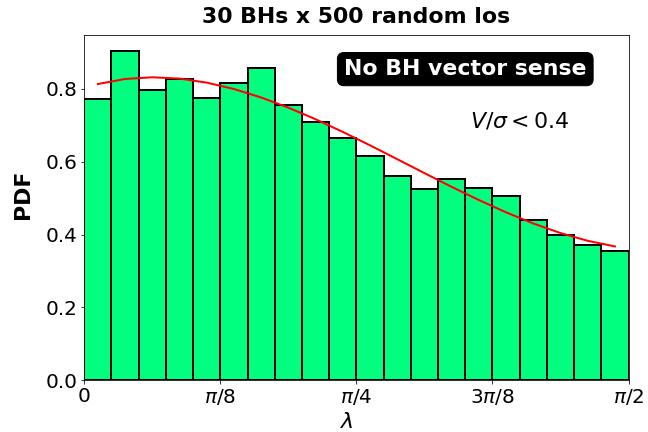}}
\rotatebox{0}{\includegraphics[width=\columnwidth]{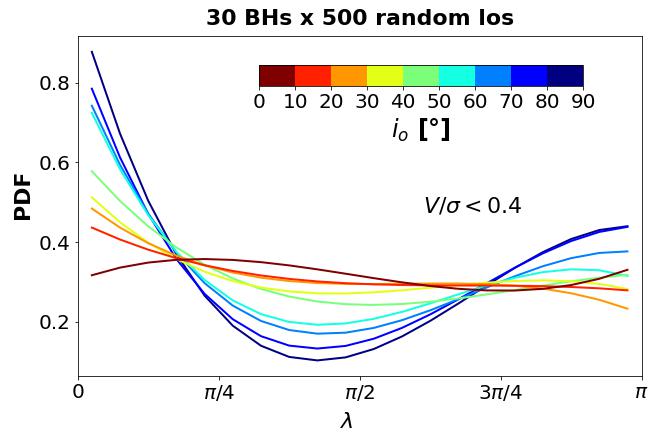}}
\rotatebox{0}{\includegraphics[width=\columnwidth]{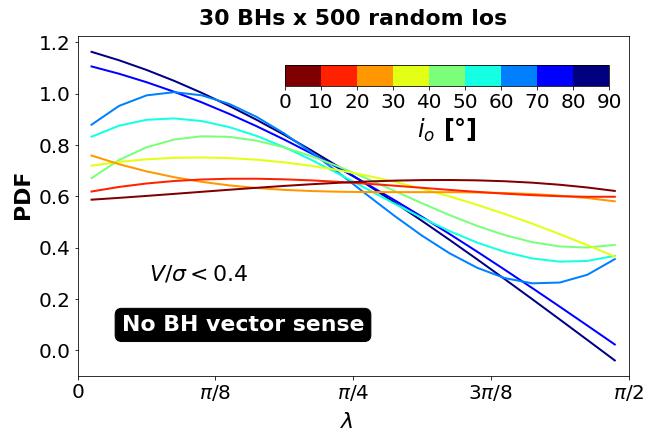}}
\caption{Same as Fig.~\ref{fig_mis1}, but selecting galaxies with 
$V/\sigma$<0.4, i.e.,
preferentially spheroidal-dominated galaxies.
 }
\label{fig_mis3}
\end{center}
 \end{figure*}

\end{document}